\documentclass[aps,prb,twocolumn,superscriptaddress,floatfix]{revtex4-2}
\usepackage{graphicx}
\usepackage{amsmath}
\usepackage{amssymb}
\usepackage{tikz}
\usepackage{braket}
\usepackage{pgfplots}
\usepackage{slashed}
\usepackage[version=4]{mhchem}
\pgfplotsset{compat=1.17}
\usepackage{subcaption}
\usetikzlibrary{shapes.geometric}
\usetikzlibrary{shadows,patterns,perspective}
\usetikzlibrary {decorations,decorations.text}
\usetikzlibrary{decorations.pathreplacing}
\usepackage{float}
\usepackage{hyperref}
\usepackage{xcolor,ulem}
\usepackage{physics}
\usepackage{mathtools}

\DeclarePairedDelimiter\fl{\lfloor}{\rfloor}
      
\definecolor{linkcolor}{RGB}{0, 0, 255}      
\definecolor{citecolor}{RGB}{0, 128, 0}     
\definecolor{urlcolor}{RGB}{255, 0, 0}       

\allowdisplaybreaks

\hypersetup{
    colorlinks=true,
    linkcolor=linkcolor,
    citecolor=citecolor,
    urlcolor=urlcolor,
    linktoc=all,  
    pdfborder={0 0 0}  
}

\DeclareCaptionJustification{justified}{\leftskip=0pt \rightskip=0pt \parfillskip=0pt plus 1fil}

\usepackage[outline]{contour}
\contourlength{1.2pt}

\usetikzlibrary{
  shapes.geometric,
  shadows,
  patterns,
  perspective,
  decorations,
  decorations.text
  }

\usetikzlibrary{decorations.markings,calc,decorations.pathreplacing}
\usetikzlibrary{positioning,arrows.meta}

\tikzset{
  baseline/.style = {line width=0.8pt, draw=black!70, line cap=round},
  tick/.style     = {line width=0.9pt, draw=black, line cap=round},
  site/.style     = {fill=black, draw=none},
  Jbond/.style    = {line width=1.4pt, draw=green!70!black, dotted, line cap=round},
  gbond/.style    = {line width=1.4pt, draw=cyan!80!black, line cap=round},
  Jlab/.style     = {font=\bfseries\footnotesize, text=green!60!black},
  glab/.style     = {font=\bfseries\footnotesize, text=cyan!80!black},
  upA/.style      = {-stealth, line width=0.5mm, draw=red!80!black},
  dnA/.style      = {-stealth, line width=0.5mm, draw=blue!80!black}
}

\newcommand{\nicechain}[2]{%
  \begin{scope}[shift={#1}]
    \def\dx{1.20}%
    \def\LarrSmall{0.85}%
    \pgfmathsetmacro{\LarrLarge}{2*\LarrSmall}%
    \def\rsiteSmall{0.10}%
    \pgfmathsetmacro{\rsiteLarge}{2*\rsiteSmall}%
    \def\ticklen{0.17}%

    \newcount\Np \Np=0
    \foreach \s [count=\j] in {#2} {\global\Np=\j}
    \newcount\N \N=\numexpr\Np-1\relax

    \ifnum\Np>1
      \draw[Jbond] ({1*\dx},0) -- ({2*\dx},0);
      \path ($({1*\dx},0)!0.5!({2*\dx},0)$) ++(0,0.24) node[Jlab]{J};
    \fi

    \ifnum\N>2
      \foreach \j in {2,...,\numexpr\N\relax}{
        \pgfmathtruncatemacro{\jp}{\j+1}
        \draw[gbond] ({\j*\dx},0) -- ({\jp*\dx},0);
        \path ($({\j*\dx},0)!0.5!({\jp*\dx},0)$) ++(0,0.20) node[glab]{g};
      }
    \fi

    \foreach \s [count=\j] in {#2}{
      \ifnum\j=1
        \edef\LarrSite{\LarrSmall}%
        \edef\rsite{\rsiteSmall}%
      \else
        \edef\LarrSite{\LarrLarge}%
        \edef\rsite{\rsiteLarge}%
      \fi

      \pgfmathsetmacro{\HL}{0.5*\LarrSite}

      \draw[tick] ({\j*\dx},-\ticklen) -- ({\j*\dx},\ticklen);
      \fill[site] ({\j*\dx},0) circle (\rsite);

      \ifnum\s=1
        \draw[upA] ({\j*\dx},-\HL) -- ++(0,\LarrSite);
      \else
        \draw[dnA] ({\j*\dx},\HL) -- ++(0,-\LarrSite);
      \fi
    }
  \end{scope}
}

\usepackage{xcolor}

\definecolor{linkcolor}{RGB}{0,0,255}      
\definecolor{citecolor}{RGB}{0,128,0}     
\definecolor{urlcolor}{RGB}{255,0,0}      

\definecolor{dy}{rgb}{0.9,0.9,0.4}
\definecolor{dr}{rgb}{0.95,0.65,0.55}
\definecolor{db}{rgb}{0.5,0.8,0.9}
\definecolor{dg}{rgb}{0.2,0.9,0.6}
\definecolor{BrickRed}{rgb}{0.8,0.3,0.3}
\definecolor{Navy}{rgb}{0.2,0.2,0.6}
\definecolor{DarkGreen}{rgb}{0.1,0.4,0.1}
\definecolor{phaseK}{HTML}{CC79A7}
\definecolor{phaseY}{HTML}{56B4E9}
\definecolor{phaseU}{HTML}{E69F00}
\definecolor{phaseF}{HTML}{23EB91}
\definecolor{phaseA}{HTML}{C25B31}

\newcommand{\mr}[1]{\mathrm{#1}}

\providecommand{\mc}[1]{\mathcal{#1}}
\providecommand{\Bar}[1]{\bar{#1}}

\begin{document}
\definecolor{dy}{rgb}{0.9,0.9,0.4}
\definecolor{dr}{rgb}{0.95,0.65,0.55}
\definecolor{db}{rgb}{0.5,0.8,0.9}
\definecolor{dg}{rgb}{0.2,0.9,0.6}
\definecolor{BrickRed}{rgb}{0.8,0.3,0.3}
\definecolor{Navy}{rgb}{0.2,0.2,0.6}
\definecolor{DarkGreen}{rgb}{0.1,0.4,0.1}

\title{Boundary phases and thermodynamics of the Kondo spin-$s$  chain: \\ from overscreened Kondo to boundary-bound states  }

\author{Abay Zhakenov}
\email{abay.zhakenov@rutgers.edu}
\affiliation{Department of Physics and Astronomy, Center for Materials Theory, Rutgers University,
Piscataway, New Jersey 08854, United States of America} 
\author{Pradip Kattel}
\affiliation{Department of Quantum Matter Physics, University of Geneva, Quai Ernest-Ansermet 24, 1211 Geneva, Switzerland}
 
\author{Andreas Gleis}
\affiliation{Department of Physics and Astronomy, Center for Materials Theory, Rutgers University,
Piscataway, New Jersey 08854, United States of America} 
\author{Natan Andrei}
\affiliation{Department of Physics and Astronomy, Center for Materials Theory, Rutgers University,
Piscataway, New Jersey 08854, United States of America}

\begin{abstract}

We study a spin-$\frac12$ impurity coupled to the boundary of a strongly correlated spin-$s$ Takhtajan--Babujian chain, an integrable model whose low-energy physics is described by a perturbed $SU(2)_{2s}$ Wess--Zumino--Witten conformal field theory. While boundary conformal field theory determines the low-energy universality class of the weak-coupling regime, exact Bethe Ansatz methods reveal a sequence of boundary quantum phase transitions in which impurity-bound states emerge and reorganize the Hilbert space into multiple excitation towers built on distinct boundary configurations. This tower restructuring provides the organizing principle for a rich boundary phase diagram extending beyond the conventional Kondo regime. Weak antiferromagnetic coupling realizes the overscreened $2s$-channel Kondo universality class, whereas stronger couplings generate localized boundary modes and qualitatively new screening mechanisms. To describe the resulting thermodynamics, we develop a generalized thermodynamic Bethe Ansatz framework that captures the multi-tower structure across all regimes. The impurity entropy reproduces the boundary conformal field theory prediction in the overscreened Kondo regime but develops pronounced nonmonotonic temperature dependence once boundary-bound states appear, in quantitative agreement with large-scale finite-temperature matrix-product-operator simulations. Complementary dynamical calculations reveal sharp threshold features in the impurity spectral function that directly track the underlying tower structure. Together, boundary conformal field theory, exact Bethe Ansatz, generalized thermodynamic Bethe Ansatz, and tensor-network simulations provide a unified description of impurity screening, boundary-bound-state formation, and excitation-tower reconstruction in a correlated spin-$s$ chain.
\end{abstract}

\maketitle

\section{Introduction}

Quantum impurities provide a  powerful probe of strongly correlated systems~\cite{hewson1997kondo,kondo2012physics,  andrei1983solution, tsvelick1984solution}. Their coupling to an extended many-body environment can generate emergent energy scales, nontrivial screening mechanisms, and boundary quantum phase transitions~\cite{andrei1981scales, kattel2023kondo, kattel2025anisHeisenberg}.  More generally, impurity problems provide a framework for understanding how local degrees of freedom interact with collective many-body excitations and how boundary perturbations influence interacting quantum systems. Integrable spin chains provide a rare setting in which these phenomena can be studied nonperturbatively across all energy scales.

Here, we consider quantum impurities coupled to generalized spin chains.
Spin chains constitute paradigmatic models for exploring strong correlation effects in low-dimensional quantum systems. Despite their apparent simplicity, they capture many quintessential many-body phenomena such as 
quantum criticality~\cite{pfeuty1970one,nomura1994critical}, anomalous transport~\cite{gopalakrishnan2023anomalous,gopalakrishnan2019kinetic,ljubotina2019kardar,scheie2021detection}, emergent field theories~\cite{giamarchi2003quantum,haldane1983continuum,affleck1988field}, nontrivial entanglement structure~\cite{osterloh2002scaling,laflorencie2008kondo, amico2008entanglement},
the Kondo  effect~\cite{andrei1984heisenberg,wang1997exact,frahm1997open,laflorencie2008kondo,gaines2025spin,kattel2023kondo,zhakenov2025thermodynamics}, RKKY interactions~\cite{bayat2012entanglement}, and quantum phase transitions~\cite{pfeuty1970one,giamarchi2003quantum,franchini2017introduction}. Their relative tractability allows for a broad range of analytical and numerical approaches, including bosonization~\cite{giamarchi2003quantum,gogolin2004bosonization,von1998bosonization}, conformal field theory~\cite{PhysRevB.36.5291,lukyanov1998low}, and renormalization-group techniques~\cite{haldane1983continuum,affleck1988field,cabra2008field}. In certain remarkable cases, the underlying models are exactly solvable via the Bethe Ansatz~\cite{bethe1931eigenwerte,yang1966one,yang1966two,yang1966three,baxter2016exactly}, allowing exact access to their thermodynamic~\cite{takahashi1999thermodynamics} and dynamical properties~\cite{caux2003dynamical,caux2005computation,pozsgay2021integrable}. Experimentally, spin-chain behavior is realized in quasi-one-dimensional systems such as \ce{SrCo2V2O8}, \ce{KCuF3}, \ce{CuCl2*2NC5H5}, \ce{Cu(NH3)4SO4*H2O} \textit{etc.} \cite{scheie2022quantum,wang2018experimental,steiner1976theoretical,nagler1991spin} and, more recently, in ultracold-atom platforms, where the spin, interactions, and lattice geometry can be engineered with a high degree of control, providing a promising route toward realizing higher-spin one-dimensional quantum magnets and testing predictions of integrable models~\cite{hild2014far,mogerle2025spin}.

In spin chains with open boundary conditions, the boundaries naturally act as quantum impurities, providing a controlled setting for investigating impurity screening and boundary critical phenomena. The edge spin can undergo Kondo-like screening, characterized by a length scale $l_K$ set by the Kondo temperature $T_K$~\cite{laflorencie2008kondo,kattel2023kondo,zhakenov2025thermodynamics}. Recent Bethe Ansatz analyses have shown that varying the antiferromagnetic boundary coupling induces boundary quantum phase transitions, interpolating between many-body Kondo-like screening and effectively single-particle screening mediated by an edge-localized bound mode. When the boundary coupling is ferromagnetic, the impurity remains unscreened at low temperatures, whereas screening arises only in the strong-coupling regime in high-energy excited states~\cite{kattel2023kondo, zhakenov2025thermodynamics}. These boundary quantum phase transitions are accompanied by a reorganization of the Hilbert space into distinct excitation towers built on different boundary configurations. As the boundary coupling is varied, new towers emerge, and their relative energetic hierarchy changes across the phase diagram, leading to qualitative changes in impurity thermodynamics and dynamics. Recent work further showed that, in the spin-$\frac{1}{2}$ Heisenberg chain, the boundary-bound-state transition is accompanied by the emergence of an exactly conserved quasi-local edge mode, providing an operator-space interpretation of the underlying tower restructuring~\cite{prosen2026quasi}.

In this work, we shall study the effects of a spin-$\frac{1}{2}$ impurity attached to an integrable spin-$s$ chain, generalizing the model of Ref.~\cite{zhakenov2025thermodynamics}. The system is described by the Hamiltonian
\begin{equation}
    H = J \vec{s}_0 \cdot \vec{S}_1 
    + g \sum_{i=1}^{N-1} Q_{2s} \left( \vec{S}_i \cdot \vec{S}_{i+1} \right),
    \label{ModelHam}
\end{equation}
where \( Q_{2s}(\vec{S}_i \cdot \vec{S}_{i+1}) \) is the Babujian polynomial of order \( 2s \) in the SU(2)-invariant scalar product~\cite{TAKHTAJAN1982479,BABUJIAN1982479,BABUJIAN1983317} \footnote{Setting $s=1/2$ gives the different impurity coupling from the one in \cite{zhakenov2025thermodynamics}: $J_\mathrm{Heisenberg}=J/4$. This happens due to the choice of the interaction terms written in this paper in terms of spin operators while in \cite{zhakenov2025thermodynamics} the Hamiltonian was in terms of the Pauli matrices.}. The first few Babujian polynomials are \footnote{The Babujian polynomial used in this paper comes with an additional factor of 4 when compared with the original paper \cite{BABUJIAN1983317}: $Q_{2s}(x) = 4 Q_{2s}^{\mathrm{original}}(x)$. This choice was motivated by more convenient forms of the interaction for spins $s=1/2, 1$.}:
\begin{equation}
    \begin{aligned}
    Q_1(x) &= 3+4x\;,\\
    Q_2(x) &= 6+x-x^2\;,\\
    Q_3(x) &= \frac{35}{6} -\frac{1}{4}x + \frac{2}{27}x^2 + \frac{4}{27}x^3\;.
\end{aligned}
\end{equation}

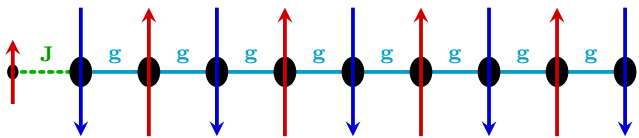
\begin{figure}[H]
    \centering
    \begin{tikzpicture}
      \begin{scope}[xscale=0.75]
        \nicechain{(0,0)}{1,-1,1,-1,1,-1,1,-1,1,-1}{}{ }
      \end{scope}
    \end{tikzpicture}
    \caption{Spin-$s$ chain with a spin-$\frac12$ boundary impurity. The impurity coupling $J$ is shown as a green dotted bond, while the bulk coupling $g$ is shown as cyan solid bonds.}
    \label{fig:spinchain}
\end{figure}

The case of periodic boundary conditions was previously studied in Ref.~\cite{Schlottmann_1991}, where integrability is preserved only for a fine-tuned impurity interaction term. In contrast, open boundary conditions allow for an exact solution for the simple $H_\mathrm{imp}=J\vec{s}_0 \cdot \vec{S}_1$ with arbitrary impurity coupling $J$~\cite{PhysRevB.60.6594} when either the impurity spin $s'=1/2$ or the bulk spin $s=1/2$. Here, we focus on a spin-$\frac{1}{2}$ impurity and show that impurity-bound states give rise to a rich boundary-phase structure. The impurity induces a hierarchy of boundary-bound states that reorganize the Hilbert space into distinct towers of bulk excitations, each built upon a different boundary configuration. The emergence of these boundary states leads to a nonmonotonic temperature dependence of the impurity entropy, which we benchmark using tensor network methods. Although we restrict our analysis to $s' = \frac{1}{2}$, the tower-TBA framework developed here can be straightforwardly generalized to arbitrary impurity spin.

When the boundary spin is antiferromagnetically coupled to the bulk ($J>0$), the impurity is overscreened throughout the antiferromagnetic regime. The nature of the screening, however, depends on the ratio $J/g$. More precisely, in the weak–antiferromagnetic regime (AF1): $J/g\in (0, \frac{4}{s(s+1)})$, the impurity is overscreened by a many-body Kondo cloud formed from the surrounding spin-$s$ bulk degrees of freedom. Increasing the boundary coupling leads to the formation of localized bound modes (phases AF2 and AF3 corresponding to $J/g\in (\frac{4}{s(s+1)}, \frac{16}{4s+1})$ and $J/g >\frac{16}{4s+1}$), thereby transforming the overscreening mechanism from one mediated by an extended many-body Kondo cloud to one governed by localized boundary-bound states.

For ferromagnetic couplings (\(J<0\)), two distinct regimes (F1 and F2) appear, characterized by qualitatively different impurity configurations. While both phases consist of three towers, the bound string tower is gapped in the F2 regime, leading to characteristic peaks in the impurity spectral function $A(\omega)$, absent in the F1, where the base states of all three towers have the same energy.

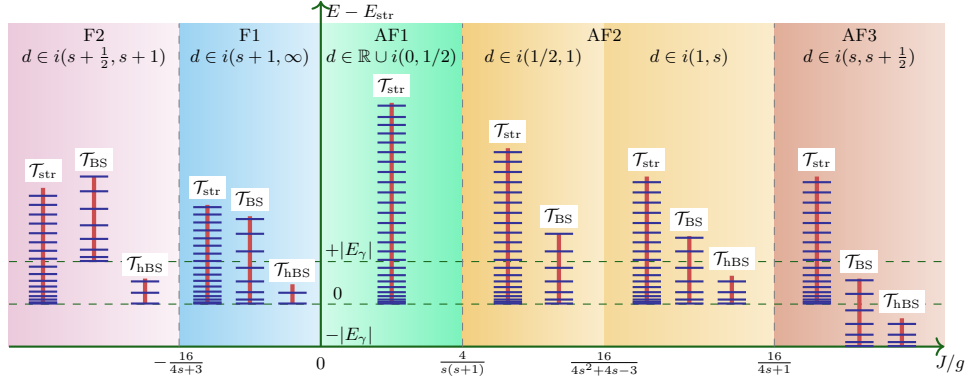
\begin{figure*}[t]
    \centering
        \begin{tikzpicture}[scale=0.75, every node/.style={scale=0.75}]

\shade[left color=phaseK!40, right color=phaseK!10] (-0.5,0) rectangle (2.5,5.7);
\shade[left color=phaseY!60, right color=phaseY!20] (2.5,0) rectangle (5,5.7);
\shade[left color=phaseF!20, right color=phaseF!60] (5,0) rectangle (7.5,5.7);
\shade[left color=phaseU!50, right color=phaseU!20] (7.5,0) rectangle (10,5.7);
\shade[left color=phaseU!40, right color=phaseU!20] (10,0) rectangle (13,5.7);
\shade[left color=phaseA!50, right color=phaseA!20] (13,0) rectangle (16,5.7);

\draw [thick,DarkGreen,->] (-0.5,0)--(16,0);
\draw [thick,DarkGreen,<-] (5,6.1)--(5,0);
\node at (5.7,5.9) {\small $E - E_\mathrm{str}$};
\node at (16.1,-0.3) {\small $J/g$};
\node at (2.5,-0.3) {\small $-\frac{16}{4s+3}$};
\node at (5.0,-0.3) {\small $0$};
\node at (7.5,-0.3) {\small $\frac{4}{s(s+1)}$};
\node at (10,-0.3) {\small $\frac{16}{4s^2+4s-3}$};
\node at (13,-0.3) {\small $\frac{16}{4s+1}$};

\draw [dashed,DarkGreen] (-0.5,0.75)--(16,0.75);
\node at (5.3,0.95) {\small 0};
\draw [dashed,DarkGreen] (-0.5,1.5)--(16,1.5);
\node at (5.5,1.7) {\small $+|E_\gamma|$};
\node at (5.5,0.2) {\small $-|E_\gamma|$};

\foreach \x in {2.5,7.5,13}
  \draw[dashed,gray] (\x,0)--(\x,5.7);

\node at (1,5.5) {\small F2};
\node at (1,5.1) {\small $d\in i(s+\tfrac12,s+1)$};
\node at (3.75,5.5) {\small F1};
\node at (3.75,5.1) {\small $d \in i(s+1,\infty$)};
\node at (6.25,5.5) {\small AF1};
\node at (6.25,5.1) {\small $d\in\mathbb{R} \cup i(0,1/2)$};
\node at (10,5.5)  {\small AF2};
\node at (8.75,5.1)  {\small $d\in i(1/2,1)$};
\node at (11.5,5.1)  {\small $d\in i(1,s)$};
\node at (14.5,5.5)  {\small AF3};
\node at (14.5,5.1)  {\small $d\in i(s,s+\tfrac12)$};

\node [fill=white,inner sep=2pt] at (6.25,4.6) {$\mathcal{T}_\mathrm{str}$};
\node [fill=white,inner sep=2pt] at (9.2,2.3) {$\mathcal{T}_\mathrm{BS}$};
\node [fill=white,inner sep=2pt] at (0.1,3.1) {$\mathcal{T}_\mathrm{str}$};
\node [fill=white,inner sep=2pt] at (3,2.8) {$\mathcal{T}_\mathrm{str}$};
\node [fill=white,inner sep=2pt] at (1,3.3) {$\mathcal{T}_\mathrm{BS}$};
\node [fill=white,inner sep=2pt] at (8.3,3.8) {$\mathcal{T}_\mathrm{str}$};
\node [fill=white,inner sep=2pt] at (3.75,2.6) {$\mathcal{T}_\mathrm{BS}$};
\node [fill=white,inner sep=2pt] at (1.9,1.45) {$\mathcal{T}_\mathrm{hBS}$};
\node [fill=white,inner sep=2pt] at (4.5,1.35) {$\mathcal{T}_\mathrm{hBS}$};
\node [fill=white,inner sep=2pt] at (12.25,1.55) {$\mathcal{T}_\mathrm{hBS}$};
\node [fill=white,inner sep=2pt] at (10.75,3.3) {$\mathcal{T}_\mathrm{str}$};
\node [fill=white,inner sep=2pt] at (11.5,2.25) {$\mathcal{T}_\mathrm{BS}$};
\node [fill=white,inner sep=2pt] at (15.25,0.8) {$\mathcal{T}_\mathrm{hBS}$};
\node [fill=white,inner sep=2pt] at (13.75,3.3) {$\mathcal{T}_\mathrm{str}$};
\node [fill=white,inner sep=2pt] at (14.5,1.5) {$\mathcal{T}_\mathrm{BS}$};

\foreach \x/\ymin/\ymax in {0.1/0.75/2.8, 1/1.5/3, 1.9/0.75/1.2,
3/0.75/2.5, 3.75/0.75/2.3, 4.5/0.75/1.1,
    6.25/0.75/4.3,
    9.2/0.75/2, 8.3/0.75/3.5, 10.75/0.75/3, 11.5/0.75/1.95, 12.25/0.75/1.25, 13.75/0.75/3, 14.5/0/1.2, 15.25/0/0.5}
  \draw[ultra thick,BrickRed] (\x,\ymin)--(\x,\ymax);

\newcommand{\plotlevelszero}[2]{
  \foreach \e in {#2} {
    \draw[thick,Navy] (#1-0.25,\e+0.75)--(#1+0.25,\e+0.75);
  }
}
\newcommand{\plotlevelsPlus}[2]{
  \foreach \e in {#2} {
    \draw[thick,Navy] (#1-0.25,\e+1.5)--(#1+0.25,\e+1.5);
  }
}

\newcommand{\plotlevelsMinus}[2]{
  \foreach \e in {#2} {
    \draw[thick,Navy] (#1-0.25,\e)--(#1+0.25,\e);
  }
}
\plotlevelszero{6.25}{0.01,0.04,0.08,0.14,0.21,0.30,0.40,0.51,0.64,0.78,0.93,1.09,1.26,
1.43,1.61,1.79,1.97,2.15,2.33,2.51,2.68,2.85,3.01,3.16,3.30,3.5}

\plotlevelszero{9.2}{0.01,0.08,0.21,0.40,0.64,0.93,1.24}
\plotlevelszero{8.3}{0.01,0.04,0.08,0.14,0.21,0.30,0.40,0.51,0.64,0.78,0.93,1.09,1.26,
1.43,1.61,1.79,1.97,2.15,2.33,2.51,2.68}

\plotlevelszero{11.5}{0.01,0.08,0.21,0.40,0.64,0.93,1.17}
\plotlevelszero{12.25}{0.01,0.08,0.21,0.40}
\plotlevelszero{10.75}{0.01,0.04,0.08,0.14,0.21,0.30,0.40,0.51,0.64,0.78,0.93,1.09,1.26,
1.43,1.61,1.79,1.97,2.15}

\plotlevelsMinus{14.5}{0.01,0.08,0.21,0.40,0.64,0.93,1.17}
\plotlevelsMinus{15.25}{0.01,0.08,0.21,0.40}
\plotlevelszero{13.75}{0.01,0.04,0.08,0.14,0.21,0.30,0.40,0.51,0.64,0.78,0.93,1.09,1.26,
1.43,1.61,1.79,1.97,2.15}

\plotlevelszero{0.1}{0.01,0.03,0.08,0.14,0.22,0.31,0.41,0.53,0.66,0.80,0.94,1.09,1.25,1.42,1.58,1.75,1.91}
\plotlevelsPlus{1}{0.01,0.08,0.21,0.40,0.64,0.93,1.24, 1.5}
\plotlevelszero{1.9}{0.01,0.2, 0.4}

\plotlevelszero{3}{0.01,0.03,0.08,0.14,0.22,0.31,0.41,0.52,0.64,0.76,0.89,
1.03,1.16,1.30,1.44,1.58,1.71}
\plotlevelszero{3.75}{0.01,0.08,0.21,0.40,0.64,0.93,1.24, 1.5}
\plotlevelszero{4.5}{0.01,0.2}

\end{tikzpicture}
    \caption{Phase diagram of the spin-$s$ Takhtajan--Babujian chain with a spin-$\tfrac12$ boundary impurity [Hamiltonian Eq.~\eqref{ModelHam}]. In the antiferromagnetic regime (\(J/g>0\)), three distinct phases (AF1–AF3) appear, corresponding to overscreened, partially localized, and locally bound impurity screening. Each regime features characteristic excitation towers—$\mathcal{T}_\mathrm{str}$, $\mathcal{T}_\mathrm{BS}$, and $\mathcal{T}_\mathrm{hBS}$—built on distinct boundary configurations. The ferromagnetic regimes (F1, F2) exhibit analogous multi-tower structures with inverted hierarchy. The boundary coupling is encoded by the impurity parameter $d$, or equivalently $\gamma=-id$, through $J/g=16/[(1+2s)^2+4d^2]$ [Eq.~\eqref{Jgamma}]. Here $d$ is real only within the weak antiferromagnetic coupling phase AF1; for both stronger antiferromagnetic and ferromagnetic coupling  $d$ is purely imaginary, so that $\gamma$ is real. A boundary-bound state (boundary string) appears once $\gamma>\tfrac12$, but the boundary gap $|E_\gamma|$ that lifts one tower above the others opens only in the gapped phases AF3 and F2 [Eq.~\eqref{eq:Egamma}]. The vertical axis shows the energy of each tower's spectrum measured from the string tower $\mathcal{T}_\mathrm{str}$. A quantitative companion listing the coupling ranges, residual entropies, and thermodynamic/dynamical signatures of each phase is given in Table~\ref{tab:overview}.}
    \label{fig:PD}
\end{figure*}

The excitation spectrum thus splits into towers—families of bulk excitations built on distinct boundary base states. There is a single tower in the AF1 regime (Kondo overscreening), two in the first part of AF2, and three in all remaining regimes.
Across these transitions, the restructuring of the excitation towers leaves characteristic fingerprints in both thermodynamic and dynamical observables. The splitting into towers manifests directly in the impurity entropy, defined as the difference between the entropies of the systems with and without the impurity,
\begin{equation}
    S_{\mathrm{imp}} = S_{\mathrm{bulk+imp}} - S_{\mathrm{bulk}}.
\end{equation}
Non-monotonic behavior of $S_{\mathrm{imp}}$ reflects the progressive quenching of impurity degrees of freedom through the formation of localized boundary modes. Dynamically, the impurity spectral function $A(\omega)$ undergoes qualitative transformations across the phase boundaries, providing a direct window into the underlying screening mechanism and tower structure. Table~\ref{tab:overview} collects these fingerprints—coupling ranges, residual entropies $S_\mathrm{imp}(0)$ and $S_\mathrm{imp}(\infty)$, and the thermodynamic and dynamical signatures—for all five phases, complementing the phase diagram of Fig.~\ref{fig:PD}.

\begin{table*}[t]
\centering
\renewcommand{\arraystretch}{1.5}
\setlength{\tabcolsep}{6pt}
\begin{tabular}{l c c c c l l}
\hline\hline
Phase & $J/g$ & $\gamma=-id$ & $S_\mathrm{imp}(0)$ & $S_\mathrm{imp}(\infty)$ & Entropy $S_\mathrm{imp}(T)$ & Spectral function $A(\omega)$\\
\hline
AF1
 & $\left(0,\ \dfrac{4}{s(s+1)}\right)$
& $i\mathbb{R}\cup(0,\tfrac12)$
 & $\ln\!\big[2\cos\tfrac{\pi}{2s+2}\big]$ & $\ln 2$
 & monotonic
 & smooth\\
AF2
 & $\left(\dfrac{4}{s(s+1)},\ \dfrac{16}{4s+1}\right)$
 & $\left(\tfrac12,\,s\right)$
 & $\ln\!\big[2\cos\tfrac{\pi}{2s+2}\big]$ & $\ln 2$
 & non-monotonic
 & smooth\\
AF3
 & $\left(\dfrac{16}{4s+1},\ \infty\right)$
 & $\left(s,\,s+\tfrac12\right)$
 & $\ln\!\big[2\cos\tfrac{\pi}{2s+2}\big]$ & $\ln 2$
 & non-monotonic 
 & threshold peak at $\omega\!\approx\!|E_\gamma|$\\
F2
 & $\left(-\infty,\ -\dfrac{16}{4s+3}\right)$
 & $\left(s+\tfrac12,\,s+1\right)$
 & $\ln 2$ & $\ln 2$
 & non-monotonic 
 & threshold peak at $\omega\!\approx\!|E_\gamma|$\\
F1
 & $\left(-\dfrac{16}{4s+3},\ 0\right)$
 & $\left(s+1,\,\infty\right)$
 & $\ln 2$ & $\ln 2$
 & non-monotonic
 & smooth\\
\hline\hline
\end{tabular}
\caption{Boundary phases of the spin-$s$ Takhtajan--Babujian chain with a spin-$\tfrac12$ boundary impurity [Hamiltonian Eq.~\eqref{ModelHam}], ordered by increasing $J/g$, as a quantitative companion to the phase diagram (Fig.~\ref{fig:PD}). The impurity parameter $\gamma=-id$ is fixed by the coupling ratio through $J/g=16/[(1+2s)^2+4d^2]$ [Eq.~\eqref{Jgamma}]; $d$ is real only in the weak-coupling part of AF1 and purely imaginary elsewhere, so $\gamma$ is real beyond weak coupling. The last two columns give the thermodynamic and dynamical fingerprints derived in Secs.~\ref{castle} and~\ref{dynamics}: the impurity entropy crosses over monotonically only in the overscreened Kondo phase AF1 and develops a non-monotonic dip once boundary-bound modes quench impurity degrees of freedom, while the $T=0$ spectral function develops a sharp tower-onset threshold near $\omega\approx|E_\gamma|$ precisely in the gapped phases AF3 and F2. Throughout, $S_\mathrm{imp}(\infty)=\ln 2$, and $S_\mathrm{imp}(0)=\ln[2\cos(\pi/(2s+2))]$ is the overscreened $2s$-channel Kondo value in the antiferromagnetic regime, reverting to the free-spin value $\ln 2$ in the ferromagnetic regime.}
\label{tab:overview}
\end{table*}

To address these questions, we combine complementary analytical and numerical approaches. Perturbative renormalization-group arguments and boundary conformal field theory establish the expected low-energy universality class and residual impurity entropy, while the exact Bethe Ansatz provides a non-perturbative description of the full boundary phase diagram and its thermodynamics across all energy scales. The resulting thermodynamic predictions are benchmarked against large-scale finite-temperature matrix-product-operator simulations, and the corresponding dynamical signatures are analyzed through the impurity spectral function. Together, these methods provide a comprehensive description of impurity screening, boundary-bound-state formation, and tower restructuring in the spin-$s$ Takhtajan--Babujian chain.

The main results of this work are as follows. (i) We construct a boundary-integrable spin-$s$ Takhtajan--Babujian chain with a spin-$\frac{1}{2}$ impurity that remains solvable by the Bethe Ansatz for arbitrary boundary coupling $J$ (Appendix~\ref{app:int}). (ii) In the weak antiferromagnetic regime, the impurity realizes the $2s$-channel Kondo universality class, reproducing both its critical exponents and residual impurity entropy (Sec.~\ref{AF1 phase}). (iii) We identify five distinct boundary phases—three antiferromagnetic and two ferromagnetic—arising from the emergence of impurity-bound states and the associated reorganization of the Hilbert space into excitation towers (Sec.~\ref{castle}). (iv) To describe the resulting thermodynamics, we develop a generalized TBA framework that incorporates contributions from multiple towers and yields impurity entropy curves in quantitative agreement with finite-temperature MPO simulations (Sec.~\ref{castle}). (v) Finally, we show that the tower structure leaves clear dynamical fingerprints in the impurity spectral function, providing a direct probe of the underlying screening mechanism (Sec.~\ref{dynamics}). We conclude in Sec.~\ref{discussion} with a summary of the results and open directions.

\section{Field-theory analysis: RG and boundary CFT}
Before describing the exact non-perturbative solution of the problem, valid at all energy scales, using the exact Bethe Ansatz method, we briefly discuss some results from perturbative analysis and boundary CFT. 

\subsection{Perturbative analysis}\label{perturbative}
Note that the low-energy and long-distance physics of the Takhtajan--Babujian integrable spin chain is described by a perturbed $SU(2)_{k=2s}$ WZW model.
Writing the long wavelength expansion of the lattice spin operator as~\cite{PhysRevB.36.5291}
\begin{equation}
    \mathbf{S}_{j}
= a_{0}
\big[
\boldsymbol{\mathcal{J}}_{L}(x)
+ \boldsymbol{\mathcal{J}}_{R}(x)
\big]
+ (-1)^{j}
\mathcal{A}
\mathrm{Tr}\big(\boldsymbol{\sigma} {g}(x)\big)
+ \ldots
\end{equation}
where $a_0$ is the lattice spacing, $\mathcal{A}$ is the non-universal amplitude and $g(x)\in SU(2)$. The chiral currents satisfy the Kac-Moody algebra~\cite{eberhardt2019wess}

\begin{align}
\mathcal{J}^{a}_{L}(z)\mathcal{J}^{b}_{L}(w)
&\sim
\frac{s\delta^{ab}}{(z-w)^{2}}
+\frac{i\varepsilon^{abc}\mathcal{J}^{c}_{L}(w)}{z-w}+\ldots,
\\[6pt]
\mathcal{J}^{a}_{R}(\bar z)\mathcal{J}^{b}_{R}(\bar w)
&\sim
\frac{s\delta^{ab}}{(\bar z-\bar w)^{2}}
+\frac{i\varepsilon^{abc}\mathcal{J}^{c}_{R}(\bar w)}{\bar z-\bar w}+\ldots,
\end{align}
where the ellipses denote the regular terms. In terms of the chiral current, the low-energy description of Hamiltonian Eq.\eqref{ModelHam} (when no impurity is present i.e., when $J=0$) becomes the familiar $SU(2)_{k=2s}$ WZW model~\cite{witten1984non} whose Hamiltonian in the Sugawara form~\cite{sugawara1968field} can be written as
\begin{equation}
    H_0=\int \mathrm{d}x \frac{2\pi v}{2s+2}
:\big(
\boldsymbol{\mathcal{J}}_{L}^{2}
+
\boldsymbol{\mathcal{J}}_{R}^{2}
\big):,
\end{equation}
with the leading $SU(2)$ invariant bulk perturbation being the current-current interaction with a Hamiltonian density of the form
\begin{equation}
    \delta H= -\lambda  {\mathcal{J}}_{L}^a {\mathcal{J}}_{R}^a,
\end{equation}
where the repeated indices are assumed to be summed over. Notice that the scaling dimension of $\delta H$ is $\Delta_s=2$, such that it is marginal with vanishing beta function at tree level and the one-loop beta function becomes $\frac{d \lambda}{d \ln \ell}=\beta(\lambda)=-\lambda^2,$ which shows that the coupling $\lambda(L)$ flows to weak coupling at longer distances. 

When an impurity is attached to the boundary, it provides yet another local marginal perturbation given by the Hamiltonian density~\cite{affleck1995conformal,parcollet1998overscreened} 
\begin{equation}
    \delta H_K= u ({\mathcal{J}}_{R}^a+{\mathcal{J}}_{L}^a)S^a\delta(x).
\end{equation}
In the absence of bulk perturbations, the renormalization group (RG) flow of the boundary coupling is governed by $\beta(u) = u^2$. When the bulk marginal interaction $\lambda$ is introduced, this flow is modified to $\beta(u) = u^2 + \lambda u$~\cite{laflorencie2008kondo}. For an antiferromagnetic boundary coupling ($u > 0$), the flow runs away from the unstable ultraviolet (UV) fixed point corresponding to an unscreened impurity. In contrast, for a ferromagnetic coupling ($u < 0$), the boundary perturbation becomes marginally irrelevant, leaving the impurity unscreened even at low temperatures.

When $\lambda = 0$, the continuum limit coincides with the spin sector of the standard multichannel Kondo problem. From Bethe Ansatz, boundary conformal field theory (CFT), and numerical RG analyses, it is well established that the boundary coupling in this case flows to an intermediate fixed point associated with overscreening of the impurity. For $\lambda \neq 0$, the one-loop beta function acquires only the additional positive term $\lambda u$, so we expect the flow to approach the same intermediate-coupling fixed point. The resulting critical behavior should therefore be identical to that of the multichannel Kondo problem with $n = 2s$ screening channels~\cite{andrei1984solution,tsvelick1984solution,schlottmann1993multichannel}.

Crucially, however, the presence of a bulk marginally irrelevant perturbation $\lambda$ fundamentally reshapes the scaling of the characteristic energy scale, $T_K$. In a standard Kondo system where the bulk is purely critical ($\lambda=0$), the integration of the boundary flow $\beta(u) = u^2$ yields the familiar exponential scaling $T_K \propto \exp(-1/u)$. In the Takhtajan--Babujian chain, the bulk coupling is itself flowing as $\lambda(\ell) \sim 1/\ln \ell$. This forces the boundary coupling to ``track'' the slow logarithmic decay of the bulk. Following the logic of Ref.~\cite{laflorencie2008kondo}, the competition between these flows modifies the scaling from a simple exponential to one involving the inverse square root of the coupling:
\begin{equation}\label{eq:TKperturb}
    T_K \propto \exp\left(-\frac{\pi}{\sqrt{\lambda}}\right).
\end{equation}
This non-standard scaling is a hallmark of the interplay between bulk and boundary marginality, indicating that while the fixed point might be the same, the path taken to reach it, and the temperature at which the crossover occurs, is qualitatively different.

Thus, from this rudimentary perturbative analysis, we expect the impurity to be overscreened when $s > \frac{1}{2}$ and the lattice boundary coupling $J > 0$, whereas the impurity remains unscreened at low temperatures with a residual entropy of $\ln 2$ when $J < 0$. We shall show, however, that while this naive result is qualitatively correct, it does not capture the full structure of the problem. In particular, via an exact solution on the lattice using the Bethe Ansatz method, we find three distinct boundary phases when the boundary coupling is antiferromagnetic, and two distinct boundary phases when it is ferromagnetic, thereby revealing that the actual phase diagram is considerably richer than suggested by the simple perturbative argument.

\subsection{Boundary CFT and the Affleck--Ludwig $g$-function}
\label{sec:gfunction}

While the perturbative RG analysis identifies the expected infrared universality class, boundary conformal field theory provides a more powerful description of the corresponding fixed point and allows one to determine universal quantities such as the residual impurity entropy. The perturbative analysis of Sec.~\ref{perturbative} identifies the low-energy bulk theory with the
$SU(2)_{k=2s}$ Wess--Zumino--Witten conformal field theory perturbed by a marginally irrelevant current-current interaction.
The impurity appears as a local boundary perturbation of a conformal theory, so the infrared physics may be organized using boundary conformal field theory (BCFT) \cite{cardy2004boundary,affleck1991universal,affleck1995conformal}.

A central quantity in BCFT is the Affleck--Ludwig boundary degeneracy, or $g$-function. For a critical system of length $L$ at temperature $T=1/\beta$ with conformal boundary conditions $a$ and $b$ at its two ends, the free energy has the asymptotic form
\begin{equation}
F(T)=f_{\rm bulk}L-T\ln g_a-T\ln g_b+\ldots,
\end{equation}
where the omitted terms vanish as $L\to\infty$.
The corresponding boundary entropy is
\begin{equation}
S_{\rm bdy}=\ln g.
\end{equation}
This quantity measures the effective number of boundary degrees of freedom at a conformal fixed point and need not be an integer.

The BCFT interpretation becomes transparent from the strip partition function, which may be written in two equivalent channels. In the open-channel description, Euclidean time runs along the strip and
\begin{equation}
Z_{ab}
=
{\rm Tr}_{\mathcal H_{ab}}
e^{-\beta H_{\rm open}},
\end{equation}
where $\mathcal H_{ab}$ is the Hilbert space compatible with boundary conditions $a$ and $b$.
After exchanging the roles of space and Euclidean time, the same partition function can be written in the closed-channel representation as
\begin{equation}
Z_{ab}
=
\langle a|
e^{-L H_{\rm closed}}
|b\rangle,
\end{equation}
where $|a\rangle$ and $|b\rangle$ are conformal boundary states.
The equivalence of these two descriptions is implemented by the modular transformation $\tau\to -1/\tau$, under which the conformal characters transform as
\begin{equation}
\chi_i(-1/\tau)=\sum_j S_{ij}\chi_j(\tau),
\end{equation}
with $S_{ij}$ the modular $S$-matrix.
Cardy's consistency conditions then imply that the boundary degeneracy associated with a Cardy state labeled by the primary field $a$ is
\begin{equation}
g_a=\langle 0|a\rangle=\frac{S_{a0}}{\sqrt{S_{00}}},
\label{eq:gCardy}
\end{equation}
where $|0\rangle$ is the conformal vacuum. Thus, the $g$-function is the overlap of the boundary state with the vacuum sector.

For the $SU(2)_k$ WZW model, the modular $S$-matrix is
\begin{equation}
S_{jj'}
=
\sqrt{\frac{2}{k+2}}
\sin\left(
\frac{(2j+1)(2j'+1)\pi}{k+2}
\right),
\end{equation}
where $j,j'=0,\frac12,\ldots,\frac{k}{2}$ label the primary fields.
For a spin-$\frac12$ boundary impurity, the infrared conformal boundary condition is obtained by fusion of the free boundary condition with the spin-$\frac12$ primary field of the $SU(2)_k$ WZW theory \cite{affleck1991universal,affleck1995conformal}. The resulting change in boundary entropy is
\begin{equation}
\Delta S_{\rm bdy}
=
\ln\frac{g_{\rm IR}}{g_{\rm UV}}
=
\ln\frac{S_{\frac12,0}}{S_{0,0}}
=
\ln\left[
2\cos\left(\frac{\pi}{k+2}\right)
\right].
\end{equation}
For the present spin-$s$ Takhtajan--Babujian chain, the low-energy theory has $k=2s$, and therefore
\begin{equation}
\Delta S_{\rm bdy}
=
\ln\left[
2\cos\left(\frac{\pi}{2s+2}\right)
\right].
\label{eq:ALentropy}
\end{equation}
Since the impurity entropy in this work is defined as the difference between the system with and without the impurity, Eq.~\eqref{eq:ALentropy} is the BCFT prediction for the residual impurity contribution at the overscreened fixed point.\footnote{In a microscopic open spin chain, there is also an ordinary open-boundary contribution to the total $O(1)$ entropy. In the Bethe-Ansatz/TBA calculation below, that contribution is present in both the impurity and impurity-free systems and therefore cancels in the difference defining $S_{\rm imp}$. The BCFT result above isolates the impurity-induced change in boundary entropy.}

The result Eq.~\eqref{eq:ALentropy} coincides exactly with the Affleck--Ludwig prediction for the overscreened $2s$-channel Kondo fixed point, providing a nontrivial consistency check between the BCFT description and the exact Bethe Ansatz solution. However, the exact lattice solution reveals a richer structure than anticipated from the continuum theory. The BCFT analysis assumes that the boundary perturbation flows to a scale-invariant infrared fixed point and therefore applies only as long as no additional boundary scale is generated. As we shall show below, once the boundary coupling exceeds the threshold $J/g=4/[s(s+1)]$, a localized boundary-bound state emerges and introduces a new energy scale. The Hilbert space then reorganizes into multiple excitation towers built on distinct boundary configurations, leading to boundary phases that lie beyond the scope of the simple overscreened Kondo BCFT. As a result, the finite-temperature impurity entropy is no longer expected to follow the monotonic crossover implied by the standard Affleck--Ludwig picture and develops the non-monotonic behavior discussed later. Nevertheless, as we shall see below, the exact solution reveals that the zero-temperature impurity entropy throughout the antiferromagnetic regime remains given by Eq.~\eqref{eq:ALentropy}, despite the emergence of boundary-bound states and the associated restructuring of the excitation spectrum. For ferromagnetic couplings, on the other hand, the perturbative analysis predicts that the boundary coupling flows to zero, so the impurity remains asymptotically free and retains the entropy $\ln 2$ in both the ultraviolet and infrared limits. Having established the perturbative and BCFT expectations for the antiferromagnetic and ferromagnetic regimes, we now turn to the exact Bethe Ansatz solution. As we shall see, while the exact solution reproduces the universal fixed-point properties predicted above, it also reveals a considerably richer boundary phase structure than is apparent from the continuum description alone.

To uncover this richer structure, one must go beyond the low-energy continuum description and solve the lattice model exactly. The Bethe Ansatz formulation allows us to access all energy scales and coupling strengths on equal footing, providing a complete description of the boundary spin dynamics and thermodynamics across all phases. In the following, we derive the exact Bethe Ansatz equations for the open chain and use them to analyze the impurity contribution to the free energy and entropy, as well as the crossover behavior between the distinct phases. As shown in Fig.~\ref{fig:PD}, the model exhibits five distinct boundary phases governed by the ratio of the bulk and boundary couplings. We now summarize their key features and the corresponding impurity-screening mechanisms in the lattice Hamiltonian Eq.~\eqref{ModelHam}.
\section{AF1 Regime: Overscreened Kondo Phase}
\label{AF1 phase}

We now turn to the weak antiferromagnetic regime (\(0<J/g<\frac{4}{s(s+1)}\)), where the boundary spin is overscreened by the surrounding spin-\(s\) bulk degrees of freedom. This regime generalizes the conventional Kondo phase to higher-spin Takhtajan--Babujian chains and serves as the reference point for the remaining phases. Our objective is to obtain the impurity free energy, critical exponents, and zero-temperature impurity entropy \(S_{\mathrm{imp}}(T=0)\) for arbitrary bulk spin \(s\). Beyond its own interest, this section sets up the Bethe-Ansatz and TBA machinery used throughout the paper; the genuinely new boundary phases, absent in the conventional Kondo picture, appear once boundary-bound states emerge in Sec.~\ref{castle}.

To this end, we start from the Bethe Ansatz equations (BAE) and the corresponding energy function. Invoking the string hypothesis~\cite{takahashi1999thermodynamics}, we classify physical states as configurations of bound \(n\)-strings and describe them by the ratios of hole to root densities \(\eta_n=\rho_n^{h}/\rho_n\). This formulation leads directly to the Thermodynamic Bethe Ansatz (TBA) equations for the equilibrium distribution and allows us to compute the impurity free energy and residual entropy~\cite{yang1969thermodynamics,takahashi1999thermodynamics}. The resulting critical exponents and \(S_{\mathrm{imp}}(T=0)\) coincide with those of the \(n=2s\)–channel Kondo model, confirming the universality of the AF1 phase.

We now turn to the explicit Bethe Ansatz equations and the derivation of the corresponding TBA relations.

\subsection{Bethe Ansatz Equations and Energy Spectrum}

The exact solution of the Hamiltonian~\eqref{ModelHam} can be obtained using the boundary Quantum Inverse Scattering Method. The derivation of the corresponding Bethe Ansatz equations, together with the underlying integrability construction, is presented in Appendix~\ref{app:int}. Here we summarize the resulting equations and their implications for thermodynamics. The eigenstates of the Hamiltonian~\eqref{ModelHam} are determined by the Bethe Ansatz equations (BAE), which for an open spin-\(s\) chain with a spin-\(1/2\) boundary impurity take the form
\begin{equation}
\begin{aligned}\label{BAE p}
    \left(\frac{\mu_j-is}{\mu_j+is}\right)^{2N} \frac{\mu_j-d-i/2}{\mu_j-d+i/2}\frac{\mu_j+d-i/2}{\mu_j+d+i/2}\\=\prod_{l\neq j}^M \frac{(\mu_j-\mu_l-i)(\mu_j+\mu_l-i)}{(\mu_j-\mu_l+i)(\mu_j+\mu_l+i)}  ,
    \end{aligned}
\end{equation}
where the rapidities \(\mu_j\) parameterize the spin-wave excitations \footnote{Due to open-boundary conditions, $\mu_j$ and $-\mu_j$ correspond to the same excitation.}. Here, $M$ is the total number of rapidities $\mu_j$ ranging from $M_{\mathrm{min}}=0$ to $M_{\mathrm{max}}=\lfloor Ns+1/2 \rfloor$. Each solution, a set of Bethe roots \(\{\mu_j\}\), corresponds to an eigenstate of the Hamiltonian, with the corresponding energy
\begin{equation}\label{Energy}
    E_s(\{\mu_j\})=E_{0}-\sum_j \frac{4gs}{\mu_j^2+s^2},
\end{equation}
Here $E_0=Js/2 + 4g (N-1) \sum_{l=1}^{2s} l^{-1}$ is the energy of the Bethe vacuum $\ket{\uparrow_{1/2} \uparrow_s \ldots \uparrow_s}$ which corresponds to  $M=0$. While important for numerical checks, being a constant term, it doesn't affect the physics of the system and is ignored in the rest of the text.
The parameter \(d\) encodes the boundary to bulk coupling ratio 
\begin{equation}
J/g = \frac{16}{(1+2s)^2 + 4d^2}.
\label{Jgamma}
\end{equation}
As $d$ ranges over real and purely imaginary values, the resulting boundary coupling 
$J$ spans the full range from $-\infty$ to $\infty$:
\begin{equation}
    J/g \in \begin{cases}
  (-\infty,\, 0)\, ,                                             & d \in i\!\left(s+\tfrac{1}{2},\,\infty\right)\, , \\[4pt]
  \left(0,\, \tfrac{16}{(1+2s)^2}\right)\, ,                   & d \in \mathbb{R}\, , \\[4pt]
  \left(\tfrac{16}{(1+2s)^2},\, \infty\right)\, ,              & d \in i\!\left(0,\,s+\tfrac{1}{2}\right)
\end{cases}\,.
\end{equation}

In the AF1 regime, corresponding to weak antiferromagnetic coupling \(0<J/g<\tfrac{4}{s(s+1)}\), the impurity parameter \( d\) is real or purely imaginary, \( d \in \mathbb{R}\cup i(0,1/2)\). In the latter case, we shall use $d=i\gamma$ with $\gamma \in (0,1/2)$. 

\subsection{String Hypothesis and Integral Equations}
To extract the thermodynamics, we now pass from individual root configurations to their thermodynamic distributions. In the thermodynamic limit, solutions of Eq.~\eqref{BAE p} organize into \(n\)-strings of the form
\(\mu_{\alpha,j}^{(n)} = \lambda_\alpha^{(n)} + \frac{i}{2}(n + 1 - 2j)\) with $j=1,\ldots,n$,
where \(\lambda_\alpha^{(n)}\) are real string centers.
Writing the BAE \eqref{BAE p} for the $n$-strings and taking the logarithm yields
\begin{multline}\label{log BAE}
    2N X_{n,2s}(\lambda_j^n) + \Theta_n(\lambda_j^n)+\sum_{\pm} \Theta_{n}(\lambda_j^n\pm d)\\= 2\pi J_j^n +\sum_{m=1}^{\infty} \sum_{i=1}^{\zeta_m} \sum_\pm \Theta_{nm}(\lambda_j^n \pm \lambda_i^m),
\end{multline}
with $J_j^n$ being distinct integers fixing the root positions of the $n$-strings. The last term on the right-hand side couples the BAE for an $n$-string $\lambda_j^n$ with all $\zeta_m$ of the $m$-strings. 
In the equations above, the functions $X_{n,m}$ , $\Theta_{nm}$, and $\Theta_n$ are
\begin{widetext}
\begin{equation}
        \begin{aligned}
    \Theta_n(x)&=2\arctan \frac{2x}{n}\,,\quad  X_{nm}(x)=\sum_{\ell=1}^{\min(n,m)}\Theta_{n+m+1-2\ell}(x),
\\
\Theta_{nm}(x)&=
\begin{cases}
\Theta_{|n-m|}(x)+2\Theta_{|n-m|+2}(x)+\cdots+2\Theta_{n+m-2}(x)+\Theta_{n+m}(x),
& n\neq m,\\[2pt]
2\Theta_2(x)+2\Theta_4(x)+\cdots+2\Theta_{2n-2}(x)+\Theta_{2n}(x),& n=m,
\end{cases}
\end{aligned}
\end{equation}
\end{widetext} which are of the standard Bethe Ansatz form.

In the thermodynamic limit, the logarithmic BAE can be written as coupled integral equations for the densities of roots \(\rho_n(\lambda)\) and holes \(\rho_n^h(\lambda)\),
\begin{equation}
\rho_n^h(\lambda) = f_n(\lambda) - \sum_{m=1}^\infty (A_{n,m} * \rho_m)(\lambda),
\label{rhoeq}
\end{equation}
where \((f * g)(\lambda) = \int d\lambda' f(\lambda - \lambda') g(\lambda')\) is the standard convolution.
The kernel 
\begin{equation}
    \begin{aligned}A_{n,m}=K_{|n-m|}(\mu)+2K_{|n-m|+2}(\mu)+\ldots\\+2K_{n+m-2}(\mu)+K_{n+m}(\mu),\end{aligned}
\end{equation}
encodes scattering between strings, and
\begin{equation}
    f_n(\lambda)=2N Y_{n,2s}(\lambda)+K_n(\lambda)+\sum_\pm K_{n}(\lambda\pm d)
\end{equation}
with \(Y_{m,n}=\frac{1}{2\pi}\dv{X_{m,n}}{\lambda}=\sum_{l=1}^{\mathrm{min}(m,n)}K_{m+n+1-2l}\), contains the boundary driving term entering the free-energy functional. Importantly, \(f_n(\lambda)\) affects only the free-energy functional and not the algebraic structure of the TBA equations for the ratios
\(\eta_n(\lambda) = \rho_n^h(\lambda)/\rho_n(\lambda)\). These density equations provide the starting point for the thermodynamic description of the model. In the equations above, $K_n$ are Lorentzians
\begin{equation}
    K_n(\lambda)\equiv \frac{1}{2\pi}\dv{\Theta_n}{\lambda}=\frac{1}{\pi}\frac{(n/2)}{(n/2)^2+\lambda^2} .
\end{equation}

\subsection{Free Energy and TBA Equations}
\label{sec:TBA}

The thermodynamics of the system follow from the Bethe root distributions obtained in the previous subsection. 
Each configuration of $n$-strings contributes an energy
\begin{equation}
E = \frac{1}{2}\sum_{n=1}^\infty \int d\lambda\, \epsilon_n(\lambda)\, \rho_n(\lambda),
\end{equation}
where the single–$n$-string dispersion $\epsilon_n(\lambda) = -4\pi Y_{n,2s}(\lambda)$ follows from Eq.~\eqref{Energy}. 
The prefactor $1/2$ accounts for the mirror symmetry of the spectrum in open chains, since rapidities $\mu$ and $-\mu$ correspond to the same excitation.

At finite temperature \(T\) and magnetic field \(h\), the total free energy reads
\begin{equation}
    F = E - S_z h - T S,
\end{equation}
where the Yang–Yang entropy,
\[
S = \frac{1}{2}\sum_{n=1}^\infty \int d\lambda\,
\big[(\rho_n + \rho_n^h)\ln(\rho_n + \rho_n^h)
- \rho_n \ln \rho_n - \rho_n^h \ln \rho_n^h \big],
\]
is expressed through the root and hole densities of $n$-strings.

Minimizing \(F\) with respect to \(\rho_n\) under the constraints of Eq.~\eqref{rhoeq} yields the standard TBA relations for the ratios
\(\eta_n = \rho_n^h/\rho_n\):
\begin{equation}
\ln \eta_n(\lambda)
= \frac{g_n(\lambda)}{T}
- \sum_{m=1}^\infty \big(A_{n,m} * \ln[1 + \eta_m^{-1}]\big)(\lambda),
\label{TBA p}
\end{equation}
with the effective $n$-string energies \(g_n(\lambda)=\epsilon_n(\lambda)+nh\).
The coupled equations~\eqref{TBA p} define the equilibrium distribution of strings at temperature~$T$.

At the saddle point, the free energy can be written in terms of the solutions \(\eta_n\) of Eq.~\eqref{TBA p} as
\begin{equation}
 F_{\mathrm{saddle}}
 = -\frac{T}{2}\sum_{n=1}^\infty
 \int d\lambda\, f_n(\lambda)\,
 \ln[1+\eta_n^{-1}(\lambda)]
 - h\left(sN+\frac{1}{2}\right).
\label{Fsaddle}
\end{equation}
The infinite sum over $n$ can be simplified using the Takahashi identity,
\begin{equation}
\sum_n Y_{m,n} * \ln(1+\eta_n^{-1})
 = G* \left[\ln(1+\eta_m) - \frac{g_m}{T}\right],
\label{Takahashi}
\end{equation}
where
\(G(\lambda)=\tfrac{1}{2\cosh(\pi \lambda)}\).

Applying Eq.~\eqref{Takahashi} to Eq.~\eqref{Fsaddle} separates the total free energy into bulk and impurity parts:
\begin{equation}
\begin{aligned}
F_{\mathrm{saddle}}
&= \int  d\lambda\, \frac{N}{2\cosh(\pi \lambda)}
  \big[\epsilon_{2s} - T \ln(1+\eta_{2s})\big]\\
&\quad +\frac{1}{2} \int  d\lambda\,
   \frac{g_1 - T\ln(1+\eta_1)}{2\cosh[\pi\lambda]}\\
&\quad
 + \frac{1}{2} \int  d\lambda\, 
   \sum_{\pm}
   \frac{\epsilon_1 - T\ln(1+\eta_1)}{2\cosh[\pi(\lambda \pm  d)]}\;.
\end{aligned}
\label{Fbulkimp}
\end{equation}
In the absence of the impurity, the free energy would consist only of the first two lines. For this reason, we identify them with the bulk  \(F_{\mathrm{saddle,bulk}}\), 
and the last line, arising from the boundary term in \(f_n(\lambda)\), with the impurity  \(F_{\mathrm{saddle,imp}}\). Having separated the free energy into bulk and impurity contributions, we now focus on the impurity component and the information it carries about boundary screening.

\subsection{Impurity Contribution and Residual Entropy}
\label{sec:ImpurityEntropy}

Beyond the saddle-point term discussed above, the free energy acquires universal \(O(1)\) corrections associated with the open boundaries of the chain. These arise from the nontrivial density of states of the Bethe Ansatz solution together with fluctuations around the stationary TBA configuration and can be incorporated exactly through the Fredholm-determinant representation \cite{pozsgay2010mathcal,he2024exact}
\begin{equation}
Z = e^{-\beta F_{\text{saddle}}}
    \frac{\det(\mathbf{1}-\widehat{\mathcal Q}^{-})}
         {\det(\mathbf{1}-\widehat{\mathcal Q}^{+})},
    \label{Zexact}
\end{equation}
and
\begin{equation}
F = F_{\text{saddle}}
    -T \left[
    \log\det(\mathbf{1}-\widehat{\mathcal Q}^{-})
    -\log\det(\mathbf{1}-\widehat{\mathcal Q}^{+})
    \right].
    \label{trueF}
\end{equation}

Since our focus is on the impurity contribution, we do not evaluate this universal open-edge factor\footnote{
The thermodynamics in this work is formulated in the open-channel picture, where the tower decomposition emerges naturally from the Bethe-Ansatz spectrum of the open chain. It would be interesting to understand how the same tower structure is encoded in the corresponding closed-channel formulation, where the partition function is expressed in terms of boundary-state overlaps and exact \(g\)-functions, following the approaches of Refs.~\cite{pozsgay2010mathcal,he2024exact}.
} explicitly; analytical expressions are available in Refs.~\cite{pozsgay2010mathcal,he2024exact}.  
By taking the ratio of partition functions with and without the impurity, one isolates the impurity-specific contribution without evaluating the two Fredholm determinants directly.  
The residual impurity free energy thus coincides with the saddle-point result,
\begin{equation}
    \begin{aligned}\label{F imp}
        F_\mathrm{imp}
        &= F_\mathrm{bulk+imp}-F_\mathrm{bulk}
         = F_\mathrm{saddle,imp} \\
        &= \frac{1}{2}\int \mathrm{d}\lambda\, 
           \sum_{\pm}
           \frac{\epsilon_1(\lambda) - T\ln[1+\eta_1(\lambda)]}
                {2\cosh[\pi(\lambda \pm  d)]}.
    \end{aligned}
\end{equation}
The corresponding impurity entropy is then
\(S_\mathrm{imp}(T)=-\partial_T F_\mathrm{imp}(T)\).
This quantity measures the effective number of impurity degrees of freedom and interpolates smoothly between the ultraviolet and infrared boundary fixed points.

In the following, we analyze the limiting behaviors of the impurity entropy.  
At high temperatures (\(T \to \infty\)), \(S_{\mathrm{imp}}\) reflects the total degrees of freedom introduced by the spin-$1/2$ impurity, while in the zero-temperature limit (\(T \to 0\)) it encodes the residual ground-state degeneracy of the overscreened Kondo fixed point.  
Evaluating the latter limit and first corrections provides direct access to the universal boundary $g$-factors \cite{PhysRevLett.67.161} and the critical exponents characterizing the AF1 regime.

\subsection{Zero- and High-Temperature Limits of the Impurity Entropy}
\label{sec:T0Tinfty}

In the zero- and high-temperature limits, at zero magnetic field (\(h=0\)), the TBA equations simplify considerably.  
The ratios \(\eta_n(\lambda)=\rho_n^h/\rho_n\) become constant functions \(\eta_n^{T=0,\infty}\), and Eq.~\eqref{TBA p} reduces to a set of algebraic relations.  
Solving these equations provides analytic expressions for the impurity residual entropy at the ultraviolet and infrared boundary fixed points, revealing the universal boundary $g$-factors \cite{PhysRevLett.67.161} that characterize the AF1 phase.

At \(T=0\) and \(T=\infty\), the TBA equations take the form~\cite{takahashi1999thermodynamics}
\begin{equation}
\begin{aligned}
    \ln \eta_{2s}^{T=0} &= -\infty,\\
    \ln \eta_n^{T=0} &= \tfrac{1}{2}\ln[(1+\eta_{n-1}^{T=0})(1+\eta_{n+1}^{T=0})],\\
    \lim_{n\to\infty}\frac{\ln\eta_n^{T=0}}{n} &= 0,
\end{aligned}
\end{equation}
where the driving term truncates the hierarchy at \(n=2s\), enforcing \(\ln\eta_{2s}^{T=0}=-\infty\).  
At high temperatures, the equations become
\begin{equation}
\begin{aligned}
    \ln \eta_1^{T=\infty} &= \tfrac{1}{2}\ln[1+\eta_2^{T=\infty}],\\
    \ln \eta_n^{T=\infty} &= \tfrac{1}{2}\ln[(1+\eta_{n-1}^{T=\infty})(1+\eta_{n+1}^{T=\infty})],\\
    \lim_{n\to\infty}\frac{\ln\eta_n^{T=\infty}}{n} &= 0.
\end{aligned}
\end{equation}
The well-known solutions are~\cite{takahashi1999thermodynamics}
\begin{equation}\label{eta const p}
\begin{aligned}
    \eta_{n<2s}^{T=0}&=\left( \frac{\sin[\pi(n+1)/(2s+2)]}{\sin[\pi/(2s+2)]}\right)^2-1,\\
    \eta_{n\geq 2s}^{T=0}&=(n+1-2s)^2-1,\\
    \eta_n^{T=\infty}&=(n+1)^2-1.
\end{aligned}
\end{equation}

Substituting Eq.~\eqref{eta const p} into the impurity free energy yields, at \(T\to0\) and \(h=0\),
\begin{equation}
\mathcal{S}_\mathrm{str}^{\mathrm{imp}}(T=0)
= \ln \left[2\cos \left(\frac{\pi}{2s+2}\right)\right],
\end{equation}
which coincides with the well-known result for the \(n=2s\) multichannel Kondo model \cite{andrei1984solution, tsvelick1984solution, Andrei_1995}.  
This correspondence reflects the equivalence of the low-temperature fixed point to the \(SU(2)_{2s}\) WZW theory, consistent with the perturbative analysis in Sec.~\ref{perturbative}.

At \(T=\infty\) (and \(h=0\)), Eq.~\eqref{eta const p} gives
\begin{equation*}
\lim_{T\to\infty}\frac{F_{\mathrm{imp}}(T)}{T}
= -\tfrac{1}{2}\ln \left(1+\eta_{1}^{T=\infty}\right)
= -\mathcal{S}_\mathrm{str}^{\mathrm{imp}}(T=\infty),
\end{equation*}
with
\begin{equation}
\mathcal{S}_\mathrm{str}^{\mathrm{imp}}(T=\infty)=\ln 2,
\end{equation}
reflecting the dimensionality of the impurity Hilbert space.

For completeness, the bulk contribution to the saddle-point entropy at \(T=\infty\) and \(h=0\) reads
\begin{equation}
\begin{aligned}
\mathcal{S}_\mathrm{saddle,bulk}(T=\infty)
&= -\lim_{T\to\infty}\frac{F_\mathrm{saddle,bulk}(T)}{T}
\\
&= N\ln(2s+1) + \tfrac{1}{2}\ln2.
\end{aligned}
\end{equation}
The actual bulk entropy, corresponding to the Hilbert-space dimension of \(N\) spin-\(s\) sites, is
\(\mathcal{S}_\mathrm{bulk}(T=\infty)=N\ln(2s+1)\).  
The additional \(\tfrac{1}{2}\ln2\) term therefore represents the contribution of the Fredholm determinants at infinite temperature, as anticipated from Eq.~\eqref{trueF}.

These limiting values of \(S_{\mathrm{imp}}\) confirm that the AF1 phase interpolates between a free spin with entropy \(\ln2\) at high temperature and an overscreened Kondo fixed point with residual entropy \(\ln[2\cos(\pi/(2s+2))]\) at low temperature, in full agreement with the multichannel Kondo universality class.

\subsection{Low-Temperature Behavior, Kondo Temperature, and Impurity Specific-Heat Exponent}
\label{sec:specificheat}

We now analyze the low-temperature behavior of the impurity free energy and extract the corresponding specific-heat critical exponent.  
At small temperatures \(T\ll g\) (and \(h=0\)), the TBA equations \eqref{TBA p} can be rewritten in the scaling form
\begin{equation}\label{TBA2}
    \ln \eta_n(\zeta)
    = -2\delta_{n,2s}e^{\zeta}
      + \sum_\pm G * \ln[1+\eta_{n\pm 1}(\zeta)]
      ,
\end{equation}
where \(\zeta=\pi\lambda-\ln(T/2\pi)\) is the new shifted variable introduced to make the equation above temperature-independent. The  kernel $G$ in terms of new variables is
\begin{equation*}
    G(\zeta)=\frac{1}{2\pi\cosh(\zeta)}.
\end{equation*}  
The impurity contribution to the free energy is then expressed as
\begin{equation}
\begin{aligned}
\mathcal{F}_{\mathrm{imp}}
&=\sum_\pm \frac{1}{2\pi} \int  d\zeta\,
  \frac{\epsilon_1(\zeta)-T\ln[1+\eta_1(\zeta)]}{2\cosh[\zeta+\ln(T/2\pi)\pm  \pi d]}.
\end{aligned}
\end{equation}
The integral receives its dominant contribution near
\(\zeta\approx -\ln(T \exp(\pi |d|) )\to +\infty\)
as \(T\ll \exp(-\pi |d|) \). The expression $\exp(-\pi |d|)$ is related to the Kondo temperature $T_K$ which we provide at the end of this section based on the specific heat calculations.

For \(T\ll T_K\), the free energy is well approximated by
\begin{equation}
\mathcal{F}_{\mathrm{imp}}
\simeq \mathcal{F}_{\mathrm{imp}}^0
-\frac{T}{2}\ln[1+\eta_1(+\infty)].
\end{equation}
with $\mathcal{F}_{\mathrm{imp}}^0$ being a temperature independent term.

To obtain the next-to-leading temperature correction, we introduce
\(g_n(\zeta)=\ln[1+\eta_n(\zeta)]\)
and define the deviation
\(\delta g_n(\zeta)=g_n(\zeta)-g_n(\infty)\),
which is assumed small at low \(T\).  
Linearizing Eq.~\eqref{TBA2} yields, for \(n<2s\),
\begin{equation}
\frac{f_n^2}{f_{n+1}f_{n-1}}\,
\delta g_n(\zeta)
= G \left[\delta g_{n-1}(\zeta)
           + \delta g_{n+1}(\zeta)\right],
\end{equation}
with \(f_n^2=1+\eta_n(\infty)\), \(f_0=1\), and \(\delta g_{2s}(\zeta)=g_{2s}(\zeta)\).  
After the Fourier transformation, this becomes a recurrent algebraic system whose solution reads
\begin{equation}
\delta \tilde{g}_{n<2s}(\omega)
= \hat{t}_{n,2s}(\omega)\,\tilde{g}_{2s}(\omega),
\end{equation}
where
\begin{equation}
\hat{t}_{n,2s}(\omega)
= \frac{
f_{n-1}\sinh[\omega(n+2)/2]
 - f_{n+1}\sinh[\omega n/2]
}{
2f_n\cos[\pi/(2s+2)]\,\sinh[(s+1)\omega]
}.
\end{equation}

The corresponding impurity free energy becomes
\begin{equation}
\begin{aligned}
\mathcal{F}_{\mathrm{imp}}
&=\mathcal{F}_{\mathrm{imp}}^0
-T\ln \left(
2\cos[\pi/(2s+2)]
\right)\\
&\quad
-T \int \frac{d\omega}{2\pi}\,
e^{-i\omega \ln(T/2\pi)/\pi}
\frac{\cos(\omega d)}{2\cosh(\omega/2)}\,
\hat{t}_{1,2s}(\omega)\,\tilde{g}_{2s}(\omega).
\end{aligned}
\end{equation}
Evaluating the integral by contour deformation, the dominant contribution arises from the poles of
\((\sinh[(s+1)\omega])^{-1}\) and \(\cosh(\omega/2)\).  
For \(T\ll T_K\), the leading terms originate from the poles closest to the real axis.  
The first pole at \(\omega=i\pi/(s+1)\) gives a vanishing residue, so for \(s>1\) the leading nonzero term comes from \(\omega=i2\pi/(s+1)\), yielding
\begin{equation}
\mathcal{F}_{\mathrm{imp}}
=\mathcal{F}_{\mathrm{imp}}^0
-T\,\mathcal{S}_{\mathrm{imp}}(T=0)
-\alpha\cosh(\frac{2\pi d}{s+1}) \,T^{1+\frac{2}{1+s}},
\end{equation}
where $\alpha$ is a dimensionless constant.
The specific heat, which is  dimensionless, is then a ratio 
\begin{equation}
C_{\mathrm{imp}}\propto (T/T_K)^{2/(1+s)},
\end{equation}
with the Kondo temperature
\begin{equation}
    T_K \propto \left(\sech(\frac{2\pi d}{s+1})\right)^{(1+s)/2} .
\end{equation}
For \(s=1\), the pole at \(\omega=i2\pi/(s+1)\) becomes second order, leading instead to a logarithmic correction,
\begin{equation}
C_{\mathrm{imp}}\propto \cosh(\pi d) T\ln(T_K/T)=\frac{T}{T_K}\ln(T_K/T).
\end{equation}
These critical exponents coincide with those of the \(n=2s\)–channel Kondo model, confirming that the AF1 phase belongs to the same universality class.

The Kondo temperature defined above in terms of the ratio of the couplings $J/g$ is explicitly \begin{equation}\label{eq:TK}
    T_K \propto \left(\sech(\pi\sqrt{\frac{4g}{J}-\left(s+\frac{1}{2}\right)^2})\right)^{(1+s)/2}.
\end{equation}
In the limit of small boundary coupling, $J/g \ll 1$ or, equivalently, large real impurity parameter, $d\gg 1$, Eq.~\eqref{eq:TK} simplifies to 
\begin{equation}
    T_K \propto \exp(-\pi |d|) =\exp\left(-\frac{2\pi}{\sqrt{J/g}}\right).
\end{equation}
This result is quite interesting: it recovers the modified scaling form predicted by the perturbative RG analysis in Section~\ref{perturbative}. Specifically, it confirms that the bulk marginal coupling $\lambda$ (which scales as $g/J$ in this regime) alters the standard Kondo exponent to the square-root form $\exp(-\pi/\sqrt{\lambda})$ via the mechanisms discussed in Ref.~\cite{laflorencie2008kondo}. The correspondence between this expression and the scaling found in an underscreened Kondo spin chain model~\cite{frahm1997open} points toward a common scaling behavior shared by systems where bulk marginality governs the approach to the Kondo fixed point.

\subsection{Numerical Solution of the TBA Equations}\label{sec:NTBA}

To obtain the impurity entropy at intermediate temperatures, where an analytic solution of the TBA equations is not available, we solve them numerically.  
The rapidity variable is discretized on a uniform grid  
$\lambda_j = -L + j \Delta\lambda$ $(j = 0, \ldots, N_\lambda - 1)$  
with spacing $\Delta\lambda = 2L / (N_\lambda - 1)$.  
A cutoff of $L = 40$ is sufficient because both the convolution kernel  
$K(\lambda) = [2\cosh(\pi\lambda)]^{-1}$ and the driving term
$D(\lambda) = -(2\pi/T)\,\sech(\pi\lambda)$ decay exponentially, so contributions from $|\lambda| > L$ are negligible.  
We use $N_\lambda = 1024$ grid points, which provides adequate resolution for the auxiliary functions $\eta_n(\lambda)$ and accurate quadrature of the convolution integrals.

The nonlinear integral equations are
\begin{equation}
\begin{split}
\ln \eta_n(\lambda)
&= \delta_{n,n_{\mathrm{imp}}} D(\lambda)
+ \int_{-L}^{L} K(\lambda - \lambda') \\
&\times \left[\ln(1 + \eta_{n-1}(\lambda')) 
+ \ln(1 + \eta_{n+1}(\lambda'))\right] d\lambda'
\end{split}
\end{equation}

which are truncated at $n_{\max} = 30$.  
The discrete convolutions are implemented as dense matrix products,
\begin{equation}
(K * f)(\lambda_i) \approx
\Delta\lambda \sum_j
\frac{f(\lambda_j)}{2\cosh[\pi(\lambda_i - \lambda_j)]},
\end{equation}
and for the Lorentzian integral operators used in the asymptotic closure,
\begin{equation}
([n] f)(\lambda_i) \approx
\Delta\lambda \sum_j
\frac{1}{\pi}
\frac{(n\pi / 2)}{(n\pi / 2)^2 + (\lambda_i - \lambda_j)^2}
f(\lambda_j).
\end{equation}

The fixed-point iteration starts from $\eta_0^{(1)} = 0$ and $\eta_n^{(1)} = (n + 1)^2 - 1$ for $n \ge 1$.  
At each step we update
\begin{equation}
\begin{split}
\eta_n(\lambda) \leftarrow
\exp\Big[
&\,\delta_{n,n_{\mathrm{imp}}} D(\lambda)
+ (K * \ln(1 + \eta_{n-1}))(\lambda) \\
&+ (K * \ln(1 + \eta_{n+1}))(\lambda)
\Big],
\end{split}
\end{equation}

for $1 \le n \le n_{\max}$,  
while $\eta_{n_{\max}+1}(\lambda)$ is determined self-consistently from the asymptotic closure condition that preserves the correct large-$n$ behavior.  
Specifically, for large $n$, the auxiliary functions obey the universal asymptotic relation 
$\ln(1+\eta_{n+1}) - \ln(1+\eta_{n}) \to -h/T$,
corresponding to the exponential saturation of the spin string hierarchy.  
We enforce this by evaluating
\begin{equation}
\begin{split}
\ln(1+\eta_{n_{\max}+1})
&=
\frac{
(-h/T)
+[n_{\max}+1]\ln(1+\eta_{n_{\max}})}{[n_{\max}+1]\,1} \\
&\quad
-\frac{[n_{\max}]\ln(1+\eta_{n_{\max}-1})}{[n_{\max}+1]\,1}.
\end{split}
\end{equation}
where $[n]$ denotes the Lorentzian integral operator defined above, and $[n]\,1$ its action on the constant function.  
This construction ensures that $\eta_{n_{\max}+1}$ matches smoothly onto the asymptotic tail $\eta_n \sim \exp(n h/T)$ required by the infinite hierarchy.  
The iteration continues until  
$\max_{n,\lambda} |\eta_n^{(k+1)}(\lambda) - \eta_n^{(k)}(\lambda)| < 10^{-10}$,  
which ensures convergence of all $\eta_n(\lambda)$ on the chosen grid. 

\begin{figure}[h]
    \centering
    \includegraphics[width=\linewidth]{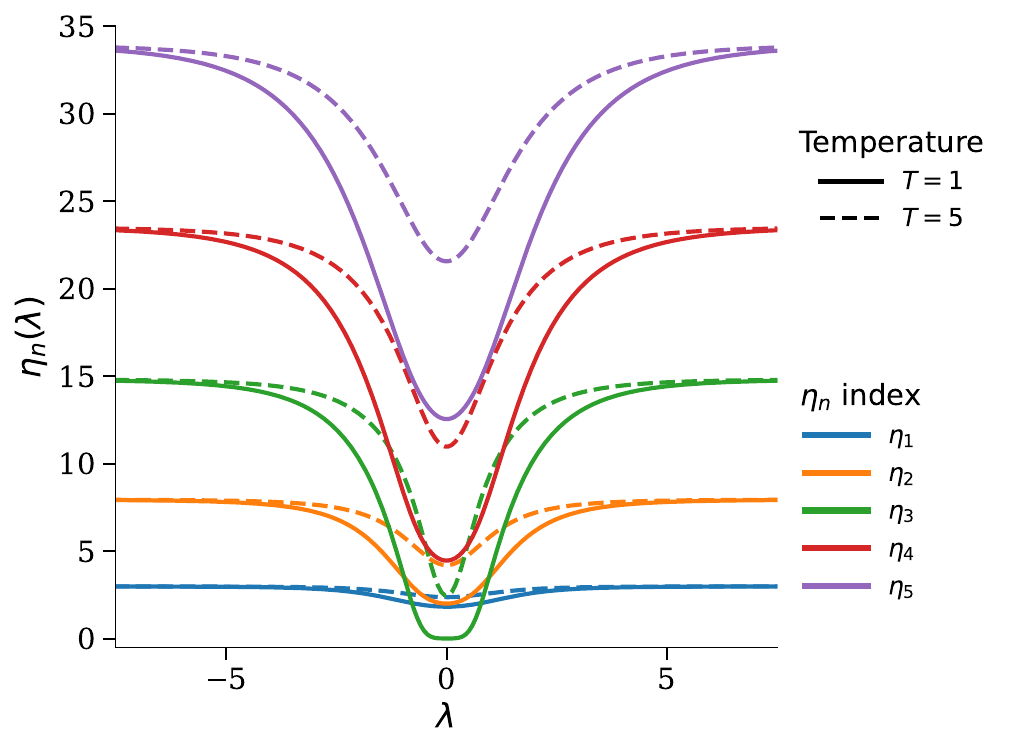}
    \caption{Functions $\eta_n(\lambda)$ obtained through the numerical solutions of the TBA equations for $S=3/2$. For finite temperature $T$ values, the functions smoothly change from the limit values $\eta^{T=\infty}_n$ to the minimum at $\lambda=0$. At small temperatures $T\ll 1$, the functions are almost flat for $|\lambda|\ll \ln(T/2\pi)$ and equal in that region to another limit value $\eta^{T=0}_n$. The flattening can be already seen to start happening to $\eta_{3}$ function at $T=1$ on the plots. As the temperature $T$ increases, the driving term in the TBA equations gets more irrelevant, and functions tend to flatten out to their at-infinity values $\eta^{T=\infty}_n$ (compare dashed and solid lines corresponding to $T=5$ and $T=1$, respectively). }
    \label{fig:etas}
\end{figure}

See Figure~\ref{fig:etas} for plots of the numerically resolved $\eta_n$ for $S=3/2$ at $T=1$ and $T=5$ with $g=1$.

\section{Other Phases: Thermodynamics in a Split Hilbert Space}
\label{castle}

For boundary couplings beyond the weak antiferromagnetic regime (\(\gamma>1/2\)), 
new impurity-bound solutions appear in the Bethe Ansatz equations, resulting in 
a qualitative reorganization of the Hilbert space.  
Instead of a single tower of bulk excitations built on one boundary configuration, the system now hosts several distinct boundary states, each serving as a base for 
its own family, or \emph{tower}, of bulk excitations.  
In this section, we formulate the Bethe Ansatz for these regimes, derive the
tower-specific contributions to the free energy, and analyze their effect on the
impurity entropy across the AF2, AF3, F1, and F2 phases.
A summary of all boundary phases, their coupling ranges, impurity-parameter ranges, active and gapped towers, and limiting impurity entropies, is collected in Table~\ref{tab:phases}.

Before specializing to individual towers, it is useful to record the general pipeline, valid for \emph{any} boundary tower of the open spin-$s$ chain. The choice of tower enters \emph{only} through the drivings $f_n^{(k)}$ [Eq.~\eqref{f k}]: its boundary string(s) dress the source term through additional $\pm$-shifted kernels $K_n(\lambda\pm ib)$, while the universal $\eta$-system [Eq.~\eqref{TBA p}]—and hence the functions $\eta_n(\lambda)$ themselves—is common to all towers. Consequently the bulk free energy is identical for every tower [first line of Eq.~\eqref{Fbulkimp}], and the entire tower dependence of the thermodynamics is carried by the impurity part of $f_n^{(k)}$ [third line of Eq.~\eqref{Fbulkimp}]. The full pipeline—from the Bethe Ansatz equations to the impurity free energy, and from there to the $T=0$ base energy and $T=\infty$ tower entropy—is summarized schematically in Figs.~\ref{fig:bae2free} and~\ref{fig:free2ent} of Appendix~\ref{app:TBA}.

\begin{table*}[ht]
\centering
\renewcommand{\arraystretch}{1.6}
\setlength{\tabcolsep}{5pt}
\begin{tabular}{l c c c c c c}
\hline\hline
Phase & $J/g$ & $\gamma=-id$ & Active towers & Gapped tower(s) & $S_\mathrm{imp}(0)$ & $S_\mathrm{imp}(\infty)$\\
\hline
AF1
 & $\left(0,\ \dfrac{4}{s(s+1)}\right)$
 & $i\mathbb{R}\cup\left(0,\tfrac12\right)$
 & $\mathcal{T}_\mathrm{str}$
 & ---
 & $\ln\!\left[2\cos\tfrac{\pi}{2s+2}\right]$ & $\ln 2$\\
AF2$'$
 & $\left(\dfrac{4}{s(s+1)},\ \dfrac{16}{4s^2+4s-3}\right)$
 & $\left(\tfrac12,1\right)$
 & $\mathcal{T}_\mathrm{str},\,\mathcal{T}_\mathrm{BS}$
 & ---
 & $\ln\!\left[2\cos\tfrac{\pi}{2s+2}\right]$ & $\ln 2$\\
AF2$''$
 & $\left(\dfrac{16}{4s^2+4s-3},\ \dfrac{16}{4s+1}\right)$
 & $\left(1,s\right)$
 & $\mathcal{T}_\mathrm{str},\,\mathcal{T}_\mathrm{BS},\,\mathcal{T}_\mathrm{hBS}$
 & ---
 & $\ln\!\left[2\cos\tfrac{\pi}{2s+2}\right]$ & $\ln 2$\\
AF3
 & $\left(\dfrac{16}{4s+1},\ \infty\right)$
 & $\left(s,s+\tfrac12\right)$
 & $\mathcal{T}_\mathrm{str},\,\mathcal{T}_\mathrm{BS},\,\mathcal{T}_\mathrm{hBS}$
 & $\mathcal{T}_\mathrm{str}$
 & $\ln\!\left[2\cos\tfrac{\pi}{2s+2}\right]$ & $\ln 2$\\
F2
 & $\left(-\infty,\ -\dfrac{16}{4s+3}\right)$
 & $\left(s+\tfrac12,s+1\right)$
 & $\mathcal{T}_\mathrm{str},\,\mathcal{T}_\mathrm{BS},\,\mathcal{T}_\mathrm{hBS}$
 & $\mathcal{T}_\mathrm{BS}$
 & $\ln 2$ & $\ln 2$\\
F1
 & $\left(-\dfrac{16}{4s+3},\ 0\right)$
 & $\left(s+1,\infty\right)$
 & $\mathcal{T}_\mathrm{str},\,\mathcal{T}_\mathrm{BS},\,\mathcal{T}_\mathrm{hBS}$
 & ---
 & $\ln 2$ & $\ln 2$\\
\hline\hline
\end{tabular}
\caption{Summary of the boundary phases of the spin-$s$ Takhtajan--Babujian chain with a spin-$\tfrac12$ boundary impurity [Hamiltonian Eq.~\eqref{ModelHam}], ordered by increasing $J/g$ (the antiferromagnetic phases AF1--AF3 for $J/g>0$ and the ferromagnetic phases F2, F1 for $J/g<0$). The impurity parameter is $\gamma=-id$, related to the couplings through Eq.~\eqref{Jgamma}. AF2 is split into two parts (AF2$'$, AF2$''$) according to whether the higher-order boundary-string tower $\mathcal{T}_\mathrm{hBS}$ is present ($\gamma>1$). ``Gapped tower(s)'' lists the tower(s) raised by the boundary gap $|E_\gamma|$ [Eq.~\eqref{eq:Egamma}] above the degenerate lowest towers; ``---'' means all active towers share the same base energy. The zero-temperature impurity entropy is $\ln[2\cos(\pi/(2s+2))]$ throughout the antiferromagnetic regime (the overscreened $2s$-channel Kondo value) and $\ln 2$ throughout the ferromagnetic regime (the free spin-$\tfrac12$ value), while $S_\mathrm{imp}(\infty)=\ln 2$ in every phase.}
\label{tab:phases}
\end{table*}

\subsection{Boundary and Higher-Order Boundary Strings}

When \(\gamma>1/2\), the Bethe Ansatz equations~\eqref{BAE p}
admit a new purely imaginary solution
\(\mu_\gamma=i(\gamma-\tfrac12)\),
known as the \emph{fundamental boundary string}.  
Being both imaginary and impurity-dependent, this root represents a boundary excitation exponentially localized next to the impurity. The corresponding localization properties were analyzed for the XX spin chain in Ref.~\cite{Kattel_2024}. For stronger couplings (\(\gamma>1\)), the BAE allow for higher-order 
imaginary roots of the form
\(\mu_{\gamma,l}^{(m)}=\mu_\gamma - i l\),
with \(l=0,\dots,m\) and
\(m=\lfloor\gamma+\tfrac12\rfloor\)~\cite{wang2015off}.  
This higher-order boundary string makes up the base state for the third family of states -- the higher-order boundary-string tower.  

To sum up, the \(n\)-strings introduced in Sec.~\ref{AF1 phase} describe 
delocalized bulk excitations.  
The BAE solutions now separate into two (for \(\gamma\in(\tfrac12,1)\)) or three (for \(\gamma>1\)) families:  
(i) configurations consisting solely of bulk \(n\)-strings, corresponding to an unexcited boundary;  
(ii) configurations containing the fundamental boundary string in addition to bulk strings, representing excitations built on a distinct boundary state; and (for $\gamma>1$) 
(iii) configurations containing a higher-order boundary string together with bulk strings, corresponding to yet another boundary configuration.  
Following Ref.~\cite{zhakenov2025thermodynamics}, we refer to each such family as a 
\emph{tower} of states, denoting the purely bulk one as 
\(\mathcal{T}_\mathrm{str}\), the boundary-string tower as 
\(\mathcal{T}_\mathrm{BS}\), and the higher-order boundary-string tower as 
\(\mathcal{T}_\mathrm{hBS}\).  
Depending on \(\gamma\), the Hilbert space thus splits into up to three 
distinct towers, each characterized by its boundary configuration and associated impurity free energy.

\subsection{Modified Bethe and TBA Equations for Multiple Towers}

The presence of boundary strings modifies the Bethe Ansatz equations by introducing tower-dependent driving terms that encode the impurity coupling.  
For each tower \(k \in \{\mathrm{str}, \mathrm{BS}, \mathrm{hBS}\}\), the logarithmic Bethe equations take the generic form
 \begin{equation}
\rho_n^h(\lambda) = f_n^{k}(\lambda) - \sum_{m=1}^\infty (A_{n,m} * \rho_m)(\lambda),
\label{rhoeq towers}
\end{equation}
analogous to Eq.~\eqref{rhoeq}, but with the impurity-dependent function 
\(f_n^{(k)}(\lambda)\) replacing the single-tower expression of the AF1 regime.  
The function \(f_n^{(k)}(\lambda)\) differs among towers only in its 
boundary contribution, while the bulk convolution structure remains the same:
\begin{equation}
\begin{aligned}\label{f k}
f_n^{\mathrm{str}}&=2N Y_{n,2s}(\lambda)+K_n(\lambda)+\sum_\pm K_{n}(\lambda\pm i \gamma)\;,\\
f_n^{\mathrm{BS}}&=2N Y_{n,2s}(\lambda)+K_n(\lambda)-\sum_\pm K_{n}(\lambda\pm i (\gamma-1))\;,\\
f_n^{\mathrm{hBS}}&=2N Y_{n,2s}(\lambda)+K_n(\lambda)-\sum_\pm K_n(\lambda\pm i(\gamma-m-1))\\&-2\sum_{l=1}^m \sum_\pm K_{n}(\lambda\pm i (\gamma-l))\;.
\end{aligned}
\end{equation}
Consequently, the bulk part of the free energy is identical for all towers, and the total free energy of each tower can be written as
\begin{equation}
F^{(k)} = F_\mathrm{bulk} + F_\mathrm{imp}^{(k)}.
\end{equation}

This decomposition implies that the total partition function can be expressed as a statistical sum over tower contributions:
\begin{equation}
Z = \sum_{k\in\mathrm{towers}} e^{-\beta F^{(k)}}
  = e^{-\beta F_\mathrm{bulk}}
    \sum_{k\in\mathrm{towers}} e^{-\beta F_\mathrm{imp}^{(k)}}\;.
\end{equation}
Using the impurity residual free energy defined in the previous section in \eqref{F imp} as $e^{-\beta F_\mathrm{imp}}=Z/e^{-\beta F_\mathrm{bulk}}$, we get
\begin{equation}
e^{-\beta F_\mathrm{imp}}
  = \sum_{k\in\mathrm{towers}} e^{-\beta F_\mathrm{imp}^{(k)}}.
\end{equation}
Each tower thus contributes independently to the impurity thermodynamics through its corresponding impurity free energy \(F_\mathrm{imp}^{(k)}\).

To evaluate \(F_\mathrm{imp}^{(k)}\), one proceeds as in Sec.~\ref{sec:TBA}, using the Takahashi identity~\eqref{Takahashi}.  
For \(\gamma>1/2\), however, the impurity-dependent part of \(f_n^{(k)}(\lambda)\) extends beyond the principal analytic strip \((\mathbb{R}-i/2,\mathbb{R}+i/2)\), requiring an appropriate continuation of the identity to include the boundary-string contributions.  
This leads to tower-specific expressions for the impurity free energy. Below, we summarize the thermodynamic expressions for each of the non-Kondo phases.

\subsubsection{Phase AF2}
For intermediate antiferromagnetic impurity coupling strength $J/g \in (\frac{4}{s(s+1)}, \frac{16}{4s+1})$, the corresponding impurity parameter is within the range $\gamma\in(1/2, s)$. For $\gamma>1/2$, in addition to the string tower $\mathcal{T}_\mathrm{str}$, the BAE \eqref{BAE p} allows for the boundary-string tower $\mathcal{T}_\mathrm{BS}$; and for $\gamma>1$, on top of the two, there appears the higher-order boundary string tower $\mathcal{T}_\mathrm{hBS}$. While base states of all three towers are of the same energy, different thermodynamic functions $f^{(k)}_n$ \eqref{f k} result in different free energy expressions.
Introducing 
$L_m(\lambda)=\ln[1+\eta_m(\lambda)]$ and
\begin{equation}
G_c(\lambda)
=
\sum_{\pm}
\frac{1}{2
\cosh\!\left[
\pi\left(\lambda\pm\frac{ic}{2}\right)
\right]}
\end{equation}
 the free energies are (up to the same constant energy term)
\begin{equation}
\begin{pmatrix}
F^{\mathrm{str}}_\mathrm{imp}\\
F^{\mathrm{BS}}_\mathrm{imp}
\end{pmatrix} =\frac{T}{2}\int \dd \lambda \begin{pmatrix}
    -G_{2\gamma -1}(\lambda) & G_{2-2\gamma}(\lambda) \\
    0 & G_{2\gamma-1}(\lambda)
\end{pmatrix} \begin{pmatrix}
    L_2 (\lambda) \\ L_1 (\lambda)
\end{pmatrix}\,.
\end{equation}

As $\gamma >1$, the higher-order boundary string tower $\mathcal{T}_\mathrm{hBS}$ is formed. Denoting
\begin{equation}
n=\lfloor2\gamma\rfloor,\qquad
\epsilon=2\gamma-n,\qquad
\bar\epsilon=1-\epsilon\,,
\end{equation}
the free energy expressions take the form 
\begin{equation}
    \begin{pmatrix}
F^{\mathrm{str}}_\mathrm{imp}\\
F^{\mathrm{BS}}_\mathrm{imp}\\
F^{\mathrm{hBS}}_\mathrm{imp}
    \end{pmatrix} 
    =
    \frac{T}{2}\int \dd \lambda 
    \begin{pmatrix}
    -G_{\epsilon} & G_{\bar\epsilon} & 0 & 0 \\
    0 & 0 & G_{\epsilon} & -G_{\bar\epsilon} \\
    0 & G_{\bar\epsilon} & G_{\epsilon} & 0
\end{pmatrix} \begin{pmatrix}
    L_{n+1} \\ L_n \\ L_{n-1} \\ L_{n-2}
\end{pmatrix}\,.
\end{equation}

\subsubsection{Phase AF3}
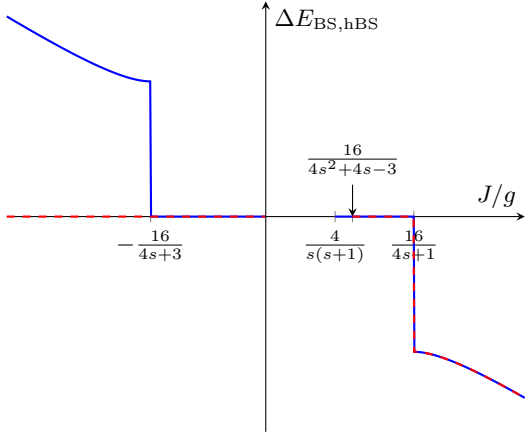
\begin{figure}
    \centering
    \begin{tikzpicture}
\begin{axis}[
    samples=300,
    axis lines=middle,
    xlabel={$J/g$},
    ylabel={$\Delta E_{\mathrm{BS,hBS}}$},
    ymin=-10, ymax=10,
    restrict y to domain=-50:50,
    xtick=\empty,
    ytick=\empty,
    extra x ticks={1.07, 1.34, 16/7, -16/9},
    extra x tick labels={
        $\frac{4}{s(s+1)}$,
        ,
        $\frac{16}{4s+1}$,                       
        $-\frac{16}{4s+3}$
    },
]
\node (L1) [anchor=south, yshift=12pt, font=\small] 
    at (axis cs:1.34, 0) {$\frac{16}{4s^2+4s-3}$};

\draw[->, >=stealth, thin] (L1.south) -- (axis cs:1.34, 0);

\addplot[blue, thick, domain=-4:-0.01]
{
(
 sqrt((16-16*x))/(2*sqrt(-x)) < 2.5
)
*
(
-(2*pi)/cos(deg(pi*(sqrt((16-16*x))/(2*sqrt(-x)) - 3/2)))
)
};

\addplot[blue, thick, domain=1.07:4]
{
(
 sqrt((16*x-16))/(2*sqrt(x)) > 1.5
)
*
(
-(2*pi)/cos(deg(pi*(sqrt((16*x-16))/(2*sqrt(x)) - 3/2)))
)
};


\addplot[red, thick, dashed, domain=-4:-0.01]
{
0
};

\addplot[red, thick, dashed, domain=1.34:4]
{
(
 sqrt((16*x-16))/(2*sqrt(x)) > 1.5
)
*
(
-(2*pi)/cos(deg(pi*(sqrt((16*x-16))/(2*sqrt(x)) - 3/2)))
)
};

\end{axis}
\end{tikzpicture}
    \caption{The energy of the base states of the boundary string and higher-order boundary string towers with respect to the string tower. The fundamental boundary-string tower (solid blue line) exists everywhere but in the phase AF1 ($J/g\in(0,\frac{4}{s(s+1)})$ and is gapped with $E_\gamma$ [see Eq.~\eqref{eq:Egamma}] in the phases AF3 and F2. The boundary string tower (dashed red line) is absent in the phase AF1 and the first part of the phase AF2, it gets gapped only in the phase AF3.}
    \label{fig:towers energy}
\end{figure}

The strong antiferromagnetic phase AF3 corresponds to the couplings ratio $J/g>\frac{16}{4s+1}$ and the impurity parameter $\gamma \in (s, s+1/2)$. 
In this phase, the spin chain prefers to form the bound modes with the impurity -- both boundary string and higher-order boundary string base states become the ground state, and the string tower carrying $(s+1)/(2s+1)$ of all the states starts with an energy gap $|E_\gamma|$:
\begin{widetext}
\begin{equation}\label{eq:Egamma}
    E_\gamma = -\frac{2\pi g}{\cos(\pi(\gamma-s))}\;.
\end{equation} 
The impurity parameter $\gamma = -id$  is given explicitly in \eqref{Jgamma} as $\gamma=\sqrt{\left(s+\frac{1}{2}\right)^2-\frac{4g}{J}}$. This implies that the energy difference $E_\gamma$ between the boundary string towers $\mathcal{T}_\mathrm{BS}$, $\mathcal{T}_\mathrm{hBS}$ and the string tower $\mathcal{T}_\mathrm{str}$  is monotonically growing with the ratio $J/g$ (see Fig.~\ref{fig:towers energy}).

The free energy contributions of the towers are
    \begin{equation}\label{Fs AF2}
    \begin{pmatrix}
F^{\mathrm{str}}_\mathrm{imp}\\
F^{\mathrm{BS}}_\mathrm{imp}\\
F^{\mathrm{hBS}}_\mathrm{imp}
    \end{pmatrix} 
    =\begin{pmatrix}
        0 \\ E_\gamma \\ E_\gamma
    \end{pmatrix}+
    \frac{T}{2}\int \dd \lambda 
    \begin{pmatrix}
    -G_{2\gamma-2s} & G_{2s+1-2\gamma} & 0 & 0 \\
    0 & 0 & G_{2\gamma-2s} & -G_{2s+1-2\gamma} \\
    0 & G_{2s+1-2\gamma} & G_{2\gamma-2s} & 0
\end{pmatrix} \begin{pmatrix}
    L_{2s+1} \\ L_{2s} \\ L_{2s-1} \\ L_{2s-2}
\end{pmatrix}\,
\end{equation}

\subsubsection{Phase F2}
When the impurity parameter crosses the value $\gamma=s+\tfrac12$, the system enters the strong ferromagnetic phase F2 $\gamma\in (s+\tfrac12,s+1)$ with the couplings ratio $J/g<-\tfrac{16}{4s+3}$. In this phase, because of the ferromagnetic nature of the impurity coupling with the spin chain, it is energetically unfavorable for the system to form a bound singlet with the impurity, and the fundamental boundary string tower $\mathcal{T}_\mathrm{BS}$ gaps out with $E_\gamma>0$ (see Fig.~\ref{fig:towers energy}), The string and the higher-order boundary string towers start at the same energy. The free energy contributions of these towers in this phase are
    \begin{equation}
    \begin{pmatrix}
F^{\mathrm{str}}_\mathrm{imp}\\
F^{\mathrm{BS}}_\mathrm{imp}\\
F^{\mathrm{hBS}}_\mathrm{imp}
    \end{pmatrix} 
    =\begin{pmatrix}
        0 \\ E_\gamma \\ 0
    \end{pmatrix}+
    \frac{T}{2}\int \dd \lambda 
    \begin{pmatrix}
    -G_{2\gamma-2s-1} & G_{2s+2-2\gamma} & 0 & 0 \\
    0 & 0 & G_{2\gamma-2s-1} & -G_{2s+2-2\gamma} \\
    0 & G_{2s+2-2\gamma} & G_{2\gamma-2s-1} & 0
\end{pmatrix} \begin{pmatrix}
    L_{2s+2} \\ L_{2s+1} \\ L_{2s} \\ L_{2s-1}
\end{pmatrix}\,.
\end{equation}
\end{widetext}

\subsubsection{Phase F1}
The last in the list is the weak ferromagnetic phase $J/g\in (-\tfrac{16}{4s+3}, 0)$. Similar to the AF2'' phase, all three towers start from the same energy, and the free energy expressions follow \eqref{Fs AF2} with $\gamma \in (s+1, \infty)$.

The closed form of the free energy contributions of each of the towers in terms of the ratios $\eta_n$ together with the limit values of these ratios (see Sec.~\ref{sec:T0Tinfty}), allow for the analytic expressions for the impurity entropy at zero and infinite temperatures. Numerical solutions (see Sec.~\ref{sec:NTBA}) of the TBA equations \eqref{TBA p} make it possible to connect the impurity entropy at the limit values and get its full temperature dependence. All of this is covered in the next section.

\subsection{Impurity Entropy Across Phases}
\label{sec:EntropyPhases}

Having expressed the total impurity free energy as a sum of independent tower contributions,
\begin{equation}
e^{-\beta F_\mathrm{imp}}
  = \sum_{k\in\mathrm{towers}} e^{-\beta F_\mathrm{imp}^{(k)}},
\end{equation}
the impurity entropy follows from the thermodynamic relation
\begin{equation}
\begin{aligned}
&S_\mathrm{imp}(T)
  = -\frac{\partial F_\mathrm{imp}}{\partial T}
  \\&= \ln \left(\sum_k e^{-\beta F_\mathrm{imp}^{(k)}}\right)
    + \frac{\sum_k \beta e^{-\beta F_\mathrm{imp}^{(k)}}
      \frac{\partial F_\mathrm{imp}^{(k)}}{\partial \beta}}
           {\sum_k e^{-\beta F_\mathrm{imp}^{(k)}}}\,.
\end{aligned}
\end{equation}
At low temperatures, the towers with the smallest base energy dominate the sum, 
while at higher temperatures, multiple towers contribute, leading to the characteristic nonmonotonic features in \(S_\mathrm{imp}(T)\).

At \(T \to 0\), the impurity entropy reflects the ground-state degeneracy of the 
lowest-energy towers.  
In the first part of the AF2 regime (\(\gamma\in(\tfrac12,1)\)), the 
\(\mathcal{T}_\mathrm{str}\) and \(\mathcal{T}_\mathrm{BS}\) towers share the same 
base energy, yielding the residual entropy  
\[
S_\mathrm{imp}(0)
 = \ln[e^{S^\mathrm{str}_\mathrm{imp}(0)}+e^{S^\mathrm{BS}_\mathrm{imp}(0)}]
 = \ln \left[2\cos \left(\frac{\pi}{2s+2}\right)\right],
\]
where
\[
S^\mathrm{str}_\mathrm{imp}(0)
 = \ln \left(
   \frac{\sin[3\pi/(2s+2)]}
        {\sin[2\pi/(2s+2)]}
 \right)\;,\]
 \[
S^\mathrm{BS}_\mathrm{imp}(0)
 = \ln \left(
   \frac{\sin[\pi/(2s+2)]}
        {\sin[2\pi/(2s+2)]}
 \right)
\]
are the respective tower contributions. 

In the second part of the AF2 phase $\gamma\in (1, s)$, there appears the higher-order boundary string tower $\mathcal{T}_\mathrm{hBS}$. The contributions of each tower can be written as
\[
S^\mathrm{str}_\mathrm{imp}(0)
 = \ln \left(
   \frac{\sin[\pi(\lfloor2\gamma\rfloor+2)/(2s+2)]}
        {\sin[\pi(\lfloor2\gamma\rfloor+1)/(2s+2)]}
 \right)\;,\]
 \[
S^\mathrm{BS}_\mathrm{imp}(0)
 = \ln \left(
   \frac{\sin[\pi(\lfloor2|\gamma-1|\rfloor+1)/(2s+2)]}
        {\sin[\pi(\lfloor2|\gamma-1|\rfloor+2)/(2s+2)]}
 \right)
\]
 \[
S^\mathrm{hBS}_\mathrm{imp}(0)
 = \ln \left(
   \frac{1}
        {\sin[\pi\tfrac{\lfloor2|\gamma-1|\rfloor}{2s+2}]\sin[\pi\tfrac{\lfloor2|\gamma-1|\rfloor+1}{2s+2}]}
 \right)
\]
Altogether, they yield the same 
\begin{multline*}
S_\mathrm{imp}(0)
 = \ln[e^{S^\mathrm{str}_\mathrm{imp}(0)}+e^{S^\mathrm{BS}_\mathrm{imp}(0)}+e^{S^\mathrm{hBS}_\mathrm{imp}(0)}]
 \\= \ln \left[2\cos \left(\frac{\pi}{2s+2}\right)\right],    
\end{multline*}

For stronger antiferromagnetic coupling (\(\gamma\in(s,s+\tfrac12)\), AF3 phase), 
the \(\mathcal{T}_\mathrm{BS}\) and \(\mathcal{T}_\mathrm{hBS}\) towers become degenerate 
and dominate the low-temperature thermodynamics, while the string tower is gapped by 
\(|E_\gamma|\).  
The zero-temperature entropy remains the same,
\[
S_\mathrm{imp}(0)
 = \ln[e^{S^\mathrm{BS}_\mathrm{imp}(0)}+e^{S^\mathrm{hBS}_\mathrm{imp}(0)}]
 = \ln \left[2\cos \left(\frac{\pi}{2s+2}\right)\right].
\]
In this phase, the impurity entropy decreases with increasing temperature as the system 
locks into localized boundary states, signaling the formation of boundary-bound modes that remove available degrees of freedom by hybridizing with the impurity.

Upon entering the ferromagnetic regime, the tower hierarchy reverses.  
In the F2 phase (\(\gamma\in(s+\tfrac12,s+1)\)), the 
\(\mathcal{T}_\mathrm{str}\) and \(\mathcal{T}_\mathrm{hBS}\) towers lie lowest in 
energy, while the \(\mathcal{T}_\mathrm{BS}\) tower becomes gapped.  
As \(\gamma\) increases further (\(\gamma>s+1\), F1 phase), all three towers become 
degenerate again, each contributing equally to the partition function from \(T=0\).  
In both ferromagnetic phases, the impurity entropy returns to 
\[
S_\mathrm{imp}(0)=\ln 2,
\]
corresponding to the twofold degeneracy of the unscreened spin-\(\tfrac12\) impurity.

At high temperatures (\(T \to \infty\)), all towers contribute equally regardless of 
their base energies, each giving
\begin{equation}
\begin{aligned}
S^\mathrm{str}_\mathrm{imp}(\infty)
 &= \ln \left(\frac{\lfloor2\gamma\rfloor+2}{\lfloor2\gamma\rfloor+1}\right),\\
S^\mathrm{BS}_\mathrm{imp}(\infty)
 &= \begin{cases}
   \ln(1/2), & \gamma \in (\tfrac12,\tfrac32),\\[2mm]
   \ln \left(\frac{\lfloor2\gamma\rfloor-1}{\lfloor2\gamma\rfloor}\right),
     & \gamma \ge \tfrac32,
   \end{cases}\\
S^\mathrm{hBS}_\mathrm{imp}(\infty)
 &= \ln \left[\frac{1}{\lfloor2\gamma\rfloor(\lfloor2\gamma\rfloor-1)}\right].
\end{aligned}
\end{equation}
These contributions combine as  
\[
S_\mathrm{imp}(\infty)
 = \ln \left(\sum_k e^{S^k_\mathrm{imp}(\infty)}\right)
 = \ln 2,
\]
independent of \(\gamma\), confirming that the total Hilbert space is complete and 
accounts for the two degrees of freedom of the spin-\(\tfrac12\) impurity.

The resulting temperature dependence of the impurity entropy is therefore 
nonmonotonic (see Fig.~\ref{fig:Simp_S1} for $s=1$ and Fig.~\ref{fig:Simp_S1.5} for $s=3/2$).  
In the non-Kondo phases, \(S_\mathrm{imp}(T)\) exhibits pronounced dips as 
boundary-bound modes remove degrees of freedom.   
These features, obtained from the numerical solution of the multi-tower TBA equations, 
are in excellent quantitative agreement with the MPO simulations discussed in the next section.  
Together, they confirm that the multi-tower decomposition accurately captures the 
thermodynamics of all boundary regimes of the model.

\section{Finite-Temperature MPO Simulation of a Spin-$s$ Chain with Impurity}

In this section, we discuss the computation of impurity contributions to thermodynamic properties of the spin $s$ chain using tensor network methods, which is feasible for small $s\leq 3/2$ and sufficiently high temperatures.
We first discuss how we obtain thermodynamic properties of a given spin chain. 
These are then computed with $N_{\mr{bulk}}$ bulk spins, both with and without an impurity, and the $O(1)$ impurity contribution is obtained by subtracting the pure-bulk result from the result with the impurity.
Since we are interested in the thermodynamic limit, $N_{\mr{bulk}} \to \infty$, we also discuss how the finite-size error due to finite $N_{\mr{bulk}}$ can be estimated without cumbersome extrapolations in $N_{\mr{bulk}}$.
All of our computations are based on the QSpace tensor network library~\cite{Weichselbaum2012,Weichselbaum2020,Weichselbaum2024}, which allows us to exploit the SU(2) spin rotation symmetry.


We work with a finite-size chain containing $N_{\mr{bulk}}$ bulk sites and represent the square root of its thermal density matrix $\rho(\beta)$ as a matrix product operator~(MPO)~\cite{Verstraete2004,Zwolak2004,Feiguin2005},
\begin{align}
\varrho(\beta) &= \frac{\mathrm{e}^{-\beta H/2}}{\sqrt{Z}} \, , \; \; Z = \mathrm{Tr} \, \mathrm{e}^{-\beta H} \, ,
\\
\rho(\beta) &= \varrho^{\dagger}(\beta) \varrho(\beta) \, ,
\end{align}
where $H$ is the Hamiltonian and $Z$ the partition function.
Since $\mathrm{Tr} \, \varrho^{\dagger}(\beta) \varrho(\beta) = 1$, $\varrho(\beta)$ behaves like a normalized quantum state in a doubled Hilbert space, and standard tensor network techniques can be used for (imaginary) time evolution or compression.

Several algorithms exist to compute $\varrho(\beta)$. 
We choose to start from $T \to \infty$, where $\varrho(0) \propto \mathbf{1}$, and use either time evolving block decimation~(TEBD)~\cite{Vidal2004} or the exponential tensor renormalization group~(XTRG)~\cite{Chen2018} to lower the temperature.

Within TEBD, we compute
\begin{align}
\tilde{\varrho}(\tau + 2\delta \tau) = \mathrm{e}^{-\delta\tau H} \varrho(\tau)
\end{align}
for small time steps $\delta\tau$, using a third-order Trotter decomposition, which has an error of $\mathcal{O} (\delta\tau^{4})$.
We then normalize,
\begin{align}
\varrho(\tau + 2 \delta \tau) &= \frac{\tilde{\varrho}(\tau + 2 \delta \tau)}{\sqrt{\delta Z(\tau + 2 \delta \tau)}} \, , 
\\
\delta Z(\tau + 2 \delta \tau) &= \mathrm{Tr} \, \tilde{\varrho}^{\dagger}(\tau + 2 \delta \tau) \tilde{\varrho}(\tau + 2 \delta \tau) \, ,
\end{align}
where $\delta Z(\tau + 2 \delta \tau)$ is the corresponding change in the partition function due to the cooling step. 
Besides the Trotter error, there is a truncation error when trimming the internal bonds of the MPO representation of $\varrho$. 
The computation time scales as $\mathcal{O}(D^3)$, where $D$ is the internal bond dimension of $\varrho$.

The partition function, free energy, and entropy after $n$ cooling steps are given by
\begin{subequations}
\label{subeq:MPO_Z_F}
\begin{align}
Z(\beta) &= Z(0) \prod_{m=1}^{n} \delta Z(2 m \delta\tau)
\\
F(\beta) &= -\frac{1}{\beta} \big[\ln Z(0) + \sum_{m=1}^{n} \ln \delta Z(2 m \delta\tau)\big] \, , 
\\
\label{eq:MPO_entropy}
S(\beta) &= \beta (F(\beta) - \langle H \rangle_{\beta}) \, , \; \; \langle H \rangle_{\beta} = \mathrm{Tr} \varrho^{\dagger}(\beta) H \varrho(\beta) \, ,
\end{align}
\end{subequations}
where $ \beta = 2 n \delta\tau$ is the inverse temperature after $n$ steps.
We choose a timestep of $\delta \tau = 0.05$ for our TEBD computations.

Within XTRG, on the other hand, one starts at some high but finite temperature $T_0 = 1/\tau_0$. 
Cooling is then achieved by squaring $\rho(\tau)$,
\begin{align}
\varrho(2\tau) \propto \varrho(\tau) \varrho(\tau) \, ,
\end{align}
followed by variational compression and normalization.
The partition function, free energy, and entropy are computed analogously to Eq.~\eqref{subeq:MPO_Z_F}. 
After $n$ steps, the inverse temperature is $2^{n} \tau_0$, where $\tau_0$ is the initial inverse temperature before the XTRG steps, i.e.\ XTRG realizes exponential cooling, though with a scaling of $\mathcal{O}(D^4)$.
To obtain $\varrho(\tau_0)$, we use TEBD with a small timestep of $\delta\tau = 0.01$, and we perform two independent computations with $\tau_0 = 0.1$ and $\tau_0 = 0.15$.

Our main objective is to compute the impurity contribution to the entropy, $S_{\mathrm{imp}}(T)$. 
For that, we perform independent computations of the entropy with a boundary impurity, $S_{\textrm{bulk+imp}}(T)$, and without a boundary impurity, $S_{\mathrm{bulk}}(T)$.
Both $S_{\textrm{bulk+imp}}(T)$ and $S_{\mathrm{bulk}}(T)$ are extensive, and their subtraction yields the intensive impurity contribution,
\begin{align*}
\label{eq:MPO_Simp}
S_{\mathrm{imp}}(T) = S_{\textrm{bulk+imp}}(T) - S_{\mathrm{bulk}}(T) \, .
\end{align*}

To estimate finite-size errors, we use the fact that a bulk chain \textit{without} boundary impurity has a unique ground state if the number of bulk sites $N_{\mr{bulk}}$ is even, while an odd number of bulk sites leads to a $(2s + 1)$-fold ground state degeneracy.
On the other hand, a bulk chain \textit{with} boundary impurity has a 2-fold ground state degeneracy if $N_{\mr{bulk}}$ is even, while $N_{\mr{bulk}}$ odd leads to a ground state degeneracy of $2s$ for $J_{\mr{imp}} > 0$ or $(2s + 2)$ for $J_{\mr{imp}} < 0$.
Therefore, for finite $N_{\mr{bulk}}$, we find 
\begin{align}
\lim_{T \to 0} S_{\mr{imp}}(T) = 
\begin{cases}
\log 2 & N_{\mr{bulk}} \;\; \text{even} \\
\log \frac{2s}{2s + 1} & N_{\mr{bulk}} \;\; \text{odd, } J_{\mr{imp}} > 0 \\
\log \frac{2s + 2}{2s + 1} & N_{\mr{bulk}} \;\; \text{odd, } J_{\mr{imp}} < 0 \, .
\end{cases}
\end{align}
For the results reported below, we set some number $N$, perform two independent computations with $N_{\mr{bulk}} = N$ and $N_{\mr{bulk}} = N+1$ bulk sites, and discard data for low temperatures where the resulting $S_{\mr{imp}}(T)$ differs by more than $10^{-2}$ between these two computations.

\begin{figure}
    \centering
    \includegraphics[width=\linewidth]{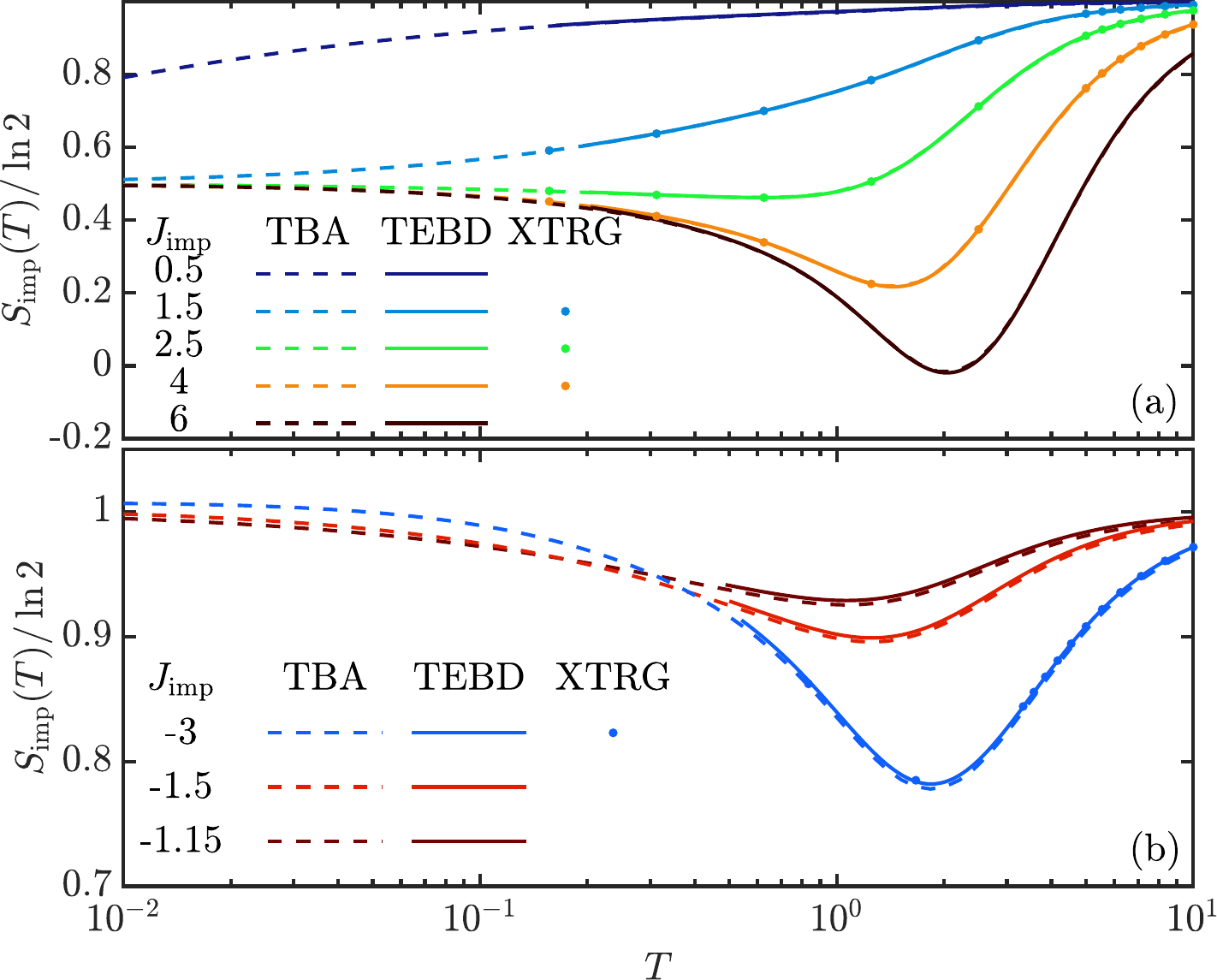}
\caption{Impurity contribution to the entropy for a spin-$s=1$
chain. We show $S_{\rm imp}(T)=S_{\textrm{bulk+imp}}(T)-S_{\rm bulk}(T)$ 
for antiferromagnetic impurity coupling in panel (a) and ferromagnetic impurity coupling in panel (b). Dashed lines show
the TBA result, solid lines show TEBD data, and dots
show XTRG data. The MPO calculations use $N_{\mr{bulk}} = 100$ bulk sites, and data are
shown only above the finite-size cutoff determined from the comparison of
$N_{\mr{bulk}}$ and $N_{\mr{bulk}}+1$ chains. For $s=1$, the positive-coupling phase boundaries
are at $J_{\rm imp}=2$ and $J_{\rm imp}=3.2$, while the ferromagnetic
F1--F2 boundary is at $J_{\rm imp}=-16/7 \simeq -2.29$. The nonmonotonic entropy dips
at stronger coupling reflect the formation of boundary-bound modes.}
    \label{fig:Simp_S1}
\end{figure}

In Fig.~\ref{fig:Simp_S1}, we show the resulting impurity contribution to the entropy for an $s=1$ bulk chain, using TEBD (solid lines), XTRG (dots), and TBA (dashed lines). 
For both $J_{\mr{imp}} > 0$ [Fig.~\ref{fig:Simp_S1}(a)] and $J_{\mr{imp}} < 0$ [Fig.~\ref{fig:Simp_S1}(b)], the TBA results agree very well with the MPO simulations. 
The latter are performed for $N = 100$, which allows us to reach temperatures somewhat below $0.2$ before finite-size effects become relevant.

For $J_{\mr{imp}} \geq 1.5$, these temperatures are low enough to see that in the thermodynamic limit, $\lim_{T\to 0} S_{\mr{imp}}(T) = 0.5 \log 2$. 
Further, for $J_{\mr{imp}} > 2$, the MPO simulations capture the non-monotonicity of $S_{\mr{imp}}(T)$ due to the bound mode formation at the boundary, which leads to a minimum of $S_{\mr{imp}}(T)$ at a temperature of order $1$.
Similarly, the bound mode formation at $J_{\mr{imp}} < 0$, which again leads to a non-monotonic $S_{\mr{imp}}(T)$ exhibiting a minimum, is captured by the MPO computations and is in good agreement with the TBA.

\begin{figure}
    \centering
    \includegraphics[width=\linewidth]{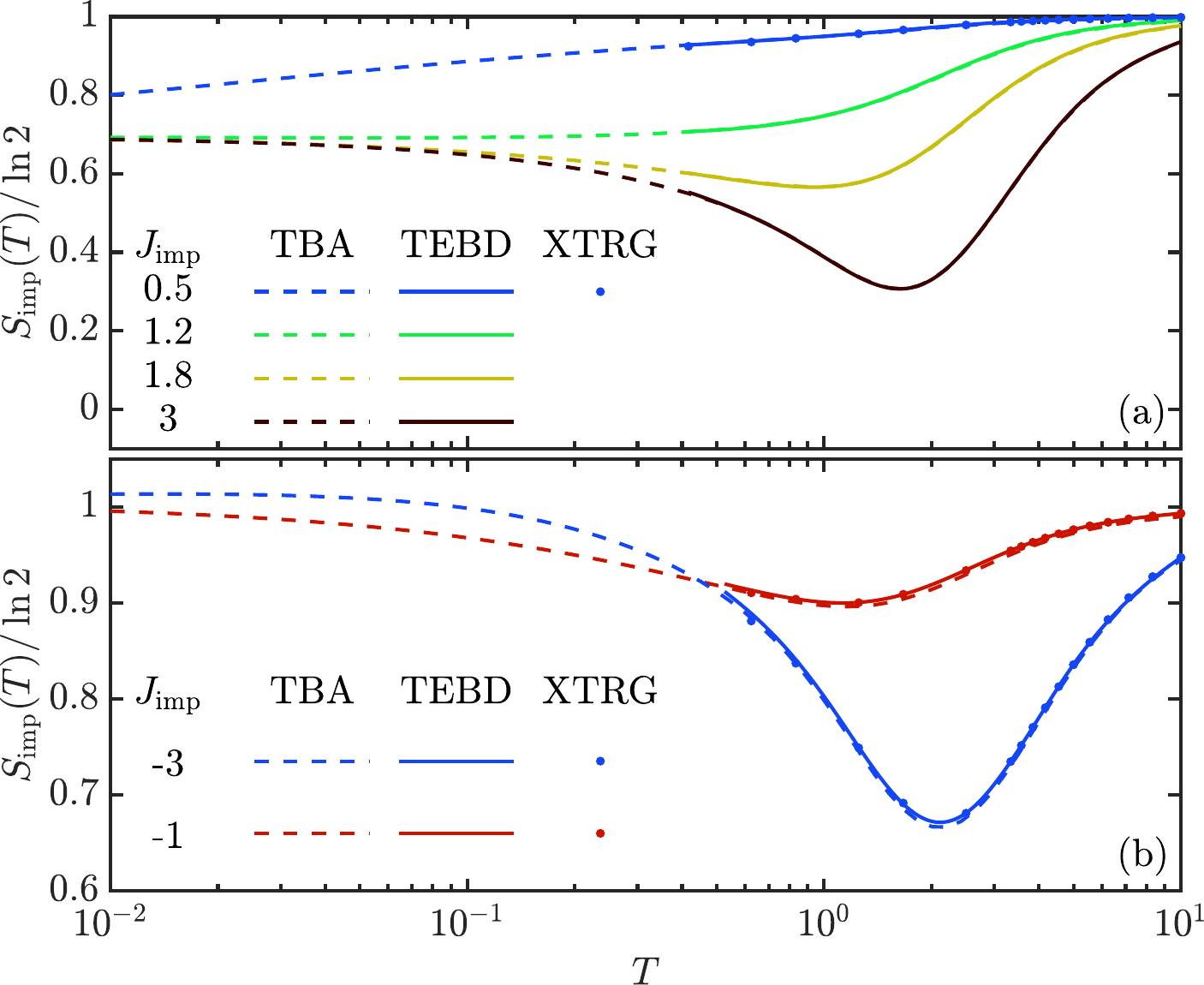}
    \caption{Impurity contribution to the entropy for a spin-$s=3/2$
chain. As in Fig.~\ref{fig:Simp_S1}, $S_{\rm imp}(T)$ is obtained by subtracting the
entropy of the pure bulk chain from that of the chain with a boundary
spin-$1/2$ impurity. Panel (a) shows
antiferromagnetic couplings and panel (b) ferromagnetic couplings.
Dashed lines denote the TBA result, solid lines TEBD, and dots XTRG. The
MPO calculations use $N_{\mr{bulk}}=100$ bulk sites and are restricted to
temperatures where finite-size effects remain below the criterion
described in the text. For $s=3/2$, the positive-coupling phase
boundaries are at $J_{\rm imp}=16/15 \simeq 1.07$ and $J_{\rm imp}=16/7 \simeq 2.29$, and the
ferromagnetic F1--F2 boundary is at $J_{\rm imp}=-16/9\simeq -1.78$.}
    \label{fig:Simp_S1.5}
\end{figure}

Figure~\ref{fig:Simp_S1.5} shows the impurity contribution to the entropy for an $s = 3/2$ bulk chain. 
For the MPO calculations, we again use $N = 100$, which allows us to reach temperatures down to around $0.4$.
We find excellent agreement between MPO and TBA for the temperatures accessible to the MPO calculations. 
For $J_{\mr{imp}} > 1.07$, we expect non-monotonic behavior of $S_{\mr{imp}}(T)$ due to the bound mode.
With our MPO computations, we capture this non-monotonic behavior for $J_{\mr{imp}} \geq 1.8$.
For $J_{\mr{imp}} = 1.2$, this non-monotonic behavior is very weak, and the temperature accessible by our MPO computations is not low enough to capture it.
The non-monotonic behavior in the ferromagnetic regime is captured by the MPO computations for both values of $J_{\mr{imp}} \in \{-1,-3\}$ considered there.

\subsection*{Summary of the Multi-Tower Thermodynamics}

The multi-tower TBA framework provides a unified description of all boundary phases: the number and relative energies of the active towers fix the low-temperature degeneracy, while their progressive thermal activation produces the nonmonotonic \(S_\mathrm{imp}(T)\) seen in the figures above. At high temperature every tower contributes and \(S_\mathrm{imp}\) recovers \(\ln2\), confirming completeness of the boundary spectrum.

Finally, while the thermodynamic analysis presented above elucidates how the boundary coupling reorganizes the Hilbert space and governs the impurity entropy, complementary insight into the underlying dynamics is provided by frequency-resolved observables.  
In particular, the impurity spectral function \(A(\omega)\) serves as a direct probe of
the spin-flip processes associated with the different boundary
towers.  The following section is devoted to this dynamical perspective, where the
behavior of \(A(\omega)\) across the phase diagram is analyzed in detail.

\section{Dynamical probes : Impurity spectral function}\label{dynamics}

\begin{figure}
    \centering
    \includegraphics[width=\linewidth]{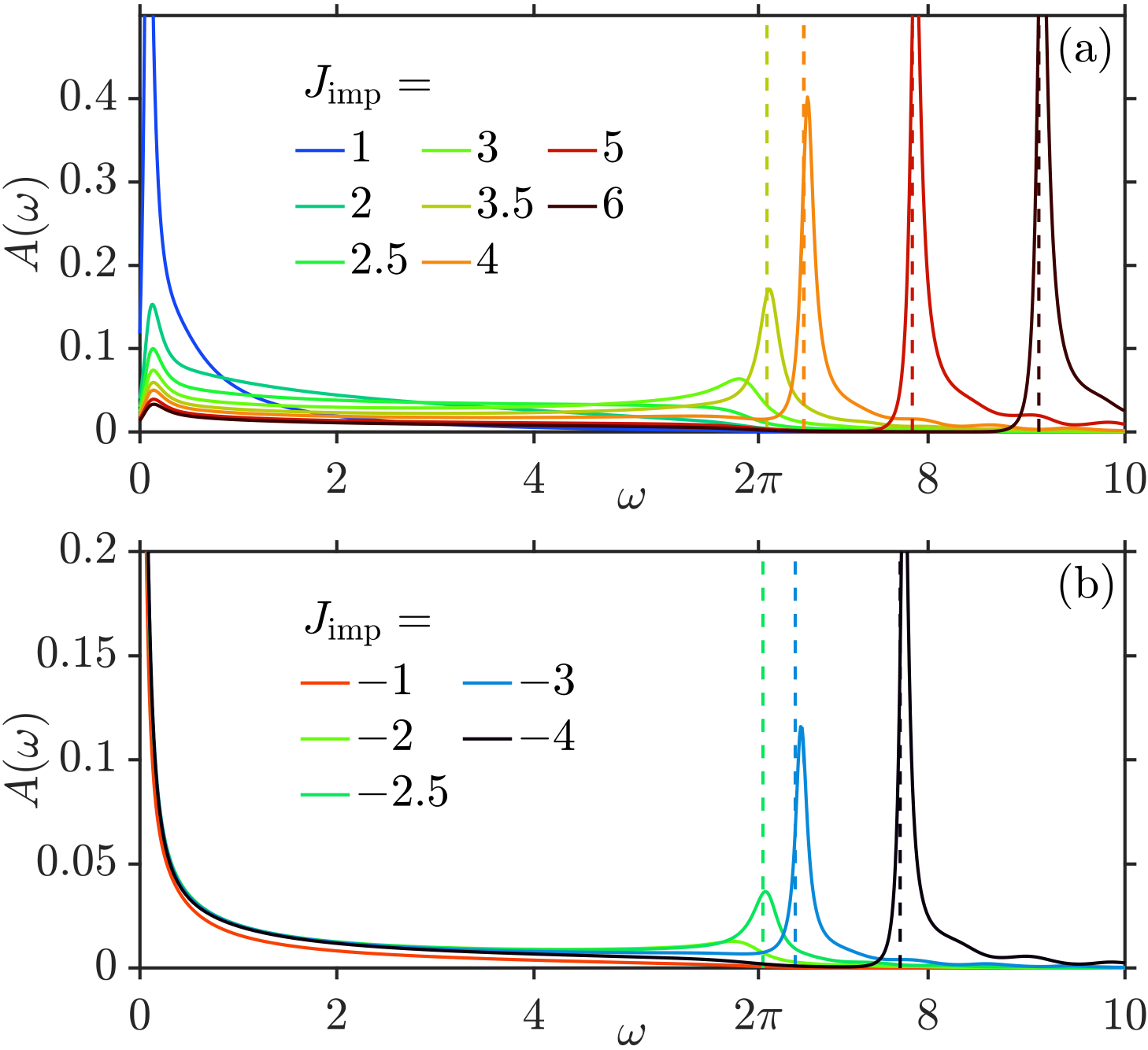}
\caption{Impurity spectral function $A(\omega)$ for the spin-$s=1$ chain at zero temperature, computed for $N_{\mr{bulk}}=199$ bulk spins using the tangent-space Krylov method. The discrete spectra obtained from $N_{\mr{kr}}=300$ Krylov iterations are broadened with a CFE+Gaussian scheme using $\sigma=0.25$ and $N_{\rm CFE}=4$. 
Panel (a) shows antiferromagnetic impurity couplings. For the gapless AF1 and AF2 phases ($0<J_{\mr{imp}}<3.2$), the spectral weight is confined within the bulk continuum $\omega\lesssim 2\pi$. In the AF3 phase ($J_{\rm imp}>3.2$), a sharp threshold appears above the continuum. This peak marks the onset of the \textit{string tower} sector; the high spectral weight reflects the extensive number of modes participating in the overscreening Kondo cloud. 
Panel (b) shows ferromagnetic couplings. In F1 ($-16/7<J_{\rm imp}<0$), weight remains in the continuum, while in F2 ($J_{\rm imp}<-16/7 \simeq -2.29$), a dynamical threshold appears at $\omega > 2\pi$. This feature signifies the activation of a \textit{boundary string tower}. The vertical dashed lines in both panels mark the boundary gap energy $E_\gamma$ predicted by the Bethe Ansatz, representing the threshold frequency at which these specific towers of states become kinematically accessible to the impurity.}
    \label{fig:Spectral_S1}
\end{figure}

%
The thermodynamics established above fixes which boundary towers are active in each phase but is insensitive to their structure. The impurity spectral function $A(\omega)$ supplies this complementary dynamical view, resolving how the different towers participate in the impurity's fluctuations. At $T=0$,
\begin{align}
\label{eq:A_def}
A(\omega) =  \langle \Psi_0 | S^{z}_{\mr{imp}} \delta(\omega - H + E_{0}) S^{z}_{\mr{imp}} | \Psi_0 \rangle \, ,
\end{align}
where $| \Psi_0 \rangle$ is the ground state of $H$ with energy $E_0$ and $S^{z}_{\mr{imp}}$ is the impurity spin-$z$ operator.

Physically, $A(\omega)$ represents the weighted density of modes at frequency $\omega$ that are coupled to the impurity. The different towers contribute distinct spectral signatures based on the spatial extent of the screening they represent. The \textit{string tower} corresponds to the formation of an extensive overscreening Kondo cloud. Since the impurity is entangled with an extensive region of the bulk, a large number of modes participate in the dynamics. Conversely, the \textit{boundary string towers} describe the formation of localized boundary-bound modes; this localization suggests a smaller set of participating modes, which should result in a lower spectral weight.

In the gapless phases (AF1, AF2, and F1), these effects are compounded as modes from multiple towers contribute starting from the lowest energy scales ($\omega \to 0$). However, in the \textbf{AF3} and \textbf{F2} phases, the emergence of a boundary gap $|E_\gamma|$ [see Eq.~\eqref{eq:Egamma}] allows us to approximately resolve the contribution of some of the towers individually. It is crucial to note that the prominent features observed in the spectral function in these gapped regimes are not isolated bound-state peaks. Rather, they represent \textit{dynamical thresholds} marking the onset of a specific tower's contribution to the impurity dynamics.

In the \textbf{AF3} regime, the threshold at $\omega \approx |E_\gamma|$ marks the activation of the string tower. This leads to a sharp rise in spectral weight, followed by a slow decay
reflecting the high density of modes available in the extensive Kondo cloud. In the \textbf{F2} phase, the threshold corresponds to the onset of a localized boundary-bound tower. As expected from the localized nature of these states, the spectral weight is suppressed compared to the AF3 case, indicating that fewer modes are available to participate in the impurity's dynamics.

We use a matrix product state~(MPS) approach to compute this spectral function for a system with $N_{\mr{bulk}} = 199$ bulk spins with $s = 1$.
Our implementation is based on the QSpace tensor library and exploits the SU(2) spin symmetry.
We first compute the ground state $|\Psi_0\rangle$ using the density matrix renormalization group~(DMRG)~\cite{White1992,Schollwock2005,Schollwock2011} 
in its single-site form with controlled bond expansion~\cite{Gleis2023,Li2024,Gleis2022} and mixing~\cite{Hubig2015,Gleis2025}, with an MPS bond dimension of $D^{\ast} = 300$ SU(2) multiplets.
Then, we compute the spectral function Eq.~\eqref{eq:A_def} using the tangent space Krylov~(TaSK)~\cite{Kovalska2025} approach, 
which approximates excitations relevant for $A(\omega)$ within the tangent space of the ground state MPS using an otherwise standard Krylov scheme.
We perform $N_{\mr{kr}} = 300$ Krylov iterations, which results in a discrete spectral function $A_{\mr{disc}}(\omega)$ that is broadened with the 
continued fraction expansion~(CFE) plus Gaussian scheme of Ref.~\cite{Kovalska2025} using $\sigma = 0.25$ as Gaussian broadening width and a CFE-depth of $N_{\mr{CFE}} = 4$.

The resulting spectral functions are shown in Fig.~\ref{fig:Spectral_S1}(a) for $J_{\mr{imp}} > 0$ (AF1-AF3) and (b) for $J_{\mr{imp}} < 0$ (F1,F2).
Irrespective of $J_{\mr{imp}}$, we find that $A(\omega)$ exhibits a continuum for $\omega < 2\pi$, which coincides with the bandwidth of the bulk continuum. 
In the antiferromagnetic regime, the upturn at low frequencies is a finite-size artifact due to a spinon zero mode of the odd finite-size bulk chain interacting with the impurity.
A similar finite-size artifact occurs for even bulk chains, where a ground state degeneracy arises once the impurity is introduced. 
We have checked that with an increasing number of bulk sites, this artifact moves to lower frequencies and becomes less pronounced.
By contrast, the low-frequency upturn in the ferromagnetic regime is a genuine feature that is expected to persist in the thermodynamic limit.
This is because the impurity decouples at low energies, which leads to a zero-frequency pole in $A(\omega)$.
In Fig.~\ref{fig:Spectral_S1}(b), this pole is smeared out due to the applied broadening.

\begin{figure}
    \centering
    \includegraphics[width=\linewidth]{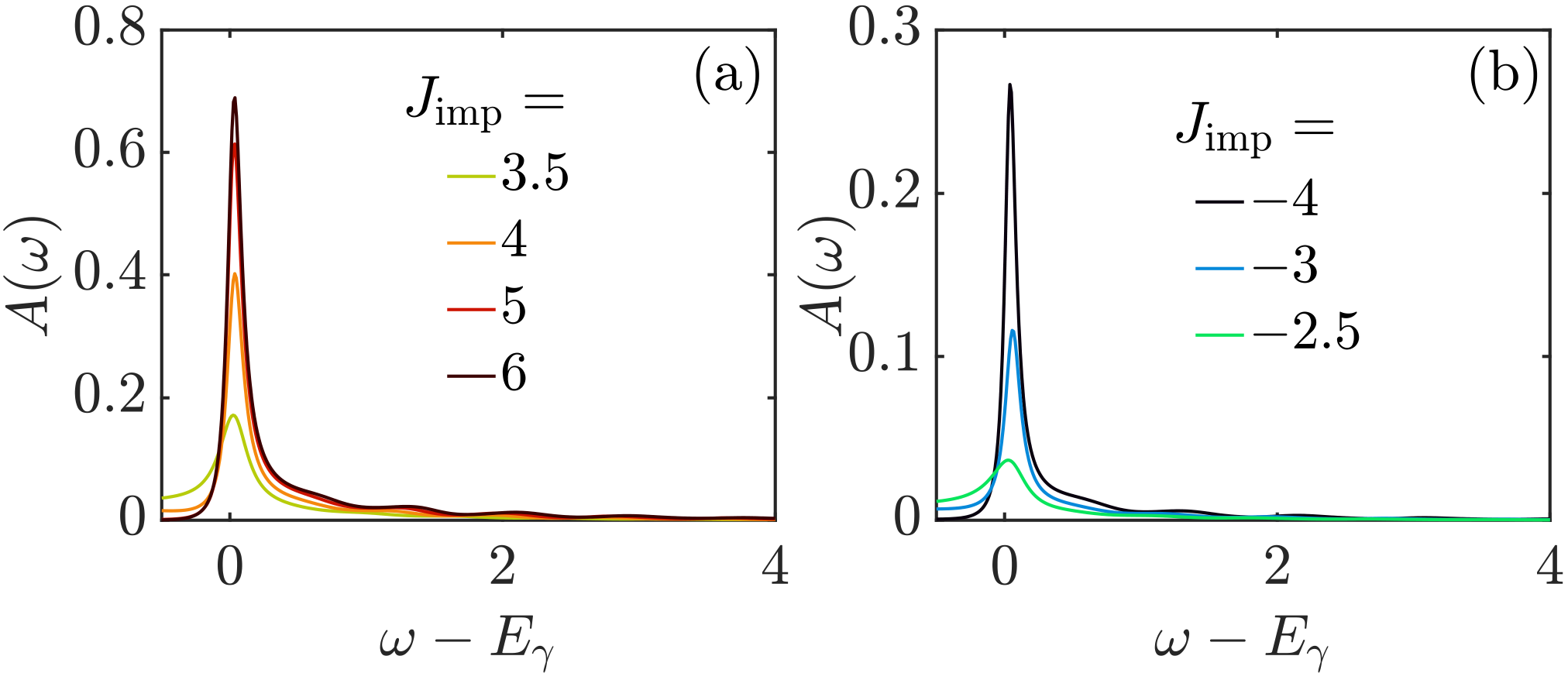}
\caption{Resolution of dynamical thresholds in the impurity spectral function for the spin-$s=1$ chain. The spectra are plotted as a function of the shifted frequency $\omega-|E_\gamma|$, where $E_\gamma$ is the Bethe-Ansatz prediction for the energy of the tower base-state (the boundary gap). 
Panel (a) shows representative couplings in the AF3 phase, where the alignment at $\omega-|E_\gamma|=0$ marks the onset of the \textit{string tower}. The substantial spectral weight and subsequent decay are characteristic of the extensive many-body modes involved in the overscreening process. 
Panel (b) shows representative couplings in the F2 phase, where the threshold marks the onset of a \textit{boundary string tower}. The lower peak intensity compared to the AF3 phase provides a dynamical signature of the localized nature of the F2 boundary configuration, which involves fewer participating modes than the extensive string tower. In both regimes, the onset at $\omega-|E_\gamma|=0$ confirms that the out-of-continuum features in Fig.~\ref{fig:Spectral_S1} are governed by the reorganization of the Hilbert space into distinct excitation towers.}
    \label{fig:Spectral_S1_BM}
\end{figure}

In the AF1 and AF2 phases at $3.2 > J_{\mr{imp}} > 0$ (for $s=1$), we find no
spectral features above the bulk bandwidth, $\omega>2\pi$. These changes
in the AF3 phase, $J_{\mr{imp}}>3.2$, where the antiferromagnetic bound
mode produces a well-separated spectral peak at $|E_\gamma|>2\pi$ [vertical
dashed lines in Fig.~\ref{fig:Spectral_S1}(a)]. A similar structure appears on the ferromagnetic
side. In F1, $-16/7<J_{\mr{imp}}<0$, the spectral weight is confined to $\omega < 2\pi$, whereas in F2,
$J_{\rm imp}<-16/7 \simeq -2.29$, a peak at $|E_\gamma|>2\pi$ signals the
ferromagnetic bound mode. 

This is made more explicit in Fig.~\ref{fig:Spectral_S1_BM}(a) and (b), where we plot
$A(\omega)$ as a function of $\omega-|E_\gamma|$ in the AF3 and F2 phases, respectively. In both cases, the high-frequency peaks align near $\omega-|E_\gamma|=0$,
showing that their position tracks the Bethe-Ansatz prediction for the
boundary gap $|E_\gamma|$.

\section{Discussion}
\label{discussion}

Combining boundary conformal field theory, the exact Bethe Ansatz, finite-temperature MPO simulations, and the impurity spectral function, we have given a unified account of how a spin-$\tfrac12$ boundary impurity is screened in the spin-$s$ Takhtajan--Babujian chain. The unifying mechanism is the reorganization of the Hilbert space, driven by impurity-dependent non-string roots, into a finite set of excitation towers. This single structure controls both the thermodynamics: the boundary phase transitions and the nonmonotonic impurity entropy confirmed by MPO; and the dynamics: the bound-state thresholds in $A(\omega)$. We close by discussing the generality of this mechanism, its experimental signatures, and its fate beyond integrability.

\subsection*{Generality of the Tower-Splitting Mechanism}

The Hilbert-space splitting observed here originates from the presence of impurity-dependent non-string roots in the Bethe Ansatz equations.  
This feature leads naturally to a tower-wise decomposition of the impurity contribution to the free energy.  
We expect this mechanism to be generic to a broad class of integrable boundary systems that (a) admit a structured description of bulk excitations via the string hypothesis and (b) contain impurity parameters giving rise to localized, non-string roots.  
Examples include the open XXX spin chain with boundary impurities~\cite{kattel2023kondo,zhakenov2025thermodynamics}, the Gross–Neveu model with boundary impurities~\cite{kattel2025thermodynamics,rg1x-bztv}, its multichannel generalizations~\cite{fzw2-xk49}, and anisotropic XXZ spin chains.  
Extensions to higher representations of $SU(N)$ integrable models and to field-theoretical limits~\cite{q6kz-2vl2} are also expected to exhibit a similar multi-tower structure, suggesting that tower thermodynamics may represent a universal feature of integrable boundary systems.

\subsection*{Experimental Signatures}

An important open question is whether the tower-splitting phenomenon can be observed experimentally.  
In principle, the coexistence of several boundary configurations should manifest through distinct thermodynamic and dynamical signatures.  
The nonmonotonic temperature dependence of the impurity entropy predicted here—characterized by dips in the antiferromagnetic phases and bumps in the ferromagnetic ones—could appear as anomalies in the specific heat or in the temperature-dependent magnetic susceptibility of quasi-one-dimensional materials such as Bechgaard salts or spin-chain compounds with engineered boundary defects.  
Similarly, in ultracold-atom implementations of spin chains with tunable boundary couplings, entropy measurements from quantum-gas microscopy or thermometry could resolve the transition between boundary phases as $\gamma$ is varied.

From a dynamical perspective, the presence of multiple boundary towers should leave clear fingerprints in spectroscopic observables.  
For instance, tunneling or neutron-scattering experiments could reveal additional low-energy peaks in the impurity spectral function \(A(\omega)\) corresponding to transitions between boundary configurations, while their temperature dependence would track the redistribution of spectral weight predicted by the multi-tower TBA.  
These combined thermodynamic and dynamical observables thus provide realistic experimental routes for detecting the boundary Hilbert-space reconstruction described in this work.

\subsection*{Stability Beyond Integrability}

A natural question concerns the robustness of the tower-splitting picture upon breaking integrability.  
Preliminary exact-diagonalization studies of nonintegrable XXX chains with inhomogeneous in-bulk links indicate that the low-energy spectrum still organizes into quasi-tower structures for both weak and strong link limits.  
This suggests that the essential physics of boundary-state formation and Hilbert-space partitioning may persist beyond strict integrability, although quantitative relations such as the exact residual entropies and critical exponents are expected to be modified.  
A systematic exploration of this stability, possibly combining numerical methods with boundary conformal field theory, is an important direction for future work.

\subsection*{Outlook}

Overall, the present study demonstrates that the interplay between integrability and boundary effects gives rise to a rich hierarchy of impurity-bound states, whose thermodynamic signatures can be captured by a multi-tower TBA description.  
Establishing the universality of this mechanism across other integrable and near-integrable systems, and its manifestations in dynamical observables, remains an open and promising avenue for further research.

\section{Acknowledgment}
We thank Yicheng Tang for fruitful discussions.  This work is supported by the Swiss National Science Foundation under grant number 200020-219400.

\bibliographystyle{unsrt}
\bibliography{ref}

\widetext

\appendix

\section{Integrability, Bethe equations, and energy formula}
\label{app:int}
 
This appendix presents the integrable construction of the open spin-$s$
Babujian--Takhtajian chain carrying a spin-$\tfrac12$ impurity at one edge.
We proceed in the logical order in which the objects are used: the
$R$-matrices and the relations they satisfy (\ref{app:R}); the double-row
monodromy and transfer matrix (\ref{app:transfer}); the extraction of the
Hamiltonian from $\partial_u\ln t_s(u)|_{u=0}$ (\ref{app:H}); the Babujian
polynomial with worked examples (\ref{app:Q}); the commutativity
$[t_j(u),t_k(v)]=0$ that establishes integrability (\ref{app:comm}); the
algebraic Bethe ansatz (ABA) with a spin-$\tfrac12$ auxiliary space and the
resulting Bethe ansatz equations (BAE) (\ref{app:ABA}); the fusion hierarchy
that connects $t_\sigma$ to $t_s$ (\ref{app:fusion}); and finally the energy
function (\ref{app:E}).
 
\subsection{$SU(2)$-invariant $R$-matrices, projectors and fundamental relations}
\label{app:R}
 
We use the $SU(2)$-invariant $R$-matrices of Babujian \cite{BABUJIAN1983317},
\begin{align}
R_{\sigma s}^{12}(u)&=\left(u+\frac{\eta}{2}\right)\mathbf{1}^{1,2}
+\eta\,\vec{\sigma}_1\cdot\vec{S}_2 ,
\label{eq:Rsigmas}\\[2pt]
R_{s}^{12}(u)&=\prod_{j=1}^{2s}(u-j\eta)\,
\sum_{\ell=0}^{2s}\ \prod_{k=1}^{\ell}\frac{u+k\eta}{u-k\eta}\,P^{\ell} ,
\label{eq:Rs}
\end{align}
where $R_{\sigma s}$ acts on $\tfrac12\otimes s$ and $R_{s}$ on $s\otimes s$.
Here $P^{\ell}$ is the projector onto the two-site sector of total spin $\ell$,
\begin{equation}
P^{\ell}\ket{j}=\delta_{\ell j}\ket{j},
\qquad
P^{\ell}=\prod_{\substack{m=0\\ m\neq \ell}}^{2s}\frac{x-x_m}{x_\ell-x_m},
\qquad
x\equiv\vec{S}_1\cdot\vec{S}_2 ,
\qquad
x_\ell=\tfrac12\big[\ell(\ell+1)-2s(s+1)\big] .
\label{eq:proj}
\end{equation}
It is useful to note that \eqref{eq:Rs} is diagonal in the
projector basis,
\begin{equation}
R_s^{12}(u)=\sum_{\ell=0}^{2s}f_\ell(u)\,P^{\ell},
\qquad
f_\ell(u)=\prod_{k=1}^{\ell}(u+k\eta)\prod_{j=\ell+1}^{2s}(u-j\eta),
\label{eq:flu}
\end{equation}
a form we use repeatedly below.
 
These $R$-matrices satisfy the Yang--Baxter equations
\begin{align}
R_{s}^{12}(\lambda)\,R_{s}^{13}(\lambda+u)\,R_{s}^{23}(u)
&=R_{s}^{23}(u)\,R_{s}^{13}(\lambda+u)\,R_{s}^{12}(\lambda),
\label{eq:YBE1}\\
R_{\sigma s}^{12}(\lambda)\,R_{s}^{13}(\lambda+u)\,R_{\sigma s}^{23}(u)
&=R_{\sigma s}^{23}(u)\,R_{s}^{13}(\lambda+u)\,R_{\sigma s}^{12}(\lambda),
\label{eq:YBE2}\\
R_{\sigma s}^{12}(\lambda)\,R_{\sigma}^{13}(\lambda+u)\,R_{\sigma s}^{23}(u)
&=R_{\sigma s}^{23}(u)\,R_{\sigma}^{13}(\lambda+u)\,R_{\sigma s}^{12}(\lambda),
\label{eq:YBE3}
\end{align}
with $R^{ij}_{\sigma}\equiv R^{ij}_{\sigma s}\big|_{s=1/2}=R^{ij}_{s}\big|_{s=1/2}$,
together with unitarity
\begin{align}
R_{\sigma s}^{12}(u)\,R_{\sigma s}^{12}(-u)
&=-\left[u^2-\left(s+\tfrac12\right)^2\eta^2\right]\mathbf{1}^{1,2},
\label{eq:unit1}\\
R_{s}^{12}(u)\,R_{s}^{12}(-u)
&=\prod_{j=1}^{2s}\left(j^2\eta^2-u^2\right)\mathbf{1}^{1,2},
\label{eq:unit2}
\end{align}
and the crossing relations
\begin{equation}
\left(R^{12}(u)\right)^{t_1}\left(R^{12}(-u-2\eta)\right)^{t_1}\propto\mathbf{1}^{1,2}.
\label{eq:cross}
\end{equation}
For the manipulations of Sec.~\ref{app:comm} it suffices to write
\eqref{eq:unit1}--\eqref{eq:cross} in the compact form
\begin{equation}
R_{12}(u)R_{12}(-u)=\mathfrak g(u)\,\mathbf{1}^{1,2},
\qquad
R_{12}^{t_1,-1,t_1}(u)=\mathfrak f(u)\,R_{12}(-u-2\eta),
\label{eq:fg}
\end{equation}
with scalar functions $\mathfrak f,\mathfrak g$; the zeros of $\mathfrak g$ sit at
$u=\pm(s+\tfrac12)\eta$ in the $\sigma s$ case and at $u=\pm j\eta$,
$1\le j\le 2s$, in the $ss$ case. Finally, since \eqref{eq:Rsigmas} and
\eqref{eq:Rs} are polynomials in the same $SU(2)$ scalar,
\begin{equation}
\comm{R^{ij}(u)}{R^{ij}(v)}=0 .
\label{eq:RR}
\end{equation}
 
\subsection{Double-row monodromy and transfer matrix}
\label{app:transfer}
 
The chain consists of $N$ bulk spin-$s$ sites $1,\dots,N$ and a spin-$\tfrac12$
impurity at site $0$, attached to the left edge. Open boundaries are
implemented by the double-row (Sklyanin-type) monodromy with auxiliary spin
$s$,
\begin{equation}
U_s^a(u)=T_s^a(u)\,\widehat{T}_s^a(u),
\label{eq:U}
\end{equation}
where the two single rows are
\begin{align}
T_s^a(u)&=R_s^{aN}(u)\cdots R_s^{a1}(u)\,R_{\sigma s}^{a0}(u-\gamma\eta),
\label{eq:T}\\
\widehat{T}_s^a(u)&=R_{\sigma s}^{a0}(u+\gamma\eta)\,R_s^{a1}(u)\cdots R_s^{aN}(u).
\label{eq:That}
\end{align}
The impurity enters only through the shifted spectral parameters
$u\mp\gamma\eta$ in the two $R_{\sigma s}^{a0}$ factors, so that $\gamma$ is a
free (continuous) parameter of the integrable family: it tunes the
impurity--host coupling without spoiling integrability. The transfer matrix is
\begin{equation}
t_s(u)=\Tr_a U_s^a(u).
\label{eq:t}
\end{equation}
A relation used repeatedly below follows directly from \eqref{eq:Rsigmas},
\begin{equation}
R_{\sigma s}^{i0}(u-\gamma\eta)\,R_{\sigma s}^{i0}(u+\gamma\eta)
=\left(R_{\sigma s}^{i0}(u)\right)^{2}-\gamma^{2}\eta^{2}\,\mathbf{1}^{i,0},
\label{eq:RRgamma}
\end{equation}
since the two $\gamma$-dependent pieces are proportional to the identity and
cancel in the cross terms.
 
\subsection{From the transfer matrix to the Hamiltonian}
\label{app:H}
 
\paragraph*{Regularity.}
Setting $u=0$ in \eqref{eq:flu} gives
$f_\ell(0)=\eta^{2s}(2s)!\,(-1)^{2s-\ell}$, hence
\begin{equation}
R_s^{k l}(0)=c\sum_{\ell=0}^{2s}(-1)^{2s-\ell}P^{\ell}_{kl}=c\,\mathbf{P}_{kl},
\qquad c\equiv\eta^{2s}(2s)!,
\label{eq:regularity}
\end{equation}
where $\mathbf{P}_{kl}$ is the permutation operator on the two spin-$s$ sites
$k,l$ (its spectral decomposition on $s\otimes s$ is exactly
$\sum_\ell(-1)^{2s-\ell}P^{\ell}$). Regularity is what makes $t_s(u)$ generate
a \emph{local} Hamiltonian.
 
\paragraph*{The transfer matrix at $u=0$.}
Using \eqref{eq:regularity} and then telescoping the permutations
($\mathbf{P}_{aN}\cdots\mathbf{P}_{a1}X_{a0}\mathbf{P}_{a1}\cdots\mathbf{P}_{aN}
=X_{10}$),
\begin{equation}
\begin{aligned}
t_s(0)&=c^{2N}\Tr_a\,\mathbf{P}_{aN}\cdots\mathbf{P}_{a1}
R_{\sigma s}^{a0}(-\gamma\eta)R_{\sigma s}^{a0}(\gamma\eta)
\mathbf{P}_{a1}\cdots\mathbf{P}_{aN}\\
&=c^{2N}\Tr_a\,R_{\sigma s}^{10}(-\gamma\eta)R_{\sigma s}^{10}(\gamma\eta)
=c^{2N}(2s+1)\,R_{\sigma s}^{10}(-\gamma\eta)R_{\sigma s}^{10}(\gamma\eta).
\end{aligned}
\end{equation}
By \eqref{eq:RRgamma} with $u=0$ and
$R_{\sigma s}(0)=\eta\big(\tfrac12+\vec\sigma_0\cdot\vec S_1\big)$, whose square
is $\tfrac{\eta^2}{4}(1+2s)^2\mathbf{1}$ because
$\vec\sigma_0\cdot\vec S_1$ takes the values $s$ and $-(s+1)$ on the two
fused sectors $j=s\pm\tfrac12$,
\begin{equation}
t_s(0)=c^{2N}(1+2s)\,\eta^{2}\left[\tfrac14(1+2s)^{2}-\gamma^{2}\right]\mathbf{1}.
\label{eq:t0}
\end{equation}
Note that $t_s(0)$ is a nonvanishing $c$-number as long as
$4\gamma^2\neq(1+2s)^2$; this is the condition for the impurity coupling below
to stay finite.
 
\paragraph*{The derivative at $u=0$.}
Differentiating \eqref{eq:t} and using \eqref{eq:regularity} for all factors
that are not differentiated, one obtains three groups of terms: the boundary term
(differentiating the two impurity $R$-matrices), and two identical groups obtained by differentiating one bulk $R_s$ in either the first or in the second row,
\begin{equation}
\begin{aligned}
t'_s(0)&=c^{2N}\Tr_a\,\mathbf{P}_{aN}\cdots\mathbf{P}_{a1}
\left.\dv{u}\left[\left(R_{\sigma s}^{a0}(u)\right)^2-\gamma^2\eta^2\right]\right|_{u=0}
\mathbf{P}_{a1}\cdots\mathbf{P}_{aN}\\
&\quad+c^{2N-1}\sum_{k=1}^{N}\Tr_a\,\mathbf{P}_{aN}\cdots
\left.\dv{u}R_s^{ak}(u)\right|_{u=0}\cdots\mathbf{P}_{a1}
R_{\sigma s}^{a0}(-\gamma\eta)R_{\sigma s}^{a0}(\gamma\eta)
\mathbf{P}_{a1}\cdots\mathbf{P}_{aN}\\
&\quad+c^{2N-1}\sum_{k=1}^{N}\Tr_a\,\mathbf{P}_{aN}\cdots\mathbf{P}_{a1}
R_{\sigma s}^{a0}(-\gamma\eta)R_{\sigma s}^{a0}(\gamma\eta)\mathbf{P}_{a1}\cdots
\left.\dv{u}R_s^{ak}(u)\right|_{u=0}\cdots\mathbf{P}_{aN}\\[4pt]
&=2c^{2N}(2s+1)\,R_{\sigma s}^{10}(0)\\
&\quad+2\,c^{-1}t_s(0)\left(\sum_{k=1}^{N-1}
\left.\dv{u}R_s^{k,k+1}(u)\right|_{u=0}\mathbf{P}_{k+1,k}
+\frac{1}{2s+1}\Tr_a\left.\dv{u}R_s^{aN}(u)\right|_{u=0}\mathbf{P}_{aN}\right),
\end{aligned}
\label{eq:tprime}
\end{equation}
where in the boundary term we used
$\dv{u}\big[(R_{\sigma s}(u))^2-\gamma^2\eta^2\big]_{u=0}=2R_{\sigma s}(0)$
(because $\dv{u}R_{\sigma s}=\mathbf{1}$), and in the bulk terms we
telescoped the permutations so that the differentiated $R$-matrix acts on the
neighbouring pair $(k,k+1)$. The final term corresponding to $k=N$, represents the free right edge and
produces only a constant, as we now check.
 
\paragraph*{Local density.}
The key local object is $c^{-1}R_s'^{\,lm}(0)\mathbf{P}_{lm}$. From
\eqref{eq:flu},
\begin{equation}
\frac{f'_\ell(0)}{f_\ell(0)}
=\sum_{k=1}^{\ell}\frac{1}{k\eta}-\sum_{j=\ell+1}^{2s}\frac{1}{j\eta}
=\frac{1}{\eta}\left[2H_\ell-H_{2s}\right],
\qquad H_n\equiv\sum_{k=1}^{n}\frac1k ,
\end{equation}
so that, using $P^{\ell}P^{\ell'}=\delta_{\ell\ell'}P^{\ell}$ and
$\left[(-1)^{2s-\ell}\right]^2=1$,
\begin{equation}
c^{-1}\left.\dv{u}R_s^{lm}(u)\right|_{u=0}\mathbf{P}_{lm}
=\frac{1}{\eta}\sum_{\ell=0}^{2s}\left[2H_\ell-H_{2s}\right]P^{\ell}_{lm}
=-\frac{H_{2s}}{\eta}
+\frac{1}{2\eta}\,Q_{2s}\!\left(\vec S_l\cdot\vec S_m\right),
\label{eq:localdensity}
\end{equation}
where the last equality \emph{defines} the Babujian polynomial
\begin{equation}
Q_{2s}(x)\equiv 4\sum_{\ell=0}^{2s}H_\ell\,P^{\ell}
=4\sum_{\ell=1}^{2s}\left(\sum_{k=1}^{\ell}\frac1k\right)
\prod_{\substack{m=0\\ m\neq\ell}}^{2s}\frac{x-x_m}{x_\ell-x_m}
\label{eq:Qdef}
\end{equation}
(the $\ell=0$ term drops out since $H_0=0$). Similarly, tracing
\eqref{eq:localdensity} over the auxiliary space with
$\Tr_a P^{n}_{aN}=\frac{2n+1}{2s+1}\mathbf{1}^{N}$ (by $SU(2)$ invariance and
$\Tr_{aN}P^n=2n+1$),
\begin{equation}
\frac{1}{2s+1}\Tr_a\, c^{-1}\left.\dv{u}R_s^{aN}(u)\right|_{u=0}\mathbf{P}_{aN}
=\frac{C_1}{2}\,\mathbf{1}^{N},
\qquad
C_1=\frac{2}{\eta}\left[-H_{2s}
+\frac{2}{(2s+1)^{2}}\sum_{n=0}^{2s}(2n+1)H_n\right],
\label{eq:C1}
\end{equation}
i.e.\ a pure constant: the free edge contributes no nontrivial operator. Collecting the
$-H_{2s}/\eta$ pieces of the $N-1$ bulk bonds gives the second constant
\begin{equation}
C_2=-\frac{2(N-1)H_{2s}}{\eta}.
\label{eq:C2}
\end{equation}
 
\paragraph*{Result.}
Substituting \eqref{eq:t0}--\eqref{eq:C2} into
$\mc{H}=\partial_u\ln t_s(u)|_{u=0}-C_1-C_2=t_s^{-1}(0)t'_s(0)-C_1-C_2$,
\begin{equation}
\mc{H}=\frac{8/\eta^{2}}{(1+2s)^{2}-4\gamma^{2}}\,R_{\sigma s}^{10}(0)
+\frac{1}{\eta}\sum_{k=1}^{N-1}Q_{2s}\!\left(\vec S_k\cdot\vec S_{k+1}\right),
\end{equation}
and finally, with
$R_{\sigma s}^{10}(0)=\tfrac{\eta}{2}+2\eta\,\vec s_0\cdot\vec S_1$ (dropping
the additive constant),
\begin{equation}
\boxed{\;
H=\underbrace{J\,\vec{s}_0\cdot\vec{S}_1}_{\text{impurity}}
+\underbrace{g\sum_{k=1}^{N-1}Q_{2s}\!\left(\vec{S}_k\cdot\vec{S}_{k+1}\right)}_{\text{bulk}},
\qquad
g=\frac{1}{\eta},
\qquad
J=\frac{16g}{(1+2s)^{2}-4\gamma^{2}} \; }
\label{eq:H-appendix}
\end{equation}
This is the Hamiltonian (Eq.~\ref{ModelHam}) used in the main text. Two features are
worth emphasizing: (i) the bulk is the standard Babujian--Takhtajian
integrable spin-$s$ chain, unaffected by the impurity; (ii) the entire effect
of $\gamma$ is to rescale the single boundary bond $J$, which spans the full
range $J\in(-\infty,\infty)$ getting stronger as $2|\gamma|\to(1+2s)$ from below
and above.
 
\subsection{The Babujian polynomial $Q_{2s}(x)$}
\label{app:Q}
 
The bulk interaction in \eqref{eq:H-appendix} is the degree-$2s$ polynomial
\eqref{eq:Qdef} in the $SU(2)$ scalar $x=\vec S_i\cdot\vec S_{i+1}$. Three
equivalent representations are useful,
\begin{align}
Q_{2s}(x)&=4\sum_{\ell=1}^{2s}H_\ell\,P^{\ell},
\label{eq:Q-projector}\\
&=4\sum_{\ell=1}^{2s}H_\ell
\prod_{\substack{m=0\\ m\neq\ell}}^{2s}\frac{x-x_m}{x_\ell-x_m},
\label{eq:Q-lagrange}\\
&=\sum_{p=0}^{2s}q_p\,x^{p},
\label{eq:Q-power}
\end{align}
the last being the expansion in powers of $x$ obtained from the Lagrange form.
Equations \eqref{eq:Q-projector}--\eqref{eq:Q-lagrange} are the statement that
$Q_{2s}$ is the unique degree-$2s$ interpolating polynomial with
\begin{equation}
Q_{2s}(x_\ell)=4H_\ell \quad (\ell\ge1),
\qquad Q_{2s}(x_0)=0 ,
\end{equation}
i.e.\ the two-site singlet is the zero of the energy scale and each higher
multiplet is weighted by a harmonic number.
 
\paragraph*{Example: $s=\tfrac12$.}
Here $2s=1$, $x_0=-\tfrac34$, $x_1=\tfrac14$, and
\begin{equation}
Q_{1}(x)=4H_1\frac{x-x_0}{x_1-x_0}=3+4x ,
\end{equation}
so that \eqref{eq:H-appendix} reduces to the Heisenberg chain
$H=J\vec s_0\cdot\vec S_1+4g\sum_k\vec S_k\cdot\vec S_{k+1}$ up to a constant.
 
\paragraph*{Example: $s=1$.}
Now $2s=2$ and $x_0=-2$, $x_1=-1$, $x_2=1$, with $H_1=1$, $H_2=\tfrac32$.
Projector form:
\begin{equation}
Q_{2}(x)=4H_1P^{1}+4H_2P^{2}=4P^{1}+6P^{2},
\end{equation}
power form:
\begin{equation}
Q_{2}(x)=6+x-x^{2}.
\end{equation}
 
\paragraph*{Example: $s=\tfrac32$.}
Here $2s=3$,
$x_0=-\tfrac{15}{4}$, $x_1=-\tfrac{11}{4}$, $x_2=-\tfrac34$, $x_3=\tfrac94$,
and $H_1=1$, $H_2=\tfrac32$, $H_3=\tfrac{11}{6}$, giving
\begin{equation}
Q_{3}(x)=4P^{1}+6P^{2}+\tfrac{22}{3}P^{3}
=\frac{4}{27}x^{3}+\frac{2}{27}x^{2}-\frac{1}{4}x+\frac{35}{6}.
\end{equation}
 
\paragraph*{Example: $s=2$.}
For completeness, $2s=4$, $x_\ell=(-6,-5,-3,0,4)$, and
\begin{equation}
Q_{4}(x)=4P^{1}+6P^{2}+\tfrac{22}{3}P^{3}+\tfrac{25}{3}P^{4}
=-\frac{x^{4}}{72}-\frac{5x^{3}}{108}+\frac{43x^{2}}{216}
+\frac{13x}{12}+\frac{22}{3}.
\end{equation}
 
Either representation may be used in \eqref{eq:H-appendix}; the projector form
is the more convenient one for analytic manipulations and for the
identification of $E_0$ in Sec.~\ref{app:E}, the power form for numerics.
 
\subsection{Commutativity of the transfer matrices}
\label{app:comm}
 
Integrability follows from $[t_j(u),t_k(v)]=0$ for arbitrary auxiliary spins
$j,k$, since expanding $t(u)$ in powers of $u$ then yields a commuting family
of charges containing $H$. We sketch the proof in three steps, following the
standard Sklyanin argument adapted to the impurity monodromy
\eqref{eq:T}--\eqref{eq:That}. (Throughout, an $R$ or $T$ without a spin label
stands for any admissible combination of $\sigma$, $s$, $\sigma s$ indices.)
 
\paragraph*{Step 1: $RTT$ relations.}
Repeated use of \eqref{eq:YBE1}--\eqref{eq:YBE3}, pushing
$R^{a\Bar a}(u-v)$ through the product site by site, gives
\begin{equation}
\begin{aligned}
R^{a\Bar a}(u-v)\,T^a(u)\,T^{\Bar a}(v)
&=R^{a\Bar a}(u-v)R^{aN}(u)R^{\Bar aN}(v)\cdots R^{a0}(u-\gamma\eta)R^{\Bar a0}(v-\gamma\eta)\\
&=R^{\Bar aN}(v)R^{aN}(u)\,R^{a\Bar a}(u-v)\cdots R^{a0}(u-\gamma\eta)R^{\Bar a0}(v-\gamma\eta)\\
&\ \ \vdots\\
&=T^{\Bar a}(v)\,T^{a}(u)\,R^{a\Bar a}(u-v),
\end{aligned}
\label{eq:RTT}
\end{equation}
where in the first line we used \eqref{eq:RR} to bring the two rows into
alternating order, and note that the impurity shifts $\mp\gamma\eta$ are
common to both rows and therefore harmless. In the same way,
\begin{equation}
R^{a\Bar a}(u-v)\,\widehat T^{a}(u)\,\widehat T^{\Bar a}(v)
=\widehat T^{\Bar a}(v)\,\widehat T^{a}(u)\,R^{a\Bar a}(u-v),
\qquad
\widehat T^{a}(u)\,R^{a\Bar a}(u+v)\,T^{\Bar a}(v)
=T^{\Bar a}(v)\,R^{a\Bar a}(u+v)\,\widehat T^{a}(u).
\label{eq:RTThat}
\end{equation}
 
\paragraph*{Step 2: reflection (boundary Yang--Baxter) relation.}
Combining \eqref{eq:RTT} and \eqref{eq:RTThat} with \eqref{eq:RR},
\begin{equation}
\begin{aligned}
R^{a\Bar a}(u-v)\,U^{a}(u)\,R^{a\Bar a}(u+v)\,U^{\Bar a}(v)
&=R^{a\Bar a}(u-v)\,T^{a}(u)\,T^{\Bar a}(v)\,R^{a\Bar a}(u+v)\,
\widehat T^{a}(u)\,\widehat T^{\Bar a}(v)\\
&=T^{\Bar a}(v)T^{a}(u)\,R^{a\Bar a}(u-v)R^{a\Bar a}(u+v)\,
\widehat T^{a}(u)\widehat T^{\Bar a}(v)\\
&=T^{\Bar a}(v)T^{a}(u)\,R^{a\Bar a}(u+v)\,\widehat T^{\Bar a}(v)\widehat T^{a}(u)\,
R^{a\Bar a}(u-v)\\
&=U^{\Bar a}(v)\,R^{a\Bar a}(u+v)\,U^{a}(u)\,R^{a\Bar a}(u-v).
\end{aligned}
\label{eq:reflection}
\end{equation}
 
\paragraph*{Step 3: tracing.}
Using \eqref{eq:fg} to move the partial transposes around and then
\eqref{eq:reflection},
\begin{equation}
\begin{aligned}
t(u)t(v)&=\Tr_{a\Bar a}U_a(u)U_{\Bar a}(v)
=\Tr_{a\Bar a}\left(R^{a\Bar a}(u+v)\right)^{t_a,-1,t_a}
U_a(u)R^{a\Bar a}(u+v)U_{\Bar a}(v)\\
&=\Tr_{a\Bar a}\left(R^{a\Bar a}(u+v)\right)^{t_a,-1,t_a}
\left(R^{a\Bar a}(u-v)\right)^{-1}
U_{\Bar a}(v)R^{a\Bar a}(u+v)U_a(u)R^{a\Bar a}(u-v)\\
&=\frac{\mathfrak f(u+v)}{\mathfrak g(u-v)}
\Tr_{a\Bar a}R^{\Bar a a}(-u-v+2\eta)R^{a\Bar a}(v-u)\,
U_{\Bar a}(v)R^{a\Bar a}(u+v)U_a(u)R^{a\Bar a}(u-v)\\
&=\Tr_{a\Bar a}\left(R^{a\Bar a}(u+v)\right)^{t_{\Bar a},-1,t_{\Bar a}}
U_{\Bar a}(v)R^{a\Bar a}(u+v)U_a(u)
=\Tr_{a\Bar a}U_{\Bar a}(v)U_a(u)=t(v)t(u).
\end{aligned}
\end{equation}
Here $u,v$ are generic, i.e.\ $u-v$ away from the zeros of $\mathfrak g$; those
special points are exactly the ones exploited in the fusion hierarchy of
Sec.~\ref{app:fusion}. Since the argument does not depend on the auxiliary spin,
\begin{equation}
\comm{t_j(u)}{t_k(v)}=0\qquad\forall\, j,k,u,v .
\end{equation}
Consequently all $t_j$ share one eigenbasis, and it suffices to diagonalize the
simplest member $t_\sigma$ (auxiliary spin $\tfrac12$) and then relate its
eigenvalues to those of $t_s$ by fusion.
 
\subsection{Algebraic Bethe ansatz with spin-$\tfrac12$ auxiliary space}
\label{app:ABA}
 
\paragraph*{Operator entries.}
Write the double-row monodromy with spin-$\tfrac12$ auxiliary space as
\begin{equation}
U_\sigma^{a}(u)=\mqty(\mc{A}(u)&\mc{B}(u)\\ \mc{C}(u)&\mc{D}(u))_a ,
\end{equation}
and the single row as $T^a_\sigma(u)=\mqty(A(u)&B(u)\\ C(u)&D(u))$. The
crossing identity
$\sigma^y_a\left(R^{ai}(u)\right)^{t_a}\sigma^y_a=-R^{ai}(-u-\eta)$
gives
\begin{equation}
\sigma^y_a\left(\widehat T^a_\sigma(u)\right)^{t_a}\sigma^y_a
=(-1)^{N+1}T^a_\sigma(-\eta-u)
\;\Longrightarrow\;
\widehat T^a_\sigma(u)=(-1)^{N+1}
\mqty(D(-u-\eta)&-B(-u-\eta)\\ -C(-u-\eta)&A(-u-\eta)),
\label{eq:crossingT}
\end{equation}
so the second row is expressed through the first, and
\begin{equation}
\mqty(\mc{A}&\mc{B}\\ \mc{C}&\mc{D})(u)
=(-1)^{N+1}\mqty(A(u)&B(u)\\ C(u)&D(u))
\mqty(D(-u-\eta)&-B(-u-\eta)\\ -C(-u-\eta)&A(-u-\eta)).
\label{eq:UUU}
\end{equation}
 
\paragraph*{Commutation relations.}
Expanding \eqref{eq:reflection} in components yields
$\comm{\mc B(u)}{\mc B(v)}=\comm{\mc C(u)}{\mc C(v)}=0$ together with
\begin{align}
\mc{A}(v)\mc{B}(u)&=\frac{(u+v)(u-v+\eta)}{(u-v)(u+v+\eta)}\mc{B}(u)\mc{A}(v)
-\frac{\eta(u+v)}{(u-v)(u+v+\eta)}\mc{B}(v)\mc{A}(u)
-\frac{\eta}{u+v+\eta}\mc{B}(v)\mc{D}(u),
\label{eq:AB}\\[4pt]
\mc{D}(u)\mc{B}(v)&=\frac{(u+v+2\eta)(u-v+\eta)}{(u-v)(u+v+\eta)}\mc{B}(v)\mc{D}(u)
-\frac{\eta(u+v+2\eta)}{(u-v)(u+v+\eta)}\mc{B}(u)\mc{D}(v)\nonumber\\
&\quad-\frac{2\eta^{2}}{(u-v)(u+v+\eta)}\mc{B}(v)\mc{A}(u)
+\frac{\eta(u-v+2\eta)}{(u-v)(u+v+\eta)}\mc{B}(u)\mc{A}(v).
\label{eq:DB}
\end{align}
As always for open chains, $\mc{A}$ and $\mc{D}$ do not by themselves satisfy a
clean exchange algebra; the combination
\begin{equation}
\widetilde{\mc{D}}(u)=\frac{2u+\eta}{\eta}\mc{D}(u)-\mc{A}(u)
\end{equation}
does. In terms of it, \eqref{eq:AB}--\eqref{eq:DB} become
\begin{align}
\widetilde{\mc D}(u)\mc B(v)&=
\frac{(u+v+2\eta)(u-v+\eta)}{(u-v)(u+v+\eta)}\mc B(v)\widetilde{\mc D}(u)
-\frac{2\eta(u+\eta)}{(u-v)(2v+\eta)}\mc B(u)\widetilde{\mc D}(v)
+\frac{4(u+\eta)v}{(u+v+\eta)(2v+\eta)}\mc B(u)\mc A(v),
\label{eq:TDB}\\[4pt]
\mc A(u)\mc B(v)&=
\frac{(u+v)(u-v-\eta)}{(u-v)(u+v+\eta)}\mc B(v)\mc A(u)
-\frac{\eta^{2}}{(u+v+\eta)(2v+\eta)}\mc B(u)\widetilde{\mc D}(v)
+\frac{2v\eta}{(u-v)(2v+\eta)}\mc B(u)\mc A(v),
\label{eq:AB2}
\end{align}
and the transfer matrix reads
\begin{equation}
t_\sigma(u)=\mc A(u)+\mc D(u)
=\frac{\eta}{2u+\eta}\widetilde{\mc D}(u)+\frac{2(u+\eta)}{2u+\eta}\mc A(u).
\label{eq:tsigma}
\end{equation}
 
\paragraph*{Reference state.}
Take all bulk spins and the impurity fully polarized,
\begin{equation}
\ket{0}=\bigotimes_{i=1}^{N}\ket{+s}_i\otimes\ket{+\tfrac12}_0 .
\end{equation}
From \eqref{eq:Rsigmas}, both $R^{ai}_{\sigma s}(u)$ and $R^{a0}_{\sigma}(u-\gamma\eta)$
are upper triangular on $\ket{0}$ in the auxiliary basis, so $C(u)\ket{0}=0$ and
\begin{equation}
A(u)\ket{0}=a(u)\ket{0},\quad
D(u)\ket{0}=d(u)\ket{0},
\qquad
a(u)=\left(\frac{\eta(1+2s)+2u}{2}\right)^{N}\!\!(u+\eta+\gamma\eta),
\quad
d(u)=\left(\frac{\eta(1-2s)+2u}{2}\right)^{N}\!\!(u+\gamma\eta).
\end{equation}
Using \eqref{eq:UUU},
\begin{equation}
\mc A(u)\ket{0}=(-1)^{N+1}A(u)D(-\eta-u)\ket{0}
=\left(\frac{\eta(1+2s)+2u}{2\eta}\right)^{2N}\left[(u+\eta)^{2}-\gamma^{2}\eta^{2}\right]\ket{0}
\equiv a^{(s)}(u)\frac{2u+\eta}{2(u+\eta)}\ket{0},
\end{equation}
while $\mc D$ requires the $RTT$ consequence
$\comm{B(u)}{C(v)}=\frac{\eta}{u-v}\left(D(v)A(u)-D(u)A(v)\right)$, giving
\begin{equation}
\mc D(u)\ket{0}=(-1)^{N+1}\frac{1}{2u+\eta}
\left[\eta\,D(-\eta-u)A(u)+2u\,D(u)A(-\eta-u)\right]\ket{0},
\end{equation}
and therefore the clean result
\begin{equation}
\widetilde{\mc D}(u)\ket{0}
=(-1)^{N+1}\frac{2u}{\eta}D(u)A(-\eta-u)\ket{0}
=\frac{2u}{\eta}\left(\frac{\eta(1-2s)+2u}{2\eta}\right)^{2N}
\left(u^{2}-\gamma^{2}\eta^{2}\right)\ket{0}.
\end{equation}
It is convenient to absorb the prefactors of \eqref{eq:tsigma} and define
\begin{equation}
a^{(s)}(u)=\frac{2(u+\eta)}{2u+\eta}
\left(\frac{\eta(1+2s)+2u}{2\eta}\right)^{2N}
\left[(u+\eta)^{2}-\gamma^{2}\eta^{2}\right],
\qquad
d^{(s)}(u)=a^{(s)}(-u-\eta),
\label{eq:ad}
\end{equation}
so that $t_\sigma(u)\ket{0}=\left[a^{(s)}(u)+d^{(s)}(u)\right]\ket{0}$. Note
that both the impurity parameter $\gamma$ and the two boundaries enter only
through the scalar factors $\left[(u+\eta)^2-\gamma^2\eta^2\right]$ and
$\left(u^2-\gamma^2\eta^2\right)$.
 
\paragraph*{Bethe states and the eigenvalue.}
Bethe states are built by acting with $\mc B$,
\begin{equation}
\ket{\lambda_1,\dots,\lambda_M}=\mc B_M\ket{0},
\qquad
\mc B_M\equiv\mc B(\lambda_1)\cdots\mc B(\lambda_M),
\qquad
\mc B_M^{j}\equiv\mc B_M\big|_{\lambda_j\to u}.
\end{equation}
Iterating \eqref{eq:TDB}--\eqref{eq:AB2} $M$ times gives
\begin{equation}
\begin{aligned}
\mc A(u)\mc B_M&=\prod_{j=1}^{M}
\frac{(u+\lambda_j)(u-\lambda_j-\eta)}{(u-\lambda_j)(u+\lambda_j+\eta)}\,
\mc B_M\mc A(u)\\
&\quad-\sum_{j=1}^{M}\frac{\eta^{2}}{(u+\lambda_j+\eta)(2\lambda_j+\eta)}
\prod_{l\neq j}^{M}\frac{(\lambda_j-\lambda_l+\eta)(\lambda_j+\lambda_l+2\eta)}
{(\lambda_j-\lambda_l)(\lambda_j+\lambda_l+\eta)}\,\mc B^{j}_M\widetilde{\mc D}(\lambda_j)\\
&\quad+\sum_{j=1}^{M}\frac{2\eta\lambda_j}{(u-\lambda_j)(2\lambda_j+\eta)}
\prod_{l\neq j}^{M}\frac{(\lambda_j+\lambda_l)(\lambda_j-\lambda_l-\eta)}
{(\lambda_j-\lambda_l)(\lambda_j+\lambda_l+\eta)}\,\mc B^{j}_M\mc A(\lambda_j),\\[6pt]
\widetilde{\mc D}(u)\mc B_M&=\prod_{j=1}^{M}
\frac{(u+\lambda_j+2\eta)(u-\lambda_j+\eta)}{(u-\lambda_j)(u+\lambda_j+\eta)}\,
\mc B_M\widetilde{\mc D}(u)\\
&\quad-\sum_{j=1}^{M}\frac{2\eta(u+\eta)}{(u-\lambda_j)(2\lambda_j+\eta)}
\prod_{l\neq j}^{M}\frac{(\lambda_j-\lambda_l+\eta)(\lambda_j+\lambda_l+2\eta)}
{(\lambda_j-\lambda_l)(\lambda_j+\lambda_l+\eta)}\,\mc B^{j}_M\widetilde{\mc D}(\lambda_j)\\
&\quad+\sum_{j=1}^{M}\frac{4\lambda_j(u+\eta)}{(u+\lambda_j+\eta)(2\lambda_j+\eta)}
\prod_{l\neq j}^{M}\frac{(\lambda_j+\lambda_l)(\lambda_j-\lambda_l-\eta)}
{(\lambda_j-\lambda_l)(\lambda_j+\lambda_l+\eta)}\,\mc B^{j}_M\mc A(\lambda_j).
\end{aligned}
\end{equation}
The first (``wanted'') lines yield the eigenvalue; the remaining
(``unwanted'') terms are proportional to $\mc B^{j}_M\ket{0}$. Hence
\begin{equation}
t_\sigma(u)\ket{\lambda_1,\dots,\lambda_M}
=\Lambda_\sigma(u)\ket{\lambda_1,\dots,\lambda_M}
+\sum_{j=1}^{M}\Lambda_j(u)\,\mc B^{j}_M\ket{0},
\end{equation}
with
\begin{equation}
\Lambda_\sigma(u)
=a^{(s)}(u)\prod_{j=1}^{M}\frac{(u+\lambda_j)(u-\lambda_j-\eta)}{(u-\lambda_j)(u+\lambda_j+\eta)}
+d^{(s)}(u)\prod_{j=1}^{M}\frac{(u+\lambda_j+2\eta)(u-\lambda_j+\eta)}{(u-\lambda_j)(u+\lambda_j+\eta)},
\label{eq:Lambdasigma}
\end{equation}
and
\begin{equation}
\begin{aligned}
\Lambda_j(u)&=\frac{4\lambda_j\eta(u+\eta)}{(\lambda_j-u)(2\lambda_j+\eta)(\lambda_j+u+\eta)}
\Bigg\{\left(\frac{\eta(1-2s)+2\lambda_j}{2}\right)^{2N}\!\!
\left(\lambda_j^{2}-\gamma^{2}\eta^{2}\right)
\prod_{l\neq j}^{M}\frac{(\lambda_j-\lambda_l+\eta)(\lambda_j+\lambda_l+2\eta)}
{(\lambda_j-\lambda_l)(\lambda_j+\lambda_l+\eta)}\\
&\hspace{4.2cm}-\left(\frac{\eta(1+2s)+2\lambda_j}{2}\right)^{2N}\!\!
\left[(\lambda_j+\eta)^{2}-\gamma^{2}\eta^{2}\right]
\prod_{l\neq j}^{M}\frac{(\lambda_j+\lambda_l)(\lambda_j-\lambda_l-\eta)}
{(\lambda_j-\lambda_l)(\lambda_j+\lambda_l+\eta)}\Bigg\}.
\end{aligned}
\label{eq:unwanted}
\end{equation}
 
\paragraph*{Bethe ansatz equations.}
Cancellation of the unwanted terms, $\Lambda_j(u)=0$ for all $j$ (assuming
$2\lambda_j+\eta\neq0$, $\lambda_j\neq u$, $\lambda_j+u+\eta\neq0$), gives
\begin{equation}
\left(\frac{\eta(1+2s)+2\lambda_j}{\eta(1-2s)+2\lambda_j}\right)^{2N}
\frac{(\lambda_j+\eta)^{2}-\gamma^{2}\eta^{2}}{\lambda_j^{2}-\gamma^{2}\eta^{2}}
=\prod_{l\neq j}^{M}\frac{(\lambda_j-\lambda_l+\eta)(\lambda_j+\lambda_l+2\eta)}
{(\lambda_j+\lambda_l)(\lambda_j-\lambda_l-\eta)} .
\end{equation}
Passing to the rapidity variables
\begin{equation}
\lambda_j/\eta=i\mu_j-\tfrac12
\qquad\Longleftrightarrow\qquad
\mu_j=-\frac{i\lambda_j}{\eta}-\frac{i}{2},
\end{equation}
this becomes the BAE quoted in the main text,
\begin{equation}
\boxed{\;
\left(\frac{\mu_j-is}{\mu_j+is}\right)^{2N}
\frac{\mu_j-\tfrac{i}{2}(1+2\gamma)}{\mu_j+\tfrac{i}{2}(1+2\gamma)}\,
\frac{\mu_j-\tfrac{i}{2}(1-2\gamma)}{\mu_j+\tfrac{i}{2}(1-2\gamma)}
=\prod_{\substack{l=1\\ l\neq j}}^{M}
\frac{(\mu_j-\mu_l-i)(\mu_j+\mu_l-i)}{(\mu_j-\mu_l+i)(\mu_j+\mu_l+i)} \;}
\label{eq:BAE-appendix}
\end{equation}
The structure is transparent: the bulk factor
$\left(\frac{\mu-is}{\mu+is}\right)^{2N}$ counts $2N$ rather than $N$ scattering events because a magnon traverses the open chain twice, and the impurity appears as
two extra boundary phases with shifted parameters
$\tfrac12(1\pm2\gamma)$, i.e.\ as two ``ghost'' sites of imaginary rapidity.
The equations are invariant under $\mu\to-\mu$, so $\{\mu_j\}$ and
$\{\pm\mu_j\}$ label the same eigenstate; the singular solution $\mu_j=0$ is
excluded, since it corresponds to $\lambda_j=-\eta/2$, a pole of
\eqref{eq:unwanted}, where the Bethe state is not normalizable.
 
\subsection{Fusion and the hierarchy relation}
\label{app:fusion}
 
The Hamiltonian is generated by $t_s(u)$, not by $t_\sigma(u)$, so we still
need the eigenvalues of the former. Rather than repeating the ABA for a
spin-$s$ auxiliary space, we use fusion \cite{Cao_2015}.
 
\paragraph*{Fused $R$-matrices.}
For two auxiliary spin-$\tfrac12$ spaces at spectral distance $\eta$, the
singlet channel decouples,
\begin{equation}
P^{0}_{12}\,R^{13}_\sigma\!\left(u-\tfrac{\eta}{2}\right)
R^{23}_\sigma\!\left(u+\tfrac{\eta}{2}\right)P^{1}_{12}=0 ,
\end{equation}
which follows from $P^{0}_{12}=-\frac{1}{2\eta}R^{12}_\sigma(-\eta)$,
$P^{1}_{12}=\frac{1}{2\eta}R^{12}_\sigma(\eta)$ and \eqref{eq:YBE3}.
Physically, scattering off a third particle cannot take the fused pair out of
the spin-$1$ subspace. Iterating,
\begin{equation}
R^{a,\{1\cdots2s\}}_{\sigma s}(u)
=\frac{1}{\prod_{k=1}^{2s-1}\left(u+\left(\tfrac12-s+k\right)\eta\right)}
P^{+}_{\{1\cdots2s\}}\prod_{k=1}^{2s}
R^{a,k}_{\sigma}\!\left(u+\left(k-\tfrac12-s\right)\eta\right)P^{+}_{\{1\cdots2s\}},
\qquad
P^{+}_{\{1\cdots2s\}}=\frac{1}{(2s)!}\prod_{k=1}^{2s}\left(\sum_{l=1}^{k}\mathbf P_{lk}\right),
\end{equation}
and similarly for $R_{s_1s_2}$, which reproduces \eqref{eq:Rs} at $s_1=s_2=s$.
The fused transfer matrices $t_j(u)=\Tr_{\{a\}}T^{\{a\}}_{j,s}(u)\widehat T^{\{a\}}_{j,s}(u)$
are built from the corresponding fused rows.
 
\paragraph*{Quantum determinant.}
Evaluating the product $t_\sigma(u-\eta)t_\sigma(u)$ at the fusion point
requires the singlet-channel scalar
\begin{equation}
\begin{aligned}
{\det}_q T_\sigma(u)&=\Tr_{a\Bar a}P^{0}\,T^{a}_\sigma(u-\eta)T^{\Bar a}_\sigma(u)
=\left(u+\tfrac{2s+1}{2}\eta\right)^{N}\left(u-\tfrac{2s+1}{2}\eta\right)^{N}
(u-\gamma\eta+\eta)(u-\gamma\eta-\eta),\\
{\det}_q \widehat T_\sigma(u)&=
\left(u+\tfrac{2s+1}{2}\eta\right)^{N}\left(u-\tfrac{2s+1}{2}\eta\right)^{N}
(u+\gamma\eta+\eta)(u+\gamma\eta-\eta).
\end{aligned}
\end{equation}
 
\paragraph*{Hierarchy.}
Decomposing $R^{a\Bar a}_\sigma(\lambda)=(\lambda+\eta)P^{1}+(\lambda-\eta)P^{0}$
inside $t_\sigma(u-\eta)t_\sigma(u)$, the $P^1$ channel reassembles the fused
$t_1$ while the $P^0$ channel factorizes into the quantum determinants above,
because $P^{0}XP^{0}=\bra{0}X\ket{0}P^{0}$ for a singlet projector. One obtains
\begin{equation}
t_1\!\left(u-\tfrac{\eta}{2}\right)
=\frac{u^{2}-\eta^{2}/4}{u^{2}\left(u^{2}-\gamma^{2}\eta^{2}\right)}
\left[t_\sigma(u-\eta)t_\sigma(u)-\delta^{(s)}(u)\right],
\label{eq:hierarchy}
\end{equation}
or, shifting $u$,
\begin{equation}
\begin{aligned}
t_1(u)&=\frac{u(u+\eta)}{(u+\tfrac{\eta}{2})^{2}\left[(u+\tfrac{\eta}{2})^{2}-\gamma^{2}\eta^{2}\right]}
t_\sigma\!\left(u-\tfrac{\eta}{2}\right)t_\sigma\!\left(u+\tfrac{\eta}{2}\right)\\
&\quad-\frac{(u+\tfrac{3\eta}{2})(u-\tfrac{\eta}{2})}
{(u+\tfrac{\eta}{2})^{2}\left[(u+\tfrac{\eta}{2})^{2}-\gamma^{2}\eta^{2}\right]}
\left(u+(s+1)\eta\right)^{2N}\left(u-s\eta\right)^{2N}
\left[\left(u+\tfrac{3\eta}{2}\right)^{2}-\gamma^{2}\eta^{2}\right]
\left[\left(u-\tfrac{\eta}{2}\right)^{2}-\gamma^{2}\eta^{2}\right],
\end{aligned}
\end{equation}
which we have verified numerically against direct diagonalization of
$t_1=\mc A_1+\mc D_1+\mc D_2$ for small $N$. In terms of eigenvalues,
\begin{equation}
\Lambda_1\!\left(u-\tfrac{\eta}{2}\right)
=\frac{u^{2}-\eta^{2}/4}{u^{2}\left(u^{2}-\gamma^{2}\eta^{2}\right)}
\left[\Lambda_\sigma(u-\eta)\Lambda_\sigma(u)-\delta^{(s)}(u)\right],
\qquad
\delta^{(s)}(u)=a^{(s)}(u)\,d^{(s)}(u-\eta),
\label{eq:hierarchyLambda}
\end{equation}
in agreement with \cite{Cao_2015} up to an overall scalar prefactor, which is
immaterial for the energy since the latter is a logarithmic derivative.
Explicitly,
\begin{equation}
\delta^{(s)}(u)=\frac{u^{2}-\eta^{2}}{u^{2}-\eta^{2}/4}
\left[\frac{\left(\tfrac{(1+2s)\eta}{2}\right)^{2}-u^{2}}{\eta^{2}}\right]^{2N}
\left[(u+\eta)^{2}-\gamma^{2}\eta^{2}\right]\left[(u-\eta)^{2}-\gamma^{2}\eta^{2}\right].
\end{equation}
 
\paragraph*{$TQ$ form.}
Introducing the Baxter polynomial
\begin{equation}
Q(u)=\prod_{j=1}^{M}(u-\lambda_j)(u+\lambda_j+\eta),
\qquad Q(u)=Q(-u-\eta),
\end{equation}
Eq.~\eqref{eq:Lambdasigma} takes the standard form
\begin{equation}
\Lambda_\sigma(u)=a^{(s)}(u)\frac{Q(u-\eta)}{Q(u)}+d^{(s)}(u)\frac{Q(u+\eta)}{Q(u)},
\label{eq:TQ}
\end{equation}
and \eqref{eq:hierarchyLambda} becomes
\begin{equation}
\begin{aligned}
\Lambda_1(u)&=\frac{(u+\eta)\,u}
{(u+\tfrac{\eta}{2})^{2}\left[(u+\tfrac{\eta}{2})^{2}-\gamma^{2}\eta^{2}\right]}\\
&\quad\times\left[
\frac{Q(u-\tfrac{3\eta}{2})}{Q(u+\tfrac{\eta}{2})}
a^{(s)}\!\left(u-\tfrac{\eta}{2}\right)a^{(s)}\!\left(u+\tfrac{\eta}{2}\right)
+\frac{Q(u+\tfrac{3\eta}{2})}{Q(u-\tfrac{\eta}{2})}
d^{(s)}\!\left(u-\tfrac{\eta}{2}\right)d^{(s)}\!\left(u+\tfrac{\eta}{2}\right)\right.\\
&\hspace{1.2cm}\left.
+\frac{Q(u-\tfrac{3\eta}{2})Q(u+\tfrac{3\eta}{2})}{Q(u-\tfrac{\eta}{2})Q(u+\tfrac{\eta}{2})}
a^{(s)}\!\left(u-\tfrac{\eta}{2}\right)d^{(s)}\!\left(u+\tfrac{\eta}{2}\right)\right].
\end{aligned}
\label{eq:Lambda1Q}
\end{equation}
 
\subsection{Energy function}
\label{app:E}
 
\paragraph*{The case $s=1$.}
For $s=1$ the bulk auxiliary spin coincides with the fused spin $1$, so
\eqref{eq:Lambda1Q} is all we need. The Hamiltonian \eqref{eq:H-appendix} reads
\begin{equation}
H_{s=1}=J\,\vec s_0\cdot\vec S_1
+g\sum_{k=1}^{N-1}\left[6+\vec S_k\cdot\vec S_{k+1}-\left(\vec S_k\cdot\vec S_{k+1}\right)^{2}\right]
=J\,\vec s_0\cdot\vec S_1+g\sum_{k=1}^{N-1}\left(4P^{1}_{k,k+1}+6P^{2}_{k,k+1}\right),
\end{equation}
and
\begin{equation}
E=\left.\dv{u}\ln\Lambda_1(u)\right|_{u=0}+\mathrm{const}
=\left.\dv{u}\ln\left\{u\left[\cdots\right]\right\}\right|_{u=0}+\mathrm{const}(\gamma,\eta),
\end{equation}
where $[\cdots]$ denotes the square bracket of \eqref{eq:Lambda1Q}: the
$\{\lambda\}$-independent prefactor contributes only to the constant. Keeping
the explicit factor $u$ together with the bracket avoids the individually
divergent quantities $a^{(s)}(-\eta/2)$, $d^{(s)}(-\eta/2)$.
 
Three simplifications do the work:
\begin{enumerate}
\item $d^{(1)}(\eta/2)=0$, since $\eta(1-2s)+2u$ vanishes at $u=\eta/2$, $s=1$.
      This kills the second and third terms of the bracket at $u=0$.
\item $Q(u)=Q(-u-\eta)$ implies $Q(-3\eta/2)=Q(\eta/2)$, so the first
      $Q$-ratio equals $1$ at $u=0$.
\item $\lim_{u\to0}u\,d^{(s)}(u-\eta/2)=-\lim_{u\to0}u\,a^{(s)}(u-\eta/2)
      =-\tfrac{\eta^{3}}{2}s^{2N}\left(\tfrac14-\gamma^{2}\right)$.
\end{enumerate}
Hence
\begin{equation}
\lim_{u\to0}u\left[\cdots\right]
=\left[\lim_{u\to0}u\,a^{(s)}\!\left(u-\tfrac{\eta}{2}\right)\right]a^{(s)}\!\left(\tfrac{\eta}{2}\right),
\end{equation}
while
\begin{equation}
\left.\dv{u}\left\{u\left[\cdots\right]\right\}\right|_{u=0}
=\underbrace{\left.\dv{u}a^{(s)}\!\left(u+\tfrac\eta2\right)\right|_{0}
\left[\lim u\,a^{(s)}\right]
+\left.\dv{u}\left\{u\,a^{(s)}\!\left(u-\tfrac\eta2\right)\right\}\right|_{0}
a^{(s)}\!\left(\tfrac\eta2\right)}_{\{\lambda\}\text{-independent}}
+\left[\lim u\,a^{(s)}\right]a^{(s)}\!\left(\tfrac\eta2\right)
\left.\dv{u}\frac{Q(u-\tfrac{3\eta}{2})}{Q(u+\tfrac{\eta}{2})}\right|_{0}.
\end{equation}
The last factor is elementary: with
$Q'/Q(u)=\sum_j\left[(u-\lambda_j)^{-1}+(u+\lambda_j+\eta)^{-1}\right]$,
\begin{equation}
\left.\dv{u}\ln\frac{Q(u-\tfrac{3\eta}{2})}{Q(u+\tfrac{\eta}{2})}\right|_{u=0}
=\sum_{j=1}^{M}\left[\frac{2}{\lambda_j-\tfrac{\eta}{2}}-\frac{2}{\lambda_j+\tfrac{3\eta}{2}}\right]
=\sum_{j=1}^{M}\frac{4\eta}{\left(\lambda_j+\tfrac32\eta\right)\left(\lambda_j-\tfrac12\eta\right)} .
\end{equation}
Therefore
\begin{equation}
E(\{\lambda_j\})=\sum_{j=1}^{M}
\frac{4\eta}{\left(\lambda_j+\tfrac32\eta\right)\left(\lambda_j-\tfrac12\eta\right)}+E_0 ,
\end{equation}
and, in the rapidity variables $\lambda_j=\eta(i\mu_j-\tfrac12)$, for which
$\left(\lambda_j+\tfrac32\eta\right)\left(\lambda_j-\tfrac12\eta\right)
=-\eta^{2}\left(\mu_j^{2}+1\right)$,
\begin{equation}
E(\{\mu_j\})=-4g\sum_{j=1}^{M}\frac{1}{\mu_j^{2}+1}+E_0,
\qquad
E_0=6g\left(N-1+\frac{4/3}{9-4\gamma^{2}}\right).
\end{equation}
 
\paragraph*{General $s$.}
One may repeat the construction above with the fusion carried to auxiliary spin
$s$. Alternatively, and this is all that is needed, note that the
$\{\mu_j\}$-dependent part of the energy originates from the bulk and is
therefore insensitive to the boundary conditions and to the impurity: the
impurity enters the BAE \eqref{eq:BAE-appendix} only through the source terms
and thereby only through the \emph{positions} of the roots, not through the
functional form of the energy. Using the bulk spin-$s$ result of
Babujian \cite{BABUJIAN1983317} in our normalization,
\begin{equation}
E_{\rm bulk}(\{\mu_j\})=-4g\sum_{j=1}^{M}\frac{s}{\mu_j^{2}+s^{2}},
\end{equation}
we obtain
\begin{equation}
\boxed{\;
E(\{\mu_j\})=-4g\sum_{j=1}^{M}\frac{s}{\mu_j^{2}+s^{2}}+E_0(s,\gamma,N),
\qquad
E_0(s,\gamma,N)=\frac{sJ}{2}+4g(N-1)H_{2s} \;}
\label{eq:E-appendix}
\end{equation}
The constant is the energy of the reference state ($M=0$, all spins up),
\begin{equation}
E_0=\bra{0}H\ket{0}
=J\braket{s^{z}_0S^{z}_1}+g\sum_{k=1}^{N-1}\bra{0}Q_{2s}(\vec S_k\cdot\vec S_{k+1})\ket{0}
=\frac{sJ}{2}+4g(N-1)H_{2s},
\end{equation}
where we used that $\ket{0}$ is the two-site multiplet $\ell=2s$ on every bond,
so $Q_{2s}\to4H_{2s}$ by \eqref{eq:Qdef}. For $s=1$, $H_2=\tfrac32$ and
$J=16g/(9-4\gamma^{2})$ reproduce
$E_0=6g\left[N-1+\tfrac{4/3}{9-4\gamma^{2}}\right]$, as above.
 
\medskip
Equations \eqref{eq:BAE-appendix} and \eqref{eq:E-appendix} are the starting
point for the thermodynamic (string hypothesis) and TBA analyses in the next section.

\section{Derivation of the TBA equations}\label{app:TBA}

We derive the TBA equations for the open spin-$s$ chain with a boundary impurity, starting from the Bethe Ansatz equations (BAE) and treating, in turn, the (i) string tower, (ii) fundamental boundary-string tower, and (iii) higher-order boundary-string tower. Throughout, convolutions are
\[
(f*g)(\lambda)=\int_{-\infty}^{\infty}  d\mu\, f(\lambda-\mu)\,g(\mu),\qquad
G(\lambda)=\frac{1}{2\cosh(\pi\lambda)}.
\]
Thermodynamic integrals carry an overall factor $1/2$ due to the open-chain parity $\mu \leftrightarrow -\mu$ (double description of roots).

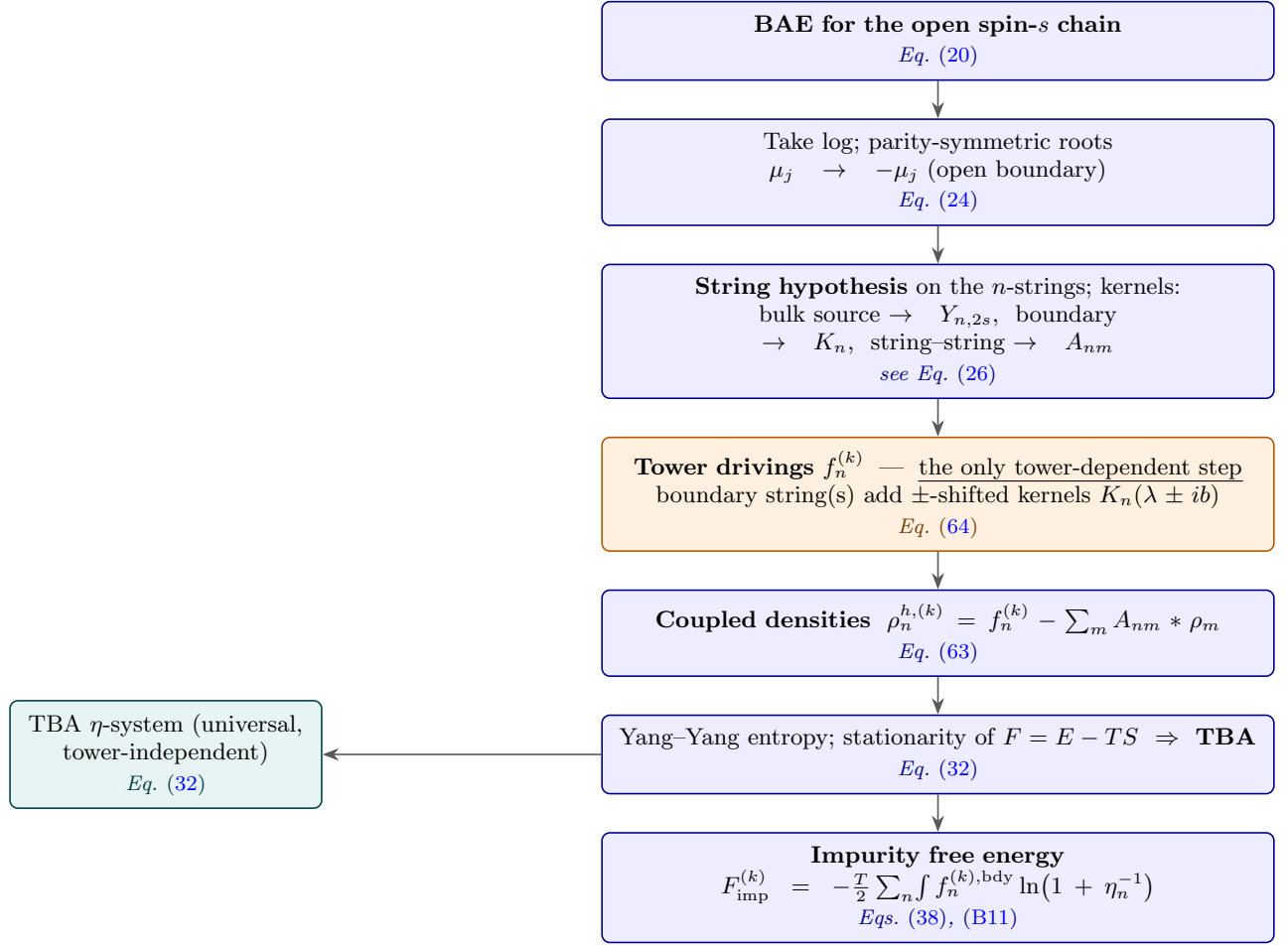
\begin{figure*}[tbp]
\centering
\begin{tikzpicture}[
  font=\small,
  >={Stealth[length=2.2mm]},
  node distance=5mm and 9mm,
  box/.style={draw=blue!55!black, semithick, rounded corners=3pt, fill=blue!7,
              align=center, text width=86mm, inner sep=5pt},
  side/.style={draw=teal!55!black, semithick, rounded corners=3pt, fill=teal!9,
               align=center, text width=38mm, inner sep=5pt},
  arr/.style={->, semithick, draw=black!65},
]
\node[box] (bae) {\textbf{BAE for the open spin-$s$ chain}\\[1pt]{\footnotesize\itshape\color{blue!50!black}Eq.~\eqref{BAE p}}};
\node[box, below=of bae] (log) {Take $\log$; parity-symmetric roots $\mu_j\to-\mu_j$ (open boundary)\\[1pt]{\footnotesize\itshape\color{blue!50!black}Eq.~\eqref{log BAE}}};
\node[box, below=of log] (str) {\textbf{String hypothesis} on the $n$-strings; kernels:\\ bulk source $\to Y_{n,2s}$,\ \ boundary $\to K_n$,\ \ string--string $\to A_{nm}$\\[1pt]{\footnotesize\itshape\color{blue!50!black}see Eq.~\eqref{rhoeq}}};
\node[box, below=of str, fill=orange!12, draw=orange!70!black] (drv) {\textbf{Tower drivings} $f_n^{(k)}$ \;---\; \emph{the only tower-dependent step}\\ boundary string(s) add $\pm$-shifted kernels $K_n(\lambda\pm ib)$\\[1pt]{\footnotesize\itshape\color{orange!45!black}Eq.~\eqref{f k}}};
\node[box, below=of drv] (den) {\textbf{Coupled densities} $\ \rho_n^{h,(k)}=f_n^{(k)}-\sum_m A_{nm}*\rho_m$\\[1pt]{\footnotesize\itshape\color{blue!50!black}Eq.~\eqref{rhoeq towers}}};
\node[box, below=of den] (pre) {Yang--Yang entropy; stationarity of $F=E-TS$ $\ \Rightarrow\ $ \textbf{TBA}\\[1pt]{\footnotesize\itshape\color{blue!50!black}Eq.~\eqref{TBA p}}};
\node[box, below=of pre] (fimp) {\textbf{Impurity free energy}\\ $F^{(k)}_{\mathrm{imp}}=-\tfrac{T}{2}\sum_{n}\!\int f_n^{(k),\mathrm{bdy}}\ln\!\big(1+\eta_n^{-1}\big)$\\[1pt]{\footnotesize\itshape\color{blue!50!black}Eqs.~\eqref{F imp}, \eqref{eq:Fimp-str}}};
\node[side] (tba) at ([xshift=-58mm]pre.west |- pre) {TBA $\eta$-system (universal, tower-independent)\\[1pt]{\footnotesize\itshape\color{teal!45!black}Eq.~\eqref{TBA p}}};
\draw[arr] (bae)--(log);
\draw[arr] (log)--(str);
\draw[arr] (str)--(drv);
\draw[arr] (drv)--(den);
\draw[arr] (den)--(pre);
\draw[arr] (pre)--(fimp);
\draw[arr] (pre.west)--(tba.east);
\end{tikzpicture}
\caption{General logic, part 1: from the Bethe Ansatz equations to the impurity free energy of a generic boundary tower. Only the driving step (highlighted, Eq.~\eqref{f k}) depends on the tower; the canonical $\eta$-system [Eq.~\eqref{TBA p}] is tower-independent.}
\label{fig:bae2free}
\end{figure*}

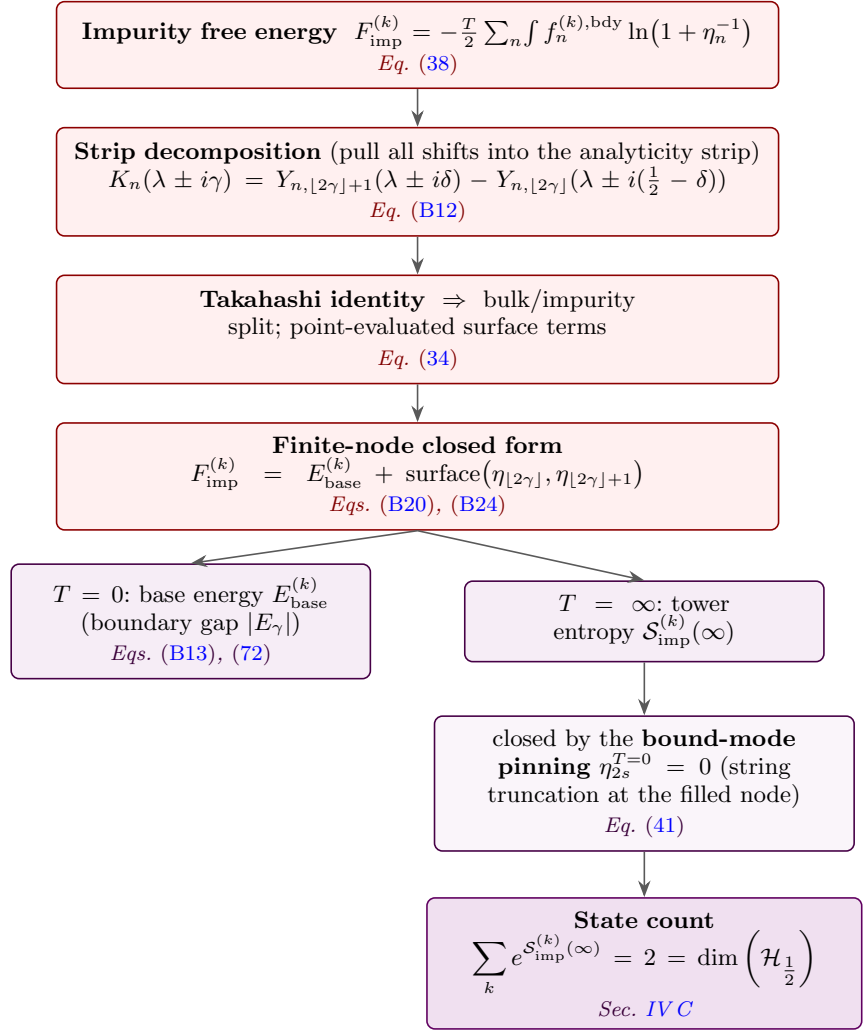
\begin{figure*}[tbp]
\centering
\begin{tikzpicture}[
  font=\small,
  >={Stealth[length=2.2mm]},
  node distance=5mm and 9mm,
  box/.style={draw=red!55!black, semithick, rounded corners=3pt, fill=red!6,
              align=center, text width=92mm, inner sep=5pt},
  term/.style={draw=violet!55!black, semithick, rounded corners=3pt, fill=violet!7,
              align=center, text width=44mm, inner sep=5pt},
  arr/.style={->, semithick, draw=black!65},
]
\node[box] (fimp) {\textbf{Impurity free energy}\ \ $F^{(k)}_{\mathrm{imp}}=-\tfrac{T}{2}\sum_{n}\!\int f_n^{(k),\mathrm{bdy}}\ln(1+\eta_n^{-1})$\\[1pt]{\footnotesize\itshape\color{red!50!black}Eq.~\eqref{F imp}}};
\node[box, below=of fimp] (strip) {\textbf{Strip decomposition} (pull all shifts into the analyticity strip)\\ $K_n(\lambda\pm i\gamma)=Y_{n,\fl{2\gamma}+1}(\lambda\pm i\delta)-Y_{n,\fl{2\gamma}}(\lambda\pm i(\tfrac12-\delta))$\\[1pt]{\footnotesize\itshape\color{red!50!black}Eq.~\eqref{eq:shift-id}}};
\node[box, below=of strip] (tak) {\textbf{Takahashi identity} $\ \Rightarrow\ $ bulk/impurity split; point-evaluated surface terms\\[1pt]{\footnotesize\itshape\color{red!50!black}Eq.~\eqref{Takahashi}}};
\node[box, below=of tak] (fin) {\textbf{Finite-node closed form}\\ $F^{(k)}_{\mathrm{imp}}=E^{(k)}_{\mathrm{base}}+\mathrm{surface}\big(\eta_{\fl{2\gamma}},\eta_{\fl{2\gamma}+1}\big)$\\[1pt]{\footnotesize\itshape\color{red!50!black}Eqs.~\eqref{eq:Fimp-BS}, \eqref{eq:Fimp-hBS}}};
\node[term] (t0) at ([xshift=-30mm,yshift=-12mm]fin.south) {$T=0$:\ base energy $E^{(k)}_{\mathrm{base}}$ (boundary gap $|E_\gamma|$)\\[1pt]{\footnotesize\itshape\color{violet!45!black}Eqs.~\eqref{eq:Estr}, \eqref{eq:Egamma}}};
\node[term] (tinf) at ([xshift=30mm,yshift=-12mm]fin.south) {$T=\infty$:\ tower entropy $\mathcal S^{(k)}_{\mathrm{imp}}(\infty)$};
\node[term, below=7mm of tinf, text width=52mm, fill=violet!4] (meth) {closed by the \textbf{bound-mode pinning} $\eta_{2s}^{T=0}=0$ (string truncation at the filled node)\\[1pt]{\footnotesize\itshape\color{violet!45!black}Eq.~\eqref{eta const p}}};
\node[term, below=6mm of meth, text width=54mm, draw=violet!75!black, fill=violet!12] (cnt) {\textbf{State count}\\ $\displaystyle\sum_{k}e^{\mathcal S^{(k)}_{\mathrm{imp}}(\infty)}=2=\dim\left(\mathcal{H}_{\tfrac12}\right)$\\[1pt]{\footnotesize\itshape\color{violet!45!black}Sec.~\ref{sec:EntropyPhases}}};
\draw[arr] (fimp)--(strip);
\draw[arr] (strip)--(tak);
\draw[arr] (tak)--(fin);
\draw[arr] (fin.south)--(t0.north);
\draw[arr] (fin.south)--(tinf.north);
\draw[arr] (tinf)--(meth);
\draw[arr] (meth)--(cnt);
\end{tikzpicture}
\caption{General logic, part 2: reducing the impurity free energy to the $T=0$ base energy and the $T=\infty$ tower entropy. Pulling the boundary shifts into the analyticity strip [Eq.~\eqref{eq:shift-id}] and applying the Takahashi identity [Eq.~\eqref{Takahashi}] yields a closed form that depends only on the two boundary nodes $\eta_{\fl{2\gamma}},\eta_{\fl{2\gamma}+1}$. The $T=\infty$ tower entropies, closed by the bound-mode pinning $\eta_{2s}^{T=0}=0$, sum to $\sum_k e^{\mathcal S^{(k)}_{\mathrm{imp}}(\infty)}=2$, the dimension of the spin-$\tfrac12$ impurity Hilbert space.}
\label{fig:free2ent}
\end{figure*}

Figures~\ref{fig:bae2free} and~\ref{fig:free2ent} summarize, in schematic form, the full derivation carried out in this appendix: the first traces the route from the Bethe Ansatz equations to the impurity free energy of a generic boundary tower, and the second its reduction to the $T=0$ base energy and the $T=\infty$ tower entropy. Only the driving step [Eq.~\eqref{f k}] is tower-dependent; the canonical $\eta$-system [Eq.~\eqref{TBA p}] is common to all towers.

\subsection{Preliminaries}
The BAE for rapidities $\{\mu_j\}$, derived in the Section~\ref{app:ABA}, read
\begin{equation}\label{eq:BAE-open}
\left(\frac{\mu_j-is}{\mu_j+is}\right)^{2N}
\frac{\mu_j-\frac{i}{2}-i\gamma}{\mu_j+\frac{i}{2}-i\gamma}
\frac{\mu_j-\frac{i}{2}+i\gamma}{\mu_j+\frac{i}{2}+i\gamma}
= \prod_{\substack{\ell=1\\ \ell\neq j}}^M
\frac{(\mu_j-\mu_\ell-i)(\mu_j+\mu_\ell-i)}{(\mu_j-\mu_\ell+i)(\mu_j+\mu_\ell+i)}.
\end{equation}
Introduce the standard phase functions
\[
\Theta_n(x)=2\arctan \frac{2x}{n},\qquad
K_n(x)=\frac{1}{2\pi}\frac{d\Theta_n}{dx}=\frac{1}{\pi}\frac{(n/2)}{(n/2)^2+x^2},
\]
and their fused combinations
\begin{align}
X_{nm}(x)&=\sum_{\ell=1}^{\min(n,m)}\Theta_{n+m+1-2\ell}(x),
\\
\Theta_{nm}(x)&=
\begin{cases}
\Theta_{|n-m|}(x)+2\Theta_{|n-m|+2}(x)+\cdots+2\Theta_{n+m-2}(x)+\Theta_{n+m}(x),
& n\neq m,\\[2pt]
2\Theta_2(x)+2\Theta_4(x)+\cdots+2\Theta_{2n-2}(x)+\Theta_{2n}(x),& n=m,
\end{cases}
\\
Y_{nm}(x)&=\frac{1}{2\pi}\frac{dX_{nm}}{dx}
=\sum_{\ell=1}^{\min(n,m)}K_{n+m+1-2\ell}(x).
\end{align}
We also use the convolution operator
\[
[n]\,f \equiv K_n * f,\qquad
A_{nm}=[|n-m|]+2[|n-m|+2]+\cdots+2[n+m-2]+[n+m].
\]

\subsection{String tower (no boundary strings)}\label{app:TBA:str}
Applying the string hypothesis, $n$-string centers $\{\lambda_j^{(n)}\}$ obey the log-BAE
\begin{equation}\label{eq:logBAE}
2N\,X_{n,2s}(\lambda_j^{(n)})+\Theta_n(\lambda_j^{(n)})
+\Theta_n(\lambda_j^{(n)}-i\gamma)+\Theta_n(\lambda_j^{(n)}+i\gamma)
=2\pi J_j^{(n)}+\sum_{m\ge 1}\sum_{i=1}^{\zeta_m}
\big[\Theta_{nm}(\lambda_j^{(n)}-\lambda_i^{(m)})
+\Theta_{nm}(\lambda_j^{(n)}+\lambda_i^{(m)})\big],
\end{equation}
with integers $J_j^{(n)}$.
Differentiating \eqref{eq:logBAE} and defining densities
\[
\frac{dJ^{(n)}}{d\mu}=\sigma_n(\mu)+\sigma_n^h(\mu),
\]
one obtains the linear integral relation
\begin{equation}\label{eq:sigma-hole}
\sigma_n^h(\mu)=f_n^{\rm str}(\mu)-\sum_{m\ge 1} A_{nm}\,\sigma_m(\mu),
\qquad
f_n^{\rm str}(\mu)=2N\,Y_{n,2s}(\mu)+K_n(\mu)+K_n(\mu + i\gamma)+K_n(\mu - i\gamma).
\end{equation}

\paragraph{Energy, magnetization, and free energy.}
The $n$-string energy is $\epsilon_n(\lambda)=-4\pi Y_{n,2s}(\lambda)$ (see eq.~\eqref{eq:E-appendix}),
so that
\[
E=-2\pi\sum_{n\ge1}\int d\lambda\,Y_{n,2s}(\lambda)\,\sigma_n(\lambda).
\]
At field $h$ the magnetization is
\(
S^z=\tfrac12+Ns-\tfrac12\sum_{n\ge1}n\int d\lambda\,\sigma_n(\lambda).
\)
The Yang–Yang entropy reads
\[
\mathcal{S}=\frac{1}{2}\sum_{n\ge1}\int d\lambda\,\Big[(\sigma_n+\sigma_n^h)\ln(\sigma_n+\sigma_n^h)
-\sigma_n\ln\sigma_n-\sigma_n^h\ln\sigma_n^h\Big].
\]
Hence, with $g_n(\lambda)=nh-4\pi Y_{n,2s}(\lambda)$,
\begin{equation}\label{eq:F-strings}
\mathcal{F}_{\rm str}=\frac{1}{2}\sum_{n\ge1}\int d\lambda\,
\Big[g_n\,\sigma_n
-T\,\sigma_n\ln(1+\eta_n^{-1})-T\,\sigma_n^h\ln(1+\eta_n)\Big]
-(Ns+\tfrac12)h,\qquad \eta_n\equiv\frac{\sigma_n^h}{\sigma_n}.
\end{equation}

\paragraph{TBA equations.}
Varying \eqref{eq:F-strings} under the constraint \eqref{eq:sigma-hole} yields
\begin{equation}\label{eq:preTBA}
\ln \big(1+\eta_n(\lambda)\big)=\frac{g_n(\lambda)}{T}
+\sum_{m\ge1} A_{nm} * \ln \big(1+\eta_m^{-1}\big)(\lambda).
\end{equation}
Acting with the standard kernel identity
\(
(\delta_{nm}-G*(\delta_{n+1,m}+\delta_{n-1,m}))A_{mk}=\delta_{nk},
\)
one arrives at the closed TBA system (hereafter $h=0$ unless stated):
\begin{equation}\label{eq:TBA}
{\;
\ln \eta_n(\lambda)
= -\frac{2\pi}{T}\,\frac{\delta_{n,2s}}{\cosh(\pi\lambda)}
+ G * \ln \big(1+\eta_{n+1}\big)(\lambda)
+ G * \ln \big(1+\eta_{n-1}\big)(\lambda).
\;}
\end{equation}
The large–$n$ boundary condition follows from $[n]K_\ell=K_{n+\ell}$:
\begin{equation}\label{eq:asymp}
\lim_{n\to\infty}\Big\{[n + 1]\ln(1+\eta_n)-[n]\ln(1+\eta_{n+1})\Big\}=-\frac{h}{T}.
\end{equation}

\paragraph{Impurity free energy of the string tower.}
At the stationary point, one finds the compact form
\begin{equation}\label{eq:Fimp-str}
{\;
\mathcal{F}_{\rm str}^{\rm imp}(T)
=-\frac{T}{2}\sum_{n\ge1} \int d\lambda\,
\big[K_n(\lambda + i\gamma)+K_n(\lambda - i\gamma)\big]\,
\ln \big(1+\eta_n^{-1}(\lambda)\big),
\;}
\end{equation}
with the bulk piece $\mathcal{F}_0$ omitted hereafter.

For later use, write $\gamma=\tfrac{\lfloor 2\gamma\rfloor}{2}+\delta$, $0\le\delta<\tfrac12$; then within the analyticity strip
\begin{equation}\label{eq:shift-id}
K_n(\lambda \pm  i\gamma)=
\sum_{r=0}^{\lfloor 2\gamma\rfloor} K_{n+r}(\lambda \pm  i\delta)
-\sum_{r=0}^{\lfloor 2\gamma\rfloor-1} K_{n+r}\Big(\lambda \pm  i\big(\tfrac12-\delta\big)\Big)
= Y_{n,\lfloor 2\gamma\rfloor+1}(\lambda \pm  i\delta)
- Y_{n,\lfloor 2\gamma\rfloor}\Big(\lambda \pm  i\big(\tfrac12-\delta\big)\Big).
\end{equation}
Using \eqref{eq:preTBA} this yields a base (zero–$T$) impurity energy
\begin{equation}\label{eq:Estr}
E_{\rm str}
=-2\pi\, G * \left[
Y_{\lfloor 2\gamma\rfloor+1,\,2s}(i\delta)
- Y_{\lfloor 2\gamma\rfloor,\,2s} \left(i\big(\tfrac12-\delta\big)\right)
\right],
\end{equation}
which can be expressed in closed form in terms of digamma functions when desired.

\paragraph{Zero- and high-temperature limits.}
At $T=0,\infty$ the ratios $\eta_n(\lambda)\to\eta_n^{T=0,\infty}$ are constants obeying
\begin{align}
\ln\eta_{2s}^{T=0}&=-\infty,\qquad
\ln\eta_n^{T=0}=\tfrac12\ln \big[(1+\eta_{n-1}^{T=0})(1+\eta_{n+1}^{T=0})\big],\\
\ln\eta_1^{T=\infty}&=\tfrac12\ln(1+\eta_2^{T=\infty}),\qquad
\ln\eta_n^{T=\infty}=\tfrac12\ln \big[(1+\eta_{n-1}^{T=\infty})(1+\eta_{n+1}^{T=\infty})\big],
\end{align}
with $\lim_{n\to\infty}\ln\eta_n^{T=0,\infty}/n=0$. The well-known solutions are
\begin{equation}\label{eq:eta-const}
\eta_{n<2s}^{T=0}=\left[\frac{\sin \frac{\pi(n+1)}{2s+2}}{\sin \frac{\pi}{2s+2}}\right]^2-1,\quad
\eta_{n\ge 2s}^{T=0}=(n+1-2s)^2-1,\qquad
\eta_{n}^{T=\infty}=(n+1)^2-1.
\end{equation}
Substituting \eqref{eq:eta-const} into \eqref{eq:Fimp-str} gives
\begin{align}
\mathcal{S}_{\rm str}^{\rm imp}(0)&=
\ln \frac{1+\eta_{\lfloor 2\gamma\rfloor+1}^{T=0}}{1+\eta_{\lfloor 2\gamma\rfloor}^{T=0}}
=\begin{cases}
\displaystyle \ln \frac{\sin \frac{\pi(\lfloor 2\gamma\rfloor+2)}{2s+2}}
{\sin \frac{\pi(\lfloor 2\gamma\rfloor+1)}{2s+2}}, & \gamma<s,\\[8pt]
\displaystyle \ln \frac{\lfloor 2\gamma\rfloor+2-2s}{\lfloor 2\gamma\rfloor+1-2s}, & \gamma\ge s,
\end{cases}
\\
\mathcal{S}_{\rm str}^{\rm imp}(\infty)&=
\ln \frac{1+\eta_{\lfloor 2\gamma\rfloor+1}^{T=\infty}}{1+\eta_{\lfloor 2\gamma\rfloor}^{T=\infty}}
=\ln \frac{\lfloor 2\gamma\rfloor+2}{\lfloor 2\gamma\rfloor+1}.
\end{align}

\subsection{Fundamental boundary-string tower ($\gamma>1/2$)}\label{app:TBA:BS}
For $\gamma>\tfrac12$ the BAE admit the impurity-dependent root
$\mu_\gamma=i(\gamma-\tfrac12)$ (fundamental boundary string). Including it shifts the driving term in \eqref{eq:sigma-hole}:
\begin{equation}\label{eq:sigma-hole-BS}
\sigma_n^h(\mu)=f_n^{\rm BS}(\mu)-\sum_{m\ge1}A_{nm}\sigma_m(\mu),\quad
f_n^{\rm BS}(\mu)=2N\,Y_{n,2s}(\mu)+K_n(\mu)-\sum_{v=\pm}K_n\big(\mu+iv(\gamma-1)\big).
\end{equation}
Thus the impurity free energy becomes
\begin{equation}\label{eq:Fimp-BS}
{\
\mathcal{F}_{\rm BS}^{\rm imp}(T)=-4\pi\,K_{2s} \left(i(\gamma-\tfrac12)\right)
+\frac{T}{2}\sum_{n\ge1}\int d\lambda\sum_{v=\pm}
K_n\big(\lambda+iv(\gamma-1)\big)\ln \big(1+\eta_n^{-1}(\lambda)\big).
\ }
\end{equation}
Using \eqref{eq:shift-id} one can write its zero–$T$ base energy $E_{\rm BS}$ in the same $G*Y$ form as \eqref{eq:Estr}; in particular $E_{\rm BS}=E_{\rm str}$ for $\gamma<s$ or $\gamma>s+1$, while for $s<\gamma<s+1$,
\(
E_{\rm BS}=E_{\rm str}-\tfrac{2\pi}{\cos\pi(\gamma-s)}.
\)
The tower entropies follow by substituting \eqref{eq:eta-const} into \eqref{eq:Fimp-BS}:
\begin{align}
\mathcal{S}_{\rm BS}^{\rm imp}(0)&=
\begin{cases}
\displaystyle \ln \frac{\sin \frac{\pi(\lfloor 2|\gamma-1|\rfloor+1)}{2s+2}}
{\sin \frac{\pi(\lfloor 2|\gamma-1|\rfloor+2)}{2s+2}}, & \gamma<s+1,\\[8pt]
\displaystyle \ln \frac{\lfloor 2\gamma\rfloor-1-2s}{\lfloor 2\gamma\rfloor-2s}, & \gamma\ge s+1,
\end{cases}
\\
\mathcal{S}_{\rm BS}^{\rm imp}(\infty)&=
\begin{cases}
\ln \tfrac12, & \gamma\in(\tfrac12,\tfrac32),\\[2pt]
\displaystyle \ln \frac{\lfloor 2\gamma\rfloor-1}{\lfloor 2\gamma\rfloor}, & \gamma\ge \tfrac32.
\end{cases}
\end{align}

\subsection{Higher-order boundary-string tower ($\gamma>1$)}\label{app:TBA:hBS}
For $\gamma>1$ there exist higher-order boundary strings
\(
\mu^{(m)}_{\gamma,l}=i(\gamma-\tfrac12-l),\ l=0,\dots,m,\ 
m=\lfloor \gamma+\tfrac12\rfloor
\)
\cite{wang2015off}. Including these roots changes the driving term to
\begin{equation}\label{eq:drv-hBS}
f_n^{\rm hBS}(\mu)=2N\,Y_{n,2s}(\mu)+K_n(\mu)
-\Bigg[
2\sum_{k=1}^{m}\sum_{v=\pm}K_n\big(\mu+iv(\gamma-k)\big)
+ \sum_{v=\pm}K_n\big(\mu+iv(\gamma-m-1)\big)
\Bigg],
\end{equation}
leading to the impurity free energy
\begin{equation}\label{eq:Fimp-hBS}
{\
\mathcal{F}_{\rm hBS}^{\rm imp}(T)=
-4\pi\sum_{k=0}^{m}K_{2s} \left(i\big(\gamma-\tfrac12-k\big)\right)
+\frac{T}{2}\sum_{n\ge1} \int d\lambda
\sum_{v=\pm} \Bigg[
2\sum_{k=1}^{m}K_n\big(\lambda+iv(\gamma-k)\big)
+K_n\big(\lambda+iv(\gamma-m-1)\big)
\Bigg]\ln \big(1+\eta_n^{-1}\big).
\ }
\end{equation}
Using the same shift/Takahashi manipulations as above, one obtains a compact $G*Y$ expression for the base energy $E_{\rm hBS}$ and, from \eqref{eq:eta-const}, the limiting entropies
\begin{align}
\mathcal{S}_{\rm hBS}^{\rm imp}(0)&=
\begin{cases}
\displaystyle \ln \frac{\sin^2 \frac{\pi}{2s+2}}{\sin \frac{\pi\lfloor 2\gamma\rfloor}{2s+2}\,
\sin \frac{\pi(\lfloor 2\gamma\rfloor+1)}{2s+2}}, & \gamma<s,\\[8pt]
\displaystyle \ln \frac{1}{(\lfloor 2\gamma\rfloor+1-2s)(\lfloor 2\gamma\rfloor-2s)}, & \gamma\ge s,
\end{cases}
\\
\mathcal{S}_{\rm hBS}^{\rm imp}(\infty)&=\ln \frac{1}{\lfloor 2\gamma\rfloor(\lfloor 2\gamma\rfloor+1)},\qquad \gamma>1.
\end{align}

\medskip
\noindent
Equations \eqref{eq:TBA}, together with the tower-specific driving terms
\eqref{eq:sigma-hole}–\eqref{eq:drv-hBS} and the analyticity/shift identity
\eqref{eq:shift-id}, provide a complete, self-contained TBA framework for all three towers. The total impurity free energy follows from the tower decomposition $e^{-\beta F_{\rm imp}}=\sum_k e^{-\beta \mathcal{F}^{\rm imp}_{(k)}}$, yielding the phase-dependent thermodynamics discussed in the main text.

\end{document}